\newcommand{\mycomment}[1]{}
\documentclass[]{agujournal2019}
\usepackage{amsmath,amssymb}
\usepackage{graphicx}
\usepackage{epstopdf}
\usepackage{lineno}
\usepackage{array}

\journalname{Water Resources Research}

\begin{document}

%
%

\title{Stochastic ecohydrological perspective on semi-distributed rainfall-runoff dynamics}


%

%
%

\authors{Mark S. Bartlett\affil{1}, Elizabeth Cultra\affil{2}, Nathan Geldner\affil{1}, Amilcare Porporato\affil{2}}



\affiliation{1}{The Water Institute, Baton Rouge, Louisiana, USA.}

\affiliation{2}{Department of Civil, Construction, and Environmental Engineering,
Princeton University, Princeton, New Jersey USA.}





\correspondingauthor{Mark S. Bartlett}{Mark.Bartlett@gmail.com}

%
%


\begin{keypoints}
\item The model unifies stochastic ecohydrology, semi-distributed hydrology, and the SCS-CN method.
\item By merging the SCS-CN method with stochastic ecohydrology, antecedent conditions in SCS-CN are linked to hydroclimate. 
\item Calibration to 81 USGS gages shows the model accurately captures runoff variability and water balance aligned with Budyko-type curves.

\end{keypoints}


%
%


\begin{abstract}

Quantifying watershed process variability consistently with climate change and ecohydrological dynamics remains a central challenge in hydrology. Stochastic ecohydrology characterizes hydrologic variability through probability distributions that link climate, hydrology, and ecology. However, these approaches are often limited to small spatial scales (e.g., point or plot level) or focus on specific fluxes (e.g., streamflow), without accounting for the entire water balance at the basin scale. While semi-distributed models account for spatial heterogeneity and upscaled hydrologic fluxes, they lack the analytical simplicity of stochastic ecohydrology or the SCS-CN method and, perhaps more importantly, do not integrate the effects of past random variability in hydroclimatic conditions. This hinders an efficient characterization of hydrological statistics at the watershed scale. To overcome these limitations, we merge stochastic ecohydrology, the spatial upscaling of semi-distributed modeling, and the SCS-CN rainfall-runoff partitioning. The resulting unified model analytically characterizes watershed ecohydrological and hydrological statistics using probability density functions (PDFs) that are functions of climate and watershed attributes---something unattainable with the Monte Carlo methods of traditional stochastic hydrology. Calibrated across 81 watersheds in Florida and southern Louisiana, the model PDFs precisely capture the long-term average water balance and runoff variance, as well as the runoff quantiles with a median normalized Nash-Sutcliffe (NNSE) efficiency of 0.95. These results also advance the SCS-CN method by providing an analytical PDF for the Curve Number (CN), explicitly linked to climate variables, baseflow, and the long-term water balance partitioning described by the Budyko curve. 

\end{abstract}

\section{Introduction}

Efficient and accurate statistical characterization of watershed hydrology is crucial for applications ranging from national-scale flood mapping to localized engineering design and long-term hydroclimate forecasting \cite{beven2012rainfall}. Traditional hydrological approaches---grounded in continuous-time simulations of historical or idealized rainfall scenarios---have proven effective in many contexts \cite{grimaldi2021continuous}. Yet, these models operate largely in isolation from ecohydrological frameworks that prioritize probabilistic dynamics over deterministic predictions \cite{rodrigueziturbe2004ecohydrology,porporato2004soil,farmer2016deterministic,beven2021issues,porporato2022ecohydrology}. Although ecohydrological models have advanced our understanding of long-term trends, risks, and the intricate interplay of climate, ecology, and hydrology \cite{porporato2022ecohydrology}, they are often limited to localized scales, particularly for soil moisture dynamics \cite{rodrigueziturbe2004ecohydrology,rigby2006simplified,porporato2001plants}.

Efforts to upscale ecohydrological insights to watershed-scale fluxes---such as runoff and evapotranspiration---have only focused on a specific aspect of the water balance. Streamflow and runoff statistics, for example, have been successfully captured within stochastic ecohydrology frameworks, yet the broader characterization of watershed-scale processes remains incomplete \cite{botter2007basin,botter2007signatures,bartlett2014excess}. This reveals a fundamental divide: traditional hydrology lacks the probabilistic rigor inherent in stochastic ecohydrology, while ecohydrology lacks the spatial resolution and operational applicability of conventional hydrological methods.

Historically, hydrology models navigate this divide by simulating state variables over time, leveraging historical or synthetic rainfall data generated via Monte Carlo sampling \cite{beven2012rainfall,beven2021issues}. These models typically employ one of three approaches: (1) an event-based, spatially lumped representation of runoff and infiltration where the temporal variability is subsumed by a prescribed antecedent condition (e.g., the SCS-CN method); (2) semi-distributed models (e.g., the VIC model); or (3) fully distributed models for an explicit spatial representation \cite{beven2012rainfall, wood1992land, moore1985probability, beven1979physically, semenova2015barriers, freeze1969blueprint}. Semi-distributed and event-based methods often expedite simulations, allowing for multiple realizations of synthetic rainfall and the corresponding hydrological fluxes \cite{beven2012rainfall, kavetski2003semidistributed}. However, the aggregated semi-distributed model outputs do not directly express the complex interplay between climate, ecology, storm events, and antecedent conditions, nor are they easily calibrated with observed statistics \cite{bloschl2010climate,duethmann2020does}. When computational efficiency is prioritized, simplified approaches such as the Soil Conservation Service (now the Natural Resources Conservation Service) curve number (SCS-CN) method often are used with median or average antecedent conditions derived from historical data \cite{ponce1996runoff,nrcs2004national}. However, such antecedent conditions neglect climate nonstationarity, leading to potential inaccuracies when the method is applied under evolving climate conditions \cite<e.g.>{pathiraja2012continuous}.

A key limitation of existing approaches, such as the SCS-CN method, is that their assumed average or median conditions often fail to align with the true average and median conditions dictated by hydroclimatic drivers, whether for antecedent conditions or for projecting meaningful future scenarios \cite{pathiraja2012continuous, bloschl2010climate}. Antecedent conditions—critical determinants of infiltration, runoff, and soil moisture dynamics—are often oversimplified in traditional models, relying on static or average representations that fail to capture their probabilistic nature \cite{hawkins2014curve, zehe2004predictability}. This inconsistency weakens the ability to link hydrological responses to climate variability and extremes \cite{bloschl2010climate, koutsoyiannis2020revisiting}. Moreover, the lack of a probabilistic framework limits the capacity to evaluate risks and adaptive design  strategies under changing hydroclimatic conditions \cite<e.g.,>{kim2019assessment, wasko2020changes}. Addressing these gaps requires a statistical approach that aligns antecedent conditions and hydroclimate with stochastic descriptions of rainfall and ecological fluxes. Such an approach also would provide a coherent foundation for both historical hydrologic analysis and future hydrologic projections \cite<e.g.,>{dooge1986looking, koutsoyiannis2020revisiting}..

To address these challenges, we propose a synthesis of traditional hydrology and stochastic ecohydrology. By reformulating semi-distributed models with equations where the unknowns are statistical properties, we derive mathematical expressions that link hydrological statistics directly to model parameters and climate characteristics. This reformulation obviates the need for computationally intensive simulations, offering a direct analytical pathway to explore hydrological variability while preserving the complexity of interactions among rainfall events, infiltration, runoff, and ecological fluxes \cite{dooge1986looking}.

Central to this framework is the modeling of rainfall as a random point process, paired with semi-distributed event-based runoff and infiltration representations, such as the SCS-CN method or analogous formulations. In turn, this statistical description of the storm event is coupled with continuous inter-storm ecohydrological fluxes (e.g., evapotranspiration) to create a complete statistical description of the system that addresses key limitations of traditional methods. The resulting model provides analytical probability distributions for antecedent conditions that are connected to both watershed hydrology variables and climate variables, including the Budyko dryness index. 

The statistical model paradigm is demonstrated using a two-layer soil watershed model (Section 2), which reinterprets the SCS-CN method in terms of hydroclimatic dynamics (Section 3). In Section 4, we integrate the SCS-CN method with Budyko-type curves, bridging the long-term hydroclimatic water balance with event-based runoff assessments. While this statistical model represents the steady-state conditions, we show that it reasonably captures a full description of the time-integrated long-term statistics (Section 5). In turn, calibrating this long-term statistical description to United States Geological Survey (USGS) gage data (Section 6) illustrates the framework's applicability, setting the stage for a discussion (Section \ref{sec:discussion}) on advancing engineering hydrology and probabilistic description of watershed process variability.

\begin{figure*}
\centering
\includegraphics[width=1.0\linewidth]{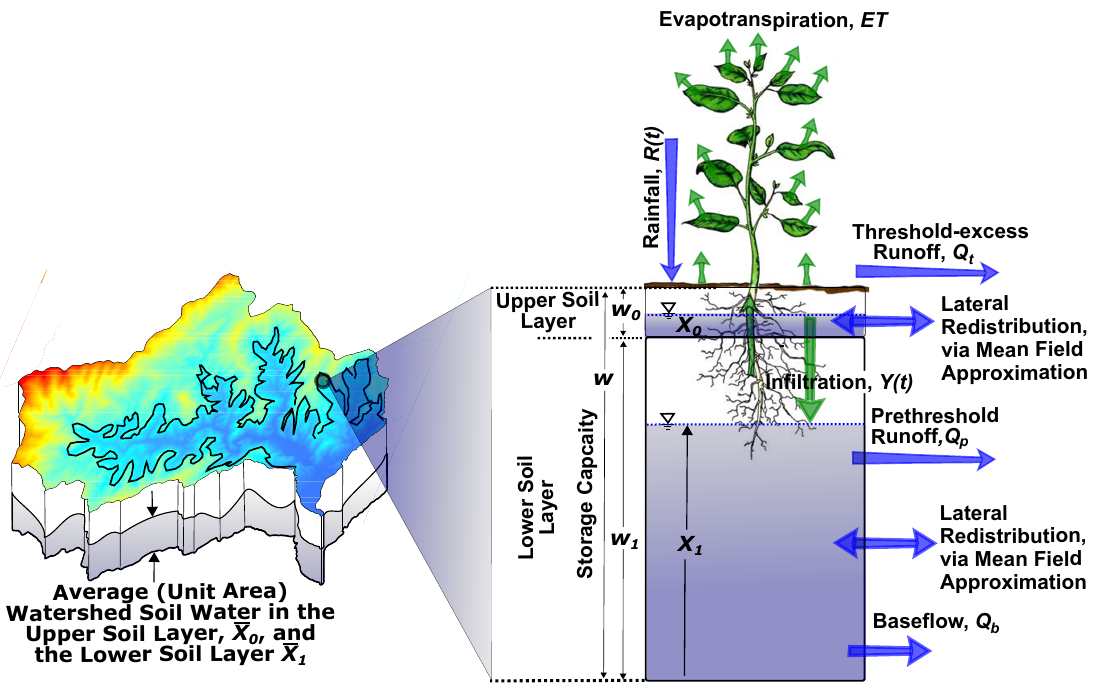}%
\label{fig:1 - upscaling}
\caption{The model describes the statistical water balance dynamics of unit-area (watershed scale) soil moisture for upper and lower soil layers, denoted as $\overline{\chi}_0$ and $\overline{\chi}_1$. These dynamics are derived by upscaling point-scale processes to unit-area averages for evapotranspiration, runoff, infiltration, and baseflow using mean-field approximations. In the upper soil layer, spatially heterogeneous rainfall redistributes over variable storage capacity, while in the lower soil layer, the spatial variability of soil moisture  and storage is described by a PDF following a semi-distributed model, such as VIC, SCS-CN, or TOPMODEL.}
\end{figure*}

\section{Model Structure}
\label{sec:model_description}

The model structure is similar to that of traditional semi-distributed hydrology models, retaining well-established principles and deterministic components. Similar to many hydrology models, it includes two soil layers---upper and lower---to capture interactions between rainfall, soil moisture, and runoff (Fig. \ref{fig:1 - upscaling}). For each soil layer, point-scale variables and fluxes are upscaled to their corresponding basin-scale unit-area counterparts, which then evolve in time (Fig. \ref{fig:1 - upscaling}). What distinguishes this model is its innovative probabilistic framework, which provides probability density functions (PDFs) for both antecedent soil moisture and runoff at the basin scale on a unit-area average basis. The framework further links these basin-scale averages (denoted by an overbar) to point-scale spatial heterogeneity, which is also represented by PDFs (see Table \ref{tab:vars_params}). This linkage, as detailed in Eq. (\ref{eq:meanfield}) in Section \ref{sec:meanfield}, allows the model PDFs, which describe basin-scale variability, to remain statistically consistent with both point-scale ecohydrological statistics of soil moisture and the long-term water balance constrained by hydroclimatic conditions.

\begin{table}
\begin{minipage}{\textwidth}
\linespread{.6}\selectfont
\caption{Variables and Parameters$^a$ 
\label{tab:vars_params} }
\begin{tabular}{c|l}
\hline
\noalign{\vskip 0.04in}
\textbf{Symbol} & \textbf{Description} \\
\hline
\noalign{\vskip 0.04in}
$\overline{\alpha}$ & Average storm total rainfall (on a unit area basis) \\
$\overline{\alpha}_0$ & Average storm total rainfall parameter of the PDF of Eq. (\ref{eq:pzm}) \\
$\overline{\alpha}_*$ & Average storm total rainfall parameter  of the PDF of Eq. (\ref{eq:pzm}) \\
$\beta$ & Fraction of watershed with nonzero prethreshold runoff\\
$\overline{\gamma}$ & Storage index, $\overline{\gamma}=\overline{w}/\overline{\alpha}$ \\
$\overline{\gamma}_1$ & Lower soil layer storage index, $\overline{\gamma}_1 = \overline{\gamma}(1-\mu)$ \\
$\vartheta$ & Fraction of dynamic soil moisture, $\overline{\chi}_1$; see Eqs. (\ref{eq:ConsistencyCondition1}) and (\ref{eq:vartheta})\\
$\lambda$ & Frequency of rainfall \\
$\mu$ & Fraction of storage in the upper soil layer \\
$\sigma^2_{(\cdot)}$ & Ensemble variance where $_{(\cdot)}$ is a variable placeholder \\
$\chi_0$ & Upper soil layer effective relative soil moisture at a point \\
$\chi_1$ & Lower soil layer effective relative soil moisture at a point \\
$\omega$ & Weight parameters of rainfall mixed exp. PDF of Eq. (\ref{eq:pzm}) \\
$\omega_\lambda$ & Censoring factor controlling the frequency of infiltration to the lower soil layer \\
$\omega_{PET}$ & Lower soil layer PET adjustment factor, $\omega_{PET} =1 - \overline{\chi}_0$ \\
$B_I$ & Baseflow index, $B_I = \frac{\overline{Q}_{b,max}}{\overline{\alpha}\lambda}$ \\
$CN$ & Curve number of Eq. (\ref{eq:CN}) \\
$CN_{\langle\overline{S}_1\rangle}$ & Curve number of Eq. (\ref{eq:CN}) based on $\langle\overline{S}_1\rangle$ \\
$D_I$ & Dryness index, $D_I = \overline{PET}/\overline{\alpha}\lambda$ \\
$ET$ & Evapotranspiration  \\
$ET_0$ & Upper soil layer evapotranspiration at a point  \\
$ET_1$ & Lower soi layer evapotranspiration at a point  \\
$F_t$ & Fraction of watershed with threshold excess runoff \\
$\overline{I}$ & Initial abstraction of the SCS-CN method,  $\overline{I} =\overline{w}\mu(1-\overline{\chi}_0)$ \\
$L_I$ & Loss index, $L_I = \frac{D_I \langle \omega_{PET}\rangle + B_I}{\langle \omega_\lambda \rangle}$ \\
$N$ & Normalization constant for PDF of  $\overline{\chi}_1$\\
$N_0$ & Normalization constant for PDF of $\overline{\chi}_0$ \\
$n$ & Vertically averaged soil porosity \\
PET &  Potential evapotranspiration \\
$P_I$ & Prethreshold runoff index, $P_I =\beta \overline{u}_1$ \\
$Q$ & Storm event runoff depth at a point \\
$Q_1$ & Lower soil layer storm event runoff depth at a point \\
$\overline{Q}_b$ & Baseflow per unit watershed area \\
$\overline{Q}_{b, max}$ & Maximum baseflow when $\overline{\chi}_1 = 1$ \\
$\overline{Q}_f$ & Stream flow per unit watershed area \\
$R$ & Storm event rainfall depth at a point \\
$\overline{S}_1$ & Maximum potential retention of the SCS-CN method \\
$s$ & Relative soil moisture at a point \\
$s_1$ & Relative soil moisture at saturation at a point \\
$s_w$ & Relative soil moisture at the wilting point \\
$t$ & Time \\
$u_1$ & Lower layer soil moisture antecedent to storm event at a point \\
$w$ &  Water storage capacity depth at a point \\
$w_0$ &  Upper soil layer water storage capacity depth at a point \\
$w_1$ &  Lower soil layer water storage capacity depth at a point \\
$Y$ & Storm event infiltration depth to the lower soil layer at a point \\
$Z_r$ & Root zone depth \\
$Z_b$ & Depth from the bottom of the upper soil layer to bedrock \\
\noalign{\vskip 0.04in}
\hline
\end{tabular}
\\
$^a$ Variables in the text with an overline bar indicate a spatial average (unit area) depth value, e.g., for the point rainfall depth, $R$, the average (unit area) value is denoted by $\overline{R}$, while angle brackets indicate an ensemble average (over the ensemble of rainfall forcing), e.g., for the spatial average rainfall, the ensemble average is $\langle\overline{R}\rangle$.
\end{minipage}
\end{table}

For the upper soil layer, each point represents an ``effective" relative soil moisture, $\chi_0 = (s - s_w)/(s_1 - s_w)$, where $s_w$ is the residual soil moisture at the wilting point, and $s_1$ is a value between the field capacity (the soil moisture level where drainage becomes negligible) and complete saturation ($s = 1$).  The total available storage capacity in the upper soil layer is $w_0 = n Z_r (s_1 - s_w)$, where $n$ is the vertically averaged soil porosity, and $Z_r$ is a depth within the rooting zone \cite{porporato2004soil}.  Similarly, for the lower soil layer, each point also represents an ``effective" relative soil moisture, $\chi_1 = (s - s_0)/(1 - s_0)$, where $s_0$ lies between the field capacity and the wilting point. The total storage capacity is $w_1 = n Z_b (1 - s_0)$, where $Z_b$ is the depth from the bottom of the upper soil layer to bedrock. For each point, the total effective storage capacity is $w = w_0 + w_1$, where the parameter $\mu$ partitions $w$ such that $w_0 = \mu w$ and $w_1 = w(1 - \mu)$. The upscaled, (unit-area) average soil moistures for the upper and lower layers are represented by $\overline{\chi}_0$ and $\overline{\chi}_1$, respectively. Similarly, the unit-area storage depth, $\overline{w}$, is divided into contributions from upper soil layer, $\overline{w}\mu$, and the lower soil layer, $\overline{w}(1 - \mu)$, with the latter primarily controlling runoff (Table \ref{tab:vars_params}). 

\begin{figure*}
\centering
\includegraphics[width=2.75in]{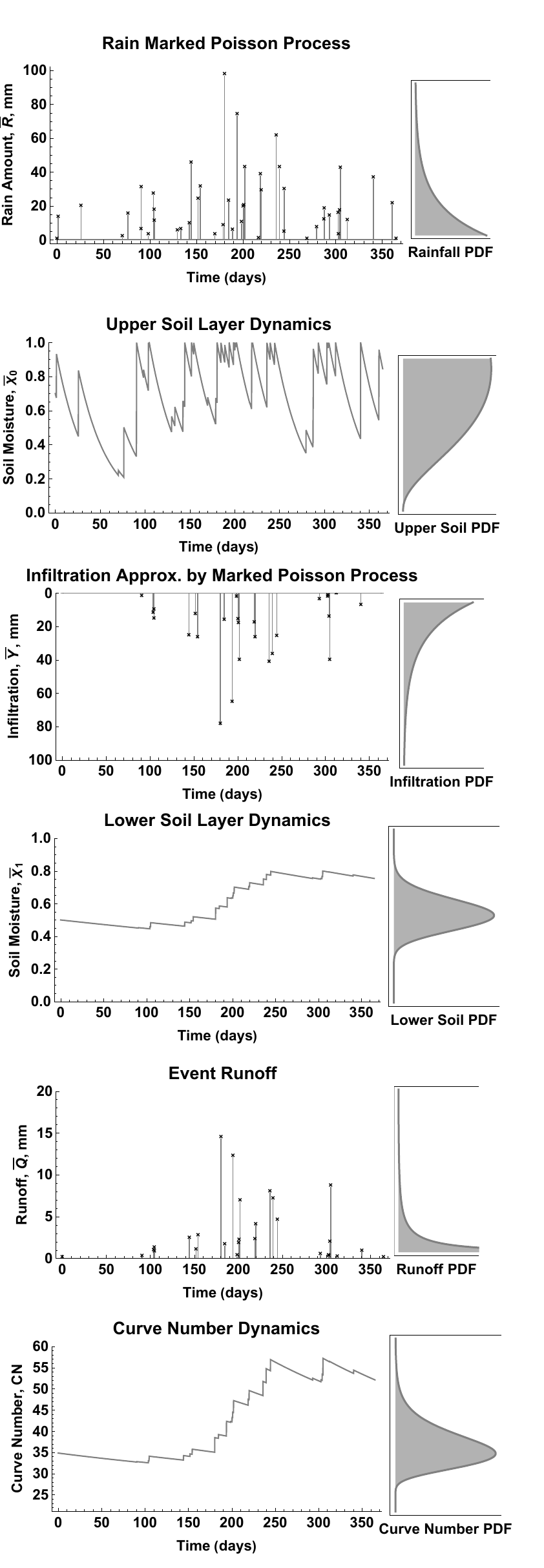}%
\label{fig:2 - process}
\caption{Sequential outline of the model over a 365-day simulation: (1) rainfall inputs, (2) upper soil layer dynamics with continuous evapotranspiration losses and (3) infiltration loses to (4) the lower soil layer with evapotranspiration and baseflow dynamics, and (5) event-based runoff magnitude described by (6) a dynamically varying Curve Number (CN) influenced by lower watershed moisture content.}
\end{figure*} 

Here, the upper soil layer follows minimalist stochastic ecohydrological PDF solutions \cite{milly1993analytic,porporato2004soil,porporato2022ecohydrology}, while the lower soil layer follows a novel formulation that accounts for the rainfall-runoff partitioning typical of semi-distributed models \cite{bartlett2018state}.  This PDF solution is based on an event-based description of semi-distributed model runoff \cite{bartlett2015unified2,bartlett2017reply}.

As outlined in Fig. \ref{fig:2 - process} on a basin average scale, the key processes and assumptions underlying the model include:
\begin{itemize} 
\item Rainfall modeling: Rainfall is treated as an event-based marked Poisson process input to the upper soil layer. 
\item Upper layer dynamics: The upper soil layer initially abstracts the rainfall, similar to the SCS-CN method, and redistributes this spatially variable rainfall within this layer until it reaches saturation. This redistribution is a mean-field approximation, as it is typical in statistical physics (see the next Section). 
\item Percolation from upper layer to lower layer: this intermittent process is approximated as a marked Poisson process having a frequency of occurrence linked to the upper layer stochastic dynamics \cite<e.g.,>{botter2007basin}.
\item Lower layer dynamics: Infiltration from upper layer increases soil moisture and initiates runoff; between storms, soil moisture is redistributed around the spatial average, following another mean-field approximation \cite<e.g.,>{bartlett2015unified2}. 
\item Analytical distributions: A probabilistic analysis yields analytical distributions (i.e., PDFs) for upper and lower soil layer soil moistures and for the hydrological fluxes such as evapotranspiration, baseflow, and runoff. 
\item CN and initial abstraction: For engineering hydrology, a CN and initial abstraction are obtained from the probabilistic description, e.g., median values, as well as 25th and 75th quantiles for respective dry and wet conditions.
\end{itemize}
As described in the next section, the probabilistic framework employs at several stages the mean-field approximation, so that the model translates spatially heterogeneous rainfall and infiltration into self-consistent spatial average fluxes, preserving local variability within the broader system dynamics.

\subsection{Mean-Field Approximation for Point Process Upscaling}
\label{sec:meanfield}
The point process upscaling approach leverages a mean-field approximation, which is commonly used in statistical physics \cite{bender2003self,aoki2014nonequilibrium} and describes the collective behavior of individual points represented by an average state \cite{Hotta2003}. The mean-field approximation---an emergent behavior resulting from interactions between local point values and the average field condition---allows the dynamics of the variables $\chi$ in the soil layers to be described in terms of the macroscopic (spatial averages), as derived from point-based values as follows:

\begin{align}
\overline{\xi}\left(\dots\right) = \iiint \xi(\dots) \,  p_{R}(R)p_{\chi w}( \chi, w \mid \overline{\chi}) \, d\chi \, dw \, dR,
\label{eq:meanfield}
\end{align}
where $p_{R}(R)p_{\chi w}(\chi, w \mid \overline{\chi})$ describes spatial variability in rainfall $R$ (or infiltration $Y$), soil moisture $\chi$, and storage capacity $w$, dependent on the average soil moisture $\overline{\chi}$. The PDF $p_{R}(R)p_{\chi w}(\chi, w \mid \overline{\chi})$ may be expanded to other variables and parameters such as potential evapotranspiration ($\text{PET}$).

When the upper soil layer is saturated such that $\overline{R}>\overline{w}\mu(1-\overline{\chi}_0)$, the excess rainfall becomes infiltration to the lower soil layer. When upscaling the upper soil layer fluxes with Eq. (\ref{eq:meanfield}),  a variety of assumptions for $p_{\chi_0w_0}(\chi_0, w_0 \mid \overline{\chi}_0)$ may be accommodated for a given semi-distributed model. For the SCS-CN model, $p_R(R)$ follows an exponential PDF, i.e., $p_R(R)=\frac{1}{\overline{R}}e^{-R/\overline{R}}$, so that infiltration (to the lower soil layer) also follows an exponential PDF, i.e., $p_Y(Y)=\frac{1}{\overline{Y}}e^{-Y/\overline{Y}}$ where  $\overline{Y} = \overline{R} -\overline{w}\mu(1-\overline{\chi}_0)$. The fact that infiltration follows an exponential PDF arises from the re-scaling property of the exponential distribution \cite<e.g.,>{bartlett2014excess}.

Once water infiltrates the lower soil layer, the spatial variability of soil moisture and storage is described by the PDF $p_{\chi_1 w_1}(\chi_1, w_1|\overline{\chi}_1)$ within the integrand of Eq. (\ref{eq:meanfield}). The dependence of this PDF on $\overline{\chi}_1$ represents another mean-field approximation that provides consistency with the model’s macroscopic behavior. This explicit dependence on the average is a common assumption with many semi-distributed models \cite{bartlett2015unified2}. When considering Eq. (\ref{eq:meanfield}) upscaling in the context of the SCS-CN method, the PDF of infiltration, $p_{Y}(Y)$, follows an exponential distribution, while the PDF $p_{\chi_1 w_1}(\chi_1, w_1|\overline{\chi}_1)$ is represented as the product of an exponential PDF for the storage capacity depth and a Dirac delta function indicating that each point value of soil moisture equals the spatial average, i.e., $p_{\chi_1 w_1}(\chi_1, w_1|\overline{\chi}_1)=\frac{1}{\overline{w}_1}e^{-w_1/\overline{w}_1}\delta(\chi_1-\overline{\chi}_1)$. 

\subsection{Upper Soil Layer Soil Moisture}
\label{sec:upperlayer}
The upper soil layer censors the rainfall input to the lower soil layer. This layer is known as the ``thin upper layer" in the VIC model and the ``root zone layer" in TOPMODEL. In both models, soil moisture in the upper layer is reduced by evapotranspiration, while water drains to the lower layer either as a moisture-dependent flux or when infiltration exceeds the upper layer's storage capacity. Similarly, the SCS-CN method has an initial abstraction layer that must saturate before runoff begins \cite{beven1979physically,wood1992land,beven2012rainfall,liang1994simple,ponce1996runoff}.

In semi-distributed models, the spatial variability of the upper soil is typically not defined. However, most formulations implicitly rely on  the previously mentioned mean-field approximation for infiltration to the lower soil layer. In this framework, we assume that evapotranspiration (at a point) decreases linearly with soil moisture. Consequently, when upscaled to a unit-area basis using Eq. (\ref{eq:meanfield}), the model again predicts a linear reduction in evapotranspiration, irrespective of the specific distribution of $p_{\chi_0 w_0}(\chi_0, w_0|\overline{\chi}_0)$. As a result, the upscaled water balance mirrors that of minimalist ecohydrological models \cite{porporato2022ecohydrology}, which focuses on fluxes for deep percolation infiltration (to a lower soil layer) and evapotranspiration \cite{porporato2004soil,milly1993analytic}. In these models, the water balance is expressed as:

\begin{align}
\overline{w}\mu\frac{d\overline{\chi}_0(t)}{dt}=\overline{R}(t)-\overline{ET}_0[\overline{\chi}_0(t)]-\overline{Y}[\overline{R}(t),\overline{\chi}_0(t)],
\label{eq:dx0/dt}
\end{align}
with soil moisture increasing with rainfall, $\overline{R}(t)$, and decreasing from evapotranspiration, $\overline{ET}_0[\overline{\chi}_0(t)]$, and infiltration to the lower soil layer, $\overline{Y}[\overline{R}(t),\overline{\chi}_0(t)]$. Following \citeA{porporato2004soil}, evapotranspiration decreases linearly with soil moisture, and deep percolation (i.e., infiltration) to the lower soil layers occurs only when storm rainfall exceeds the upper layer’s spare storage, $\overline{w}\mu(1-\overline{\chi}_0)$.  We consider rainfall on an event basis, i.e., $\overline{R}(t) = \sum_{i=1}^{N(t)} \overline{R}  \: \delta(t-t_i)$, where as indicated by the Dirac $\delta$ function, $\delta(\cdot)$, rainfall $\overline{R}$ is instantaneous at the storm event arrival times $\{t_i \}(i = 1, 2,...)$.

\begin{figure*}
\centering
\includegraphics[width=.9\linewidth]{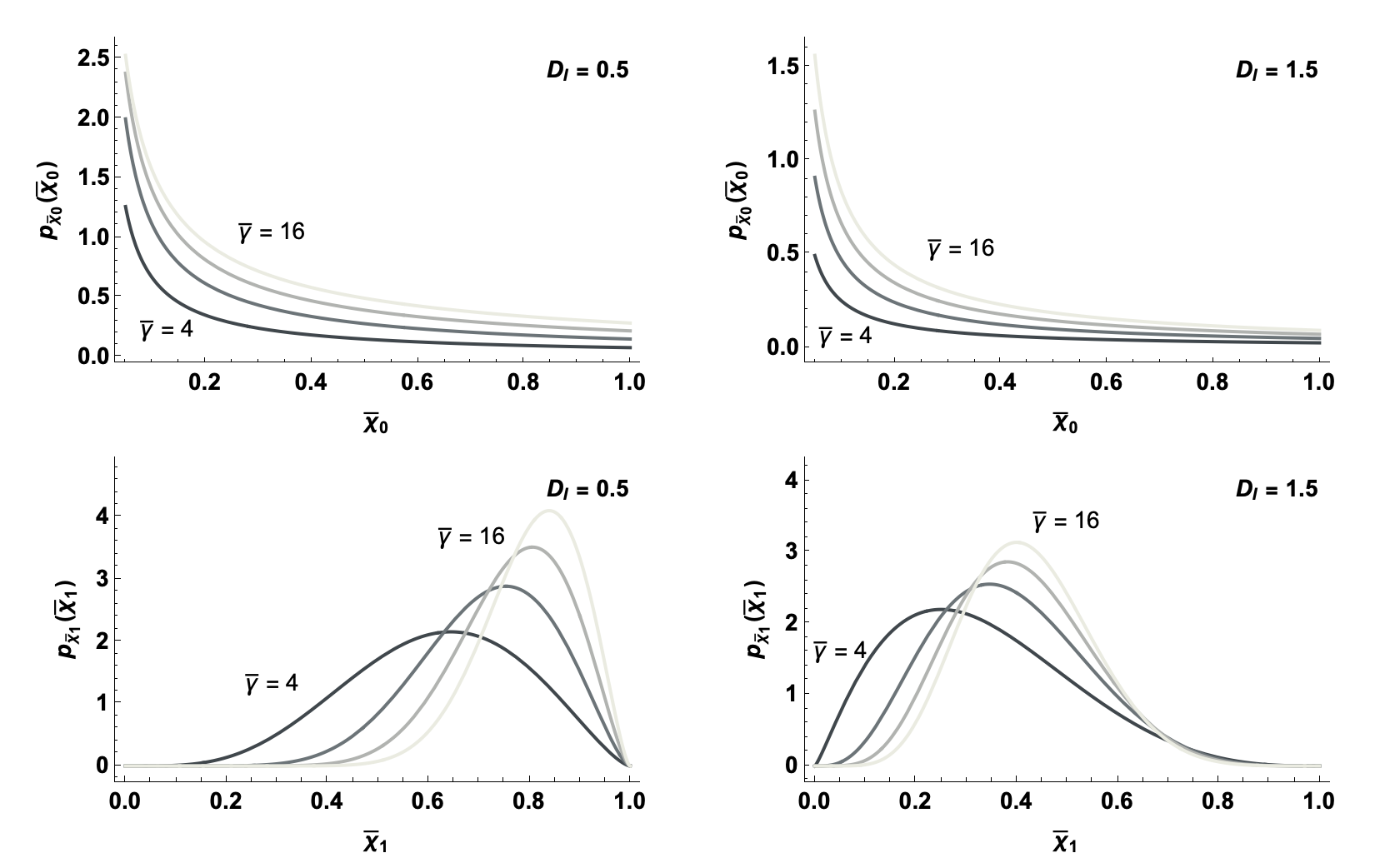}%

\caption{Comparison of the model soil moisture PDFs for different storage indices, $\overline{\gamma}$, and dryness indices, $D_I$, with the upper panels showing the upper soil layer PDF of Eq. (\ref{eq:dx0/dt}) and  the lower panels showing the lower soil PDF of Eq. (\ref{eq:PDFscsCNx}). The values of $\overline{\gamma}$ are 4, 8, 12, and 16, and the model parameters are $\mu = 0.01$, $B_I = 0.25$, and $\beta = 0.1$.
\label{fig:PDFs}}
\end{figure*}

The probabilistic description arises from modeling rainfall, $\overline{R}(t)$ as a marked Poisson process wheres storm arrivals at a rate $\lambda$ and total rainfall (per storm) amounts, $\overline{R}$, are exponentially distributed \cite{rodrigueziturbe1999probabilistic,rodrigueziturbe2004ecohydrology,bartlett2014excess}, i.e.,
\begin{align}
p_{\overline{R}}(\overline{R}) = \frac{1}{\overline{\alpha}}e^{-\frac{1}{\overline{\alpha}} \overline{R}},
\label{eq:pR}
\end{align}
where $\overline{\alpha}$ is the average rainfall total (per storm event). Similar to \ref{sec:dynamics_WC}, which describes the probabilistic dynamics of lower-layer soil moisture, the water balance equation (\ref{eq:dx0/dt}) has a corresponding master equation that governs the time evolution of the soil moisture PDF, as discussed in Chapter 7 of \citeA{porporato2022ecohydrology}. In statistical steady state, this equation yields the soil moisture probability distribution function (PDF) and the ensemble average \cite{porporato2022ecohydrology}, i.e., 
\begin{align}
p_{\overline{\chi}_0}(\overline{\chi}_0)= N_0 e^{- \overline{\gamma}\mu \overline{\chi}_0}\overline{\chi}_0^{\frac{ \overline{\gamma}\mu}{D_I}-1},
\label{eq:px0} \\
\langle \overline{\chi}_0\rangle = \frac{1}{D_I}-N_0(\overline{\gamma}\mu)^{-1}e^{-\overline{\gamma}\mu},
\label{eq:x0_avg} 
\end{align}
where the normalization constant is $N_0=\frac{( \overline{\gamma}\mu)^{\overline{\gamma}\mu /D_I}}{\Gamma\left(\frac{\overline{\gamma}\mu}{D_I}\right)-\Gamma\left(\frac{\overline{\gamma}\mu}{D_I}, \overline{\gamma} \mu \right)}$, $\Gamma(\cdot)$ is the gamma function, $\Gamma(\cdot,\cdot)$ is the upper incomplete gamma function, $D_I$ is the dryness index, and $\overline{\gamma}=\overline{w}/\overline{\alpha}$ (see Fig. \ref{fig:PDFs}). When implementing the normalization constant, $\Gamma\left(\frac{\overline{\gamma}\mu}{D_I}\right)-\Gamma\left(\frac{\overline{\gamma}\mu}{D_I},\overline{\gamma} \mu \right)$ may be replaced with the lower incomplete gamma function, $ \gamma\left(\frac{ \overline{\gamma}\mu}{D_I}, \overline{\gamma} \mu \right)$. In line with the SCS-CN initial abstraction formulation, the upper soil layer does not directly produce surface runoff---an assumption that could be adjusted in future work.

The upper soil layer acts as a censoring process, adjusting the rainfall inputs to the lower soil layer. Theoretically, the resulting frequency of deep percolation, $\overline{Y}(t)$, to the lower soil layer is the product of the rainfall frequency, $\lambda$, and a censoring factor $\omega_{\lambda}(\overline{\chi}_0) = \int_{\overline{w} \mu(1-\overline{\chi}_0)}^{\infty}p_{\overline{R}}(h)dh$. Similarly, the lower layer PET  is adjusted by  $\omega_{\text{PET}}(\overline{\chi}_0)=1-\overline{\chi}_0$, ensuring the total ET of both soil layers is less than the PET. However, for simplicity, we use an ensemble average of these factors based on the previous solutions of Eqs. (\ref{eq:px0}) and (\ref{eq:x0_avg}) \cite{rodrigueziturbe2004ecohydrology,porporato2022ecohydrology}, i.e.,
\begin{align}
\label{eq:w_lambda}
\langle\omega_{\lambda}\rangle = \frac{D_I}{ \overline{\gamma}\mu}p_{\overline{\chi}_0}(1)\\
\langle\omega_{\text{PET}}\rangle = 1-\langle \overline{\chi}_0 \rangle,
\label{eq:w_PET}
\end{align}
where the average $\langle\omega_{\lambda}\rangle$ represents the probability of rainfall exceeding the upper layer’s spare storage  \cite{rodrigueziturbe2004ecohydrology,botter2007basin}, and frequency of deep percolation to the lower soil layer, $\lambda \langle\omega_{\lambda}\rangle$  approximates a Poisson process---a reasonable assumption because the ratio of the average inputs $\lambda \langle\omega_{\lambda}\rangle$ and the characteristic time of the loss process (i.e., $\mu \overline{w}/\overline{\text{PET}}$) typically is smaller than 0.1 \cite{botter2007basin}.

\subsection{Lower Soil Layer Soil Moisture}
\label{sec:lowerlayer}
The lower soil layer determines the unit-area runoff based on the spatial variations in soil moisture and storage capacity across the watershed, characterized by a multivariate distribution as discussed in Section \ref{sec:model_description}; see Eq. (\ref{eq:meanfield}).  This distribution consistently upscales runoff and other water balance fluxes from point processes to unit-area processes. Although, for this upscaling, the spatial heterogeneity of any semi-distributed model could be adopted, here we adopt the spatial heterogeneity of the SCS-CN method---noting that the SCS-CN method may be considered a semi-distributed model as discussed in \citeA{bartlett2015unified2}. In general, the lower soil layer fluxes vary among the various semi-distributed model formulations. For example, the VIC-3L model has a flux to an even deeper layer that determines baseflow \cite{liang2001new}.

For this lower soil layer, we consider the water balance as 
\begin{align}
\overline{w}(1-\mu)\frac{d\overline{\chi}_1}{dt}= \overline{Y}(t)-\overline{Q}[\overline{Y}(t),\overline{u}_1(t)] -\overline{ET}_1\left[\overline{\chi}_1(t)\right]-Q_b[\overline{\chi}_1(t)],
\label{eq:dx1/dt}
\end{align}
where the soil moisture increases from the infiltration from the upper soil layer, $\overline{Y}(t)$ (see Section \ref{sec:upperlayer}), and decreases because of surface runoff, $\overline{Q}[\overline{Y}(t),\overline{u}_1(t)]$, evapotranspiration, $\overline{ET}_1\left[\overline{\chi}_1(t)\right]$, and baseflow, $Q_b[\overline{\chi}_1(t)]$. Note that runoff,  $\overline{Q}[\overline{Y}(t),\overline{u}_1(t)]$, depends on the soil moisture content, $\overline{u}_1$, antecedent to the storm event. In accordance with the modeling of the upper soil layer, deep percolation (infiltration) inputs and surface runoff also are on an event basis, i.e.,
$\overline{Y}(t) -  \overline{Q}(t) = \sum_{i=1}^{N(t)} (\overline{Y} - \overline{Q}) \: \delta(t-t_i)$, where as indicated by the Dirac delta function, $\delta(\cdot)$, deep percolation and runoff ($\overline{Y}$ and $\overline{Q}$) are instantaneous at the storm event arrival times $\{t_i \}(i = 1, 2,...)$.

Runoff, $\overline{Q}[\overline{Y}(t),\overline{u}_1(t)]$, is represented by an extended version of the SCS-CN method, referred to as the SCS-CNx method \cite{bartlett2016beyond}.  Unlike the original SCS-CN method, which is based on upscaling a threshold-excess runoff at a point, the SCS-CNx method is found from upscaling a threshold excess runoff complemented by a pre-threshold runoff (at a point) that arises from rainfall transferred to the stream in hydrologically connected, near-stream areas. Consequently, in the SCS-CNx method, runoff over a saturated fraction of area, $F_t$, is extended with runoff from the complementary area, i.e., $1-F_t$, where runoff---mediated by a prethreshold index, $P_I$---occurs before rainfall exceeds the storage capacity threshold at each point  \cite{bartlett2016beyond}, i.e., 

\begin{align}
\overline{Q}=\overline{Y} \: F_t+\overline{Y} (1-F_t)P_I,
\label{eq:SCSCNx}
\end{align}
where $F_t = \frac{\overline{Y}(1-P_I)}{\overline{S}_1+\overline{Y}(1-P_I)}$ depends on the prethreshold index, $P_I$, and the maximum potential retention, $\overline{S}_1=\overline{w}(1-\mu)(1-\overline{u}_1)$, which is typical of the SCS-CN method. As described by $P_I=\beta \overline{u}_1$, the prethreshold runoff occurs over a fraction of the watershed, $\beta$, in proportion to the antecedent soil moisture, $\overline{u}_1$, because  as the watershed becomes wetter, larger soil pores fill and connect, expanding the flow network capacity to transmit prethreshold runoff to the stream \cite{sidle2000stormflow, kim2005shallow, lin2012hydropedology,bartlett2016beyond}.

Similar to previous ecohydrological models, we assume linear losses (at a point) for both baseflow and evapotranspiration, i.e.,  $ET_1 = \text{PET} \langle\omega_{\text{PET}}\rangle \chi_1$ and $Q_b = Q_{b,\max} \chi_1$, where $\text{PET}$ is the potential evapotranspiration and $Q_{b,\max}$ is the maximum flux at a point that contributes to baseflow \cite{porporato2004soil,botter2007basin}. For consistency with SCS-CN unit-area runoff of Eq. (\ref{eq:SCSCNx}), these point fluxes are upscaled to a unit-area basis, so that

\begin{align}
\overline{ET}_1\left[\overline{\chi}_1(t)\right]=& \overline{\text{PET}}\langle\omega_{\text{PET}}\rangle \overline{\chi}_1(t) \\
\overline{Q}_b[\overline{\chi}_1(t)] = & \overline{Q}_{b,\max}\overline{\chi}_1(t).
\label{eq:Qb}
\end{align}
Similar to the upscaling of the point runoff to the unit area basis of Eq.  (\ref{eq:SCSCNx}),  the linear fluxes for evapotranspiration and baseflow are upscaled to a unit area basis with Eq. (\ref{eq:meanfield}), i.e., integrating the point flux over the multivariate distribution, capturing the spatial heterogeneity of the SCS-CN method \cite{bartlett2016beyond}.  The parameter $\overline{Q}_{b,\max}$ is similar to the time constant or recession constant from studies of baseflow \cite{brutsaert2005hydrology,thomas2013estimation}.

\begin{figure*}
\centering
\includegraphics[width=1\linewidth]{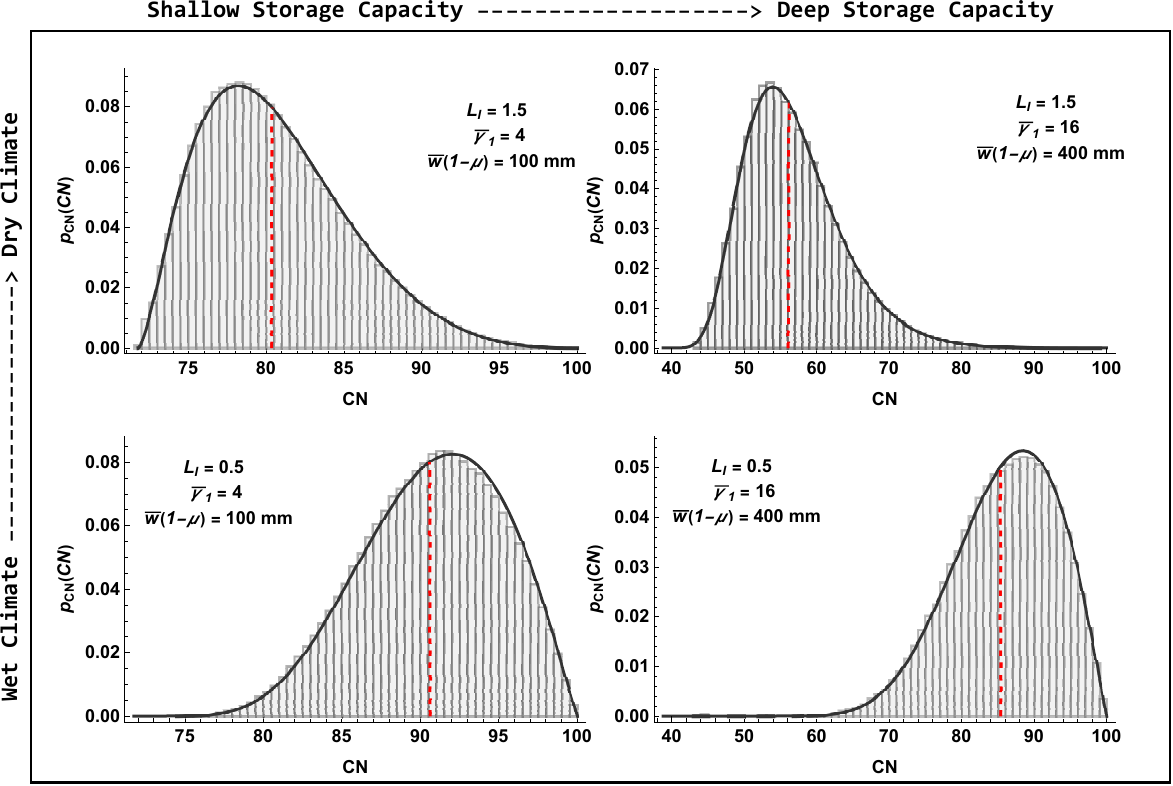}%
\caption{\label{Fig8} The PDF of the curve number, $p_{\text{CN}}(\text{CN})=\frac{25400}{\text{CN}^2 \overline{w}(1-\mu)}p_{\overline{\chi}_1}\left(\frac{\text{CN}(254+\overline{w}(1-\mu))-25400}{\text{CN}\overline{w}(1-\mu)}\right)$,  includes the approximation for the solution of $p_{\overline{\chi}_1}(\overline{\chi}_1)$ as discussed in \ref{sec:consistency}, but it still well matches the histogram from a Monte Carlo simulation of the exact system. For all cases $\beta=0.1$, and $\text{CN}_{\langle \overline{S}\rangle}$  (red line) represents the ensemble average wetness condition of $\overline{S}_1$.}
\end{figure*}

For this lower soil layer, the probabilistic description of the soil moisture arises from the statistical model of the infiltration---a marked Poisson process with infiltration events arriving at frequency $\lambda \langle\omega_{\lambda}\rangle$ with the infiltration quantities being exponentially distributed (see \ref{sec:dynamics_WC} and \ref{sec:Ft_SCS}).  For this statistical model of the infiltration inputs and water balance of Eq. (\ref{eq:dx1/dt}), the steady-state solution of the moisture content, $p_{\overline{\chi}_1}(\overline{\chi}_1)$, is derived from the general result of Eq. (46) of \cite{bartlett2018state}, i.e.,

\begin{align}
p_{\overline{\chi}_1}(\overline{\chi}_1)=N(1-\beta\vartheta\overline{\chi}_1)^{\frac{\overline{\gamma}_1(1-\beta)}{\beta(1-\vartheta\beta)}}\overline{\chi}_1^{\frac{\overline{\gamma}_1}{L_I}-1}(1-\overline{\chi}_1)^{\frac{\overline{\gamma}_1(1-\vartheta)}{1-\vartheta\beta}},
\label{eq:PDFscsCNx}
\end{align}
where the $N$ is the normalization constant (such that $\int_0^w p_{\overline{\chi}}(\overline{\chi})d\overline{\chi}=1$), and the fraction $\vartheta$ provides consistency (on average) between the exact SCS-CN method infiltration (i.e., an It\^o `jump' transition) with the approximate infiltration (i.e., the Marcus ‘jump’ transition) that is the basis of the steady state solution (see \ref{sec:consistency} and \ref{sec:ConAvgVar} for further details and analytical functions for $\vartheta$ and $N$).  The soil moisture of Eq. (\ref{eq:PDFscsCNx}) depends on three dimensionless parameters: $\overline{\gamma}_1=\overline{\gamma}(1-\mu) $, which is storage capacity normalized by the average rainfall input $\overline{\alpha}$; the fraction $\beta$, which represents the fraction of hydrologically connected, near-stream area that transmits runoff to stream before soil saturation; and the loss index,

\begin{align}
\label{eq:LI_sub}
L_I =& \frac{D_I \langle\omega_{\text{PET}}\rangle+B_I}{\langle\omega_{\lambda}\rangle},
\end{align}
where $D_I= \frac{\overline{\text{PET}}}{\overline{\alpha}\lambda}$ is the dryness index and $B_I=\frac{\overline{Q}_{b,\max}}{\overline{\alpha}\lambda}$ is the baseflow index, for which the coefficient $\overline{Q}_{b,\max}$ is the baseflow at saturation, $\overline{\chi}_1=1$. Note that $\overline{\alpha}\lambda$ equals the average rainfall intensity $\langle \overline{R}\rangle$. Figure \ref{fig:PDFs} shows the Eq. (\ref{eq:PDFscsCNx}) PDF for different values of $D_I$ and $\overline{\gamma}$. Though the PDF of the soil moisture, Eq. (\ref{eq:PDFscsCNx}) is based on a consistency condition of  \ref{sec:consistency}, this PDF of Eq. (\ref{eq:PDFscsCNx}) provides an excellent representation of the the distribution (histogram bars) derived from a Monte Carlo simulation with the exact interpretation of the SCS-CN infiltration and runoff (Fig. \ref{Fig8}).

We also may recover, the PDF of soil moisture and the associated statistics for the original SCS-CN method that only accounts for saturation excess (threshold excess) runoff. The PDF for the original SCS-CN method is retrieved from Eq. (\ref{eq:PDFscsCNx}) in the limit of $\beta\to0$, i.e.,

\begin{align}
p_{\overline{\chi}_1}(\overline{\chi}_1)=Ne^{-\overline{\gamma}_1\vartheta\overline{\chi}_1}\overline{\chi}_1^{\frac{\overline{\gamma}_1}{L_I}-1}(1-\overline{\chi}_1)^{\overline{\gamma}_1(1-\vartheta)},
\label{eq:PDFscsCN}
\end{align}
where the parameters are the same as in Eq. (\ref{eq:PDFscsCNx}), and the normalization constant and fraction $\vartheta$ are as given in \ref{sec:consistency}.

For the SCS-CN assumptions for spatial heterogeneity and runoff, Eqs. (\ref{eq:PDFscsCNx}) - (\ref{eq:PDFscsCN}) provide a comprehensive probabilistic characterization of the soil moisture, i.e., $p_{\overline{\chi}_1}(\overline{\chi}_1)$, as a function of climate described by the dryness index, $D_I$, storage relative to average rainfall, $\overline{\gamma}_1$, and the watershed drainage characteristics given by the baseflow index, $B_I$, and a fraction of watershed, $\beta$, with hydrologically connected areas that produce runoff before saturation. With this probabilistic description of the upper and lower soil layers, we may define probabilistic descriptions for the evapotranspiration, baseflow, and runoff fluxes.  

\subsection{Baseflow and Evapotranspiration Statistics}
The baseflow and evapotranspiration distributions are derived from the soil moisture distributions. The baseflow PDF is given by 
\begin{align}
p_{\overline{Q}_b}(\overline{Q}_b) = \frac{1}{\overline{Q}_{b,\max}}p_{\overline{\chi}_1}\left(\frac{\overline{Q}_b }{\overline{Q}_{b,\max}}\right), 
\label{eq:pQb}
\end{align}
which is derived with a change of variables based on the baseflow Eq. (\ref{eq:Qb}) and the PDF $p_{\overline{\chi}_1}(\overline{\chi}_1)$ of either Eq. (\ref{eq:PDFscsCNx}) or (\ref{eq:PDFscsCN}). The probability density function (PDF) of evapotranspiration is derived by modifying the potential evapotranspiration of the lower layer through the function $\omega_{\text{PET}}(\overline{\chi}_0)=1-\overline{\chi}_0$. The resulting PDF for total evapotranspiration is given by

\begin{align}
p_{\overline{ET}}(\overline{ET}) = \frac{1}{\overline{\text{PET}} \omega_{\text{PET}}(\overline{\chi}_0) }\int_{0}^{\frac{\overline{ET}}{\overline{\text{PET}}}\Theta\left[1-\frac{\overline{ET}}{\overline{\text{PET}}}\right]+1-\Theta\left[1-\frac{\overline{ET}}{\overline{\text{PET}}}\right]} p_{\overline{\chi}_0}(\overline{\chi}_0)p_{\overline{\chi}_1}\left(\frac{1}{\omega_{\text{PET}}(\overline{\chi}_0)}\left(\frac{\overline{ET}}{\overline{\text{PET}}}- \overline{\chi}_0\right)\right)d\overline{\chi}_0,
\label{eq:pET}
\end{align}
which follows from integrating the joint PDF representation of the process, as detailed in \ref{sec:PDF_derivations}. In Eq. (\ref{eq:pET}), the integration limits are explicitly expressed in terms of $\overline{ET}$ where $\Theta[\cdot]$ denotes the Heaviside step function. 

\subsection{Runoff Statistics}

The overall runoff PDF, $p_{\overline{Q}}(\overline{Q})$, consists of an atom of the probability of zero runoff (when rainfall does not infiltrate to the lower soil layer), represented with a Dirac delta, plus a continuous distribution of runoff (when rainfall does infiltrated to the lower soil layer), i.e., 
\begin{align}
p_{\overline{Q}}(\overline{Q})=P(\overline{Q}=0)\delta(\overline{Q})+(1-P(\overline{Q}=0))p_{\overline{Q}_1}(\overline{Q}),
\label{eq:pQ_all}
\end{align}
where the probability of zero runoff, $P(\overline{Q}=0)$, equals $1-\langle\omega_{\lambda}\rangle$, and the lower soil layer runoff PDF, $p_{\overline{Q}_1}(\overline{Q})$, is given by

\begin{align}
p_{\overline{Q}_1}(\overline{Q}) = &\int_0^{1}\frac{1}{2}\left(1+\frac{g(\overline{Q}
)}{v(\overline{Q})}\right)\hat{p}_{\overline{R}}\left(\frac{g(\overline{Q})-2\overline{w}(1-\overline{u})+v(\overline{Q})}{2(1-\beta\overline{u})}\right)p_{\overline{\chi}_1}(\overline{u})d\overline{u},
\label{eq:pQ}
\end{align}
where 
$v(\overline{Q})=\sqrt{4\overline{Q}\:\overline{w}(1-\overline{u})(1-\beta\overline{u})+(\overline{Q}(1-\beta\overline{u})-\overline{w}(1-\overline{u})\beta\overline{u})^2}$, $g(\overline{Q})= \overline{Q}(1-\beta\overline{u})+\overline{w}(1-\overline{u})(2-\beta\overline{u})$, and $\hat{p}_{\overline{R}}$ is a mixed exponential PDF of rainfall, i.e.,

\begin{align} 
\hat{p}_{\overline{R}}(\overline{R}) = \omega \frac{1}{\overline{\alpha}_{\circ}}e^{-\frac{1}{\overline{\alpha}_{\circ}}: \overline{R}} + (1-\omega) \frac{1}{\overline{\alpha}_{*}}e^{-\frac{1}{\overline{\alpha}_{*}}: \overline{R}}, 
\label{eq:pzm}
\end{align}
where $\omega\in(0,1)$ is the weighting factor, and $\overline{\alpha}_{\circ}$ and $\overline{\alpha}_{*}$ are the mean rainfall values of the respective exponential distributions.  Eq. (\ref{eq:pQ}) is retreived from the integrating the multivariate PDF of \ref{sec:PDF_derivations}. We choose the mixed exponential PDF of Eq. (\ref{eq:pzm}) because while the exponential PDF of Eq.  (\ref{eq:pR}) reasonably captures the rainfall that infiltrates into the soil, it fails to account for the  rainfall extremes that for the most part become runoff. Thus to better represent these rainfall extremes and runoff, the rainfall PDF of Eq. (\ref{eq:pQ}) is represented with the Eq. (\ref{eq:pzm}) mixed exponential PDF \cite{feldmann1997fitting}---an assumption that has a negligible impact on the soil moisture PDF, $p_{\overline{\chi}_1}(\overline{\chi}_1)$ because the heavier tail, representing extreme rainfall, primarily results in runoff with minimal impact on infiltration.

Based on the water balance, the ensemble average runoff, $\langle \overline{Q}\rangle$,  and lower layer ensemble average runoff, $\langle \overline{Q}_1\rangle$,  are 

\begin{align}
\label{eq:Qavg}
\langle \overline{Q}\rangle = & \langle\omega_{\lambda}\rangle\langle \overline{Q}_1\rangle \\
\langle \overline{Q}_1\rangle= & \overline{\alpha}\left(1-L_I\langle \overline{\chi}_1\rangle\right),
\label{eq:Q1avg}
\end{align}
where $\langle \overline{Q}_1\rangle$ is based on the lower soil layer water balance. Multiplying $\langle \overline{Q}\rangle$  by the frequency of storm events $\lambda$ returns an intensity (i.e., a rate) of runoff similiar to the rate of rainfall, $\lambda\overline{\alpha}$.

\section{SCS-CN Reinterpretation and Connection to Hydroclimatic Regimes}

In this novel probabilistic watershed model, the SCS-CN method is reinterpreted as part of a continuous-time framework, with the CN value redefined as a probabilistic quantity that reflects likely antecedent conditions influenced by both storm and interstorm dynamics. Thus, rather than being a static parameter determined by land use and soil type \cite{nrcs2004national}, the CN is dynamically linked to climate, interstorm variability, and ecohydrological processes. Specifically, the soil moisture PDFs, $p_{\overline{\chi}_0}(\overline{\chi}_0)$ and $p{\overline{\chi}_1}(\overline{\chi}_1)$, provide the basis for an analytical probabilistic characterization of antecedent CN and initial abstraction, $\overline{I}$, enabling these conditions to be informed by climate characteristics and interstorm watershed dynamics. This reinterpretation bridges the gap between the traditional event-based SCS-CN method and continuous soil-water interaction models by defining the likely CN and initial abstraction, $\overline{I}$, as functions of the Budyko dryness index, $D_I$, as well as parameters such as overall watershed storage capacity, $\overline{w}$, and the maximum baseflow coefficient, $\overline{Q}_{b,\max}$.

This advancement builds on the foundational principles of the SCS-CN method, which has remained a cornerstone of hydrologic engineering practice since its introduction in 1954  \cite{ponce1996runoff, rallison1982past, nrcs2004national}. To appreciate how the probabilistic reinterpretation expands the SCS-CN method capabilities, it is essential to understand the original formulation, which can be expressed synthetically as
\begin{align}
\overline{Q} = \overline{R} \: F_t,
\label{eq:SCSCN}
\end{align}
where $\overline{Q}$ represents runoff depth (per unit area),  $\overline{R}$ is the rainfall depth (per unit area), $F_t = \frac{\overline{R}}{\overline{R}+\overline{S}_1}$ denotes the fraction of area producing runoff, and $\overline{S}_1$ is the maximum potential retention depth (per unit area) \cite{nrcs2004national,bartlett2016beyond}. Over the fraction of area $F_t$, runoff is the rainfall in excess of the storage threshold at each point \cite{bartlett2016beyond}. Typically, the SCS-CN method substitutes the rainfall, $\overline{R}$, with an effective rainfall infiltration, $\overline{Y} = \overline{R}-\overline{I}$, where the initial abstraction, $\overline{I}$, represents an amount of rainfall that is retained in the watershed storage as interception, infiltration, and surface storage before runoff begins \cite{ponce1996runoff}. 
The maximum potential retention, $\overline{S}_1$ is defined by a dimensionless curve number (CN) parameter, while the initial abstraction is a fraction of the maximum potential retention, i.e., 
\begin{align}
\label{eq:CN}
\overline{S}_1=&\frac{25400}{\text{CN}}-254,\\
\overline{I}=&\mu_I \: \overline{S}_1,
\label{eq:InitialAbstraction}
\end{align}
where $\mu_I$ is the initial abstraction ratio (typically $\mu_I = 0.2$, though a value of $\mu_I = 0.05$ is often more realistic), and 25400 and 254 are in millimeters \cite{ponce1996runoff, yuan2001modified, nrcs2004national}. In practice, tables provide a CN according to soil type, hydrologic condition, antecedent moisture condition, and land use \cite{nrcs2004national}. 

In the present probabilistic reinterpretation, practitioners can now substitute the traditional CN lookup table with statistical characterizations of $\overline{I}$ and $\overline{S}_1$ using probability density functions (PDFs):
\begin{align}
    \label{eq:pI}
    p_{\overline{I}}(\overline{I}) = & \frac{1}{\overline{w}\mu}p_{\overline{\chi}_0}\left(\frac{\overline{w}\mu-\overline{I}}{\overline{w}\mu}\right) \\
    p_{\overline{S}_1}(\overline{S}_1) = & \frac{1}{\overline{w}(1-\mu)}p_{\overline{\chi}_1}\left(\frac{\overline{w}(1-\mu)-\overline{S}_1}{\overline{w}(1-\mu)}\right),
    \label{eq:pS}
\end{align}
where the distribution correspond to the original SCS-CN method when $\beta =0$ and correspond to the extended SCS-CNx method when $\beta \neq 0$. Note that Eqs. (\ref{eq:pI}) and (\ref{eq:pS})
 are derived from the upper and lower soil layer soil moisture PDFs of Eq. (\ref{eq:px0}) and  (\ref{eq:PDFscsCNx}) with a change of variables based on $\overline{I} = \overline{w}\mu(1-\overline{\chi}_0)$ and $\overline{S}_1 = \overline{w}(1-\mu)(1-\overline{\chi}_1)$, respectively.

 \begin{figure*}
\centering
\includegraphics[width=1\textwidth]{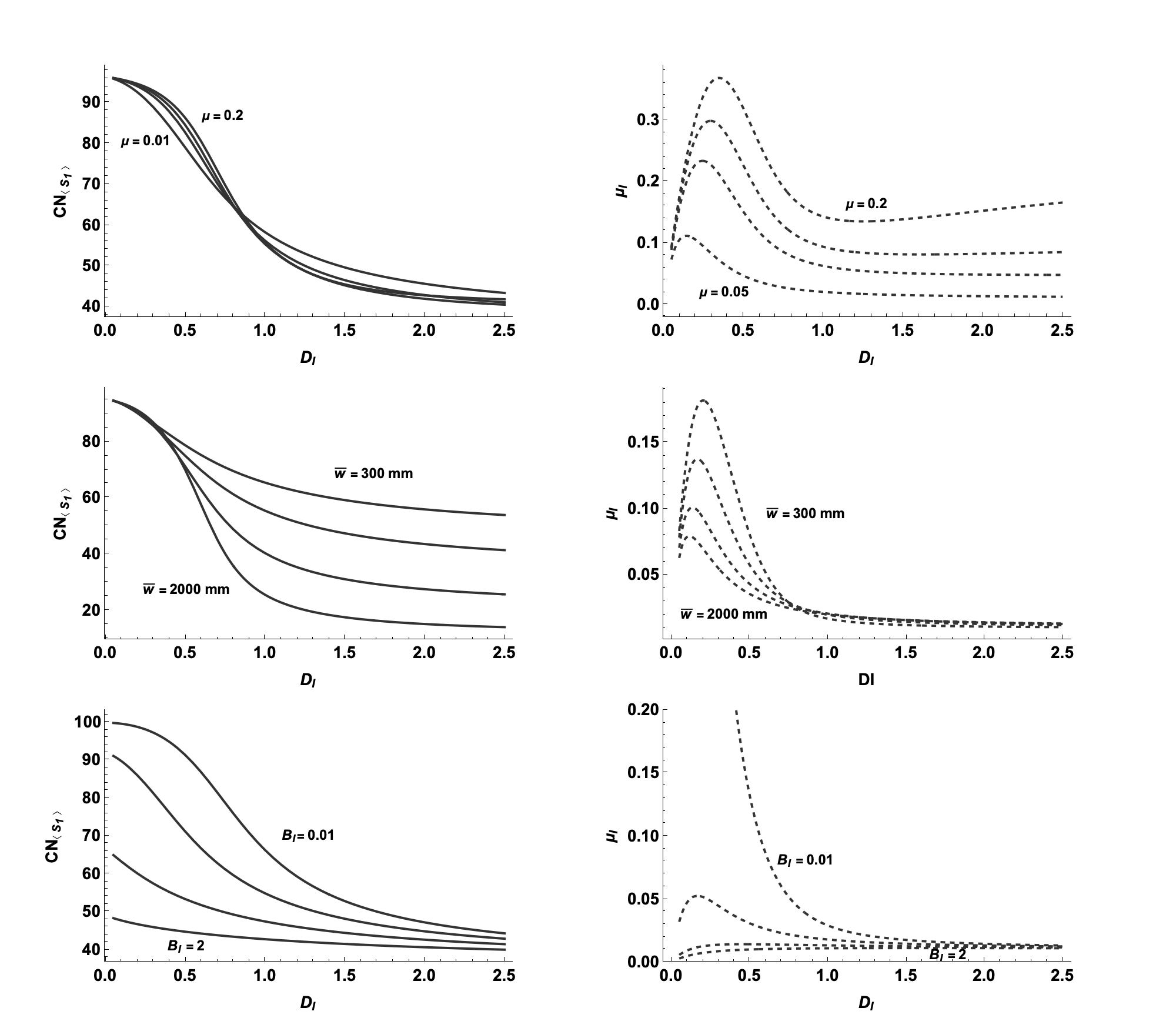}
\caption{\label{fig:5} The CN for the ensemble average of maximum potential retention, CN$_{\langle \overline{S} \rangle}$, and initial abstraction ratio, $\mu_I$, as a function of the dryness  index, with the top panels showing a varying soil layer partitioning fraction, $\mu$,  of $0.01$, $0.05$, $0.1$, and $0.2$, the middle panels showing a varying storage, $\overline{w}$ (which also varies $\gamma$), of $50$, $100$, $500$, and $1000\,\mathrm{mm}$, and the bottom panels showing a varying baseflow index, $B_I$,  of $0.01$, $0.03$, $0.9$, and $2$. Unless specified,  $\overline{w}=461\,\mathrm{mm}$, $\mu = 0.01$, $\gamma$ = 25, $B_I = 0.25$, and $\beta = 0.1$.}
\end{figure*}

The traditional lookup tables values of the CN and initial abstraction, $\overline{I}$, may be interpreted as approximating averages of a long-term continuous process. Based on Eqs. (\ref{eq:pI}) and (\ref{eq:pS}) of this reinterpretation, these respective averages are 
 \begin{align}
 \label{eq:Iavg}
\langle \overline{I}\rangle &= \overline{w}\mu(1 - \langle \overline{\chi}_0\rangle)\\
\langle \overline{S}_1\rangle &= \overline{w}(1-\mu)(1 - \langle \overline{\chi}_1\rangle).
\label{eq:Savg}
 \end{align}
In addition to these averages, the model also characterizes the variances, $\sigma^2_{\overline{I}} = \sigma^2_{\overline{\chi}_0}\:(\overline{w}(\mu))^2$ and  $\sigma^2_{\overline{S}_1} = \sigma^2_{\overline{\chi}_1}\:(\overline{w}(1-\mu))^2$. To fully exploit the predictive capabilities of this probabilistic framework, both the mean and variance estimates must be calibrated against historical observations. This, in turn, calls for a systematic approach (akin to the existing CN lookup tables) to selecting hydrologic parameters that achieve the best agreement with observations.

Departing from the classic SCS-CN method, this model explicitly links the CN and initial abstraction ratios to hydroclimatic variability, as parameterized by the Budyko dryness index, $D_I$. Here, the CN and initial abstraction ratio are derived from Eqs. (\ref{eq:CN}) and (\ref{eq:InitialAbstraction}) based on $\langle\overline{S}\rangle$ and $\langle\overline{I}\rangle$ of Eqs. (\ref{eq:Iavg}) and (\ref{eq:Savg}).  While variations in the soil layer partitioning parameter, $\mu$, exert minimal influence on CN, they play a crucial role in modulating the initial abstraction ratio, $\mu_I$, with a nearly linear dependence for $D_I>1$  (Fig. \ref{fig:5}). The CN dynamics exhibit a dual dependency: in wetter climates, i.e.,  $D_I < 1$, CN is predominantly regulated by the baseflow index, $B_I$, whereas in drier conditions, i.e., $D_I >1$, it is primarily controlled by storage depth $\overline{w}$ (Fig. \ref{fig:5}). In both cases for a specific $\mu$, the initial abstraction ratio converges to a single value as $D_I$ increases (Fig. \ref{fig:5}).

\begin{figure*}[h]
\centering
\includegraphics[width=1\linewidth]{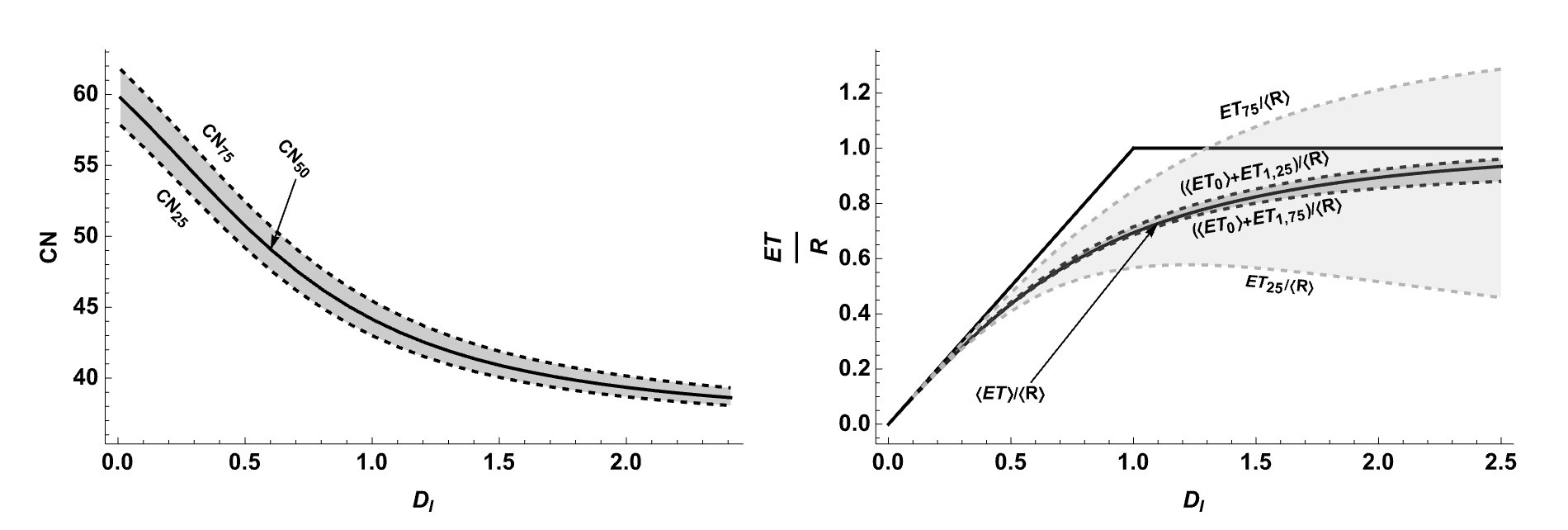}%
\caption{
\label{fig:6}
Based on the CN PDF of Eq. (\ref{eq:pCN}), the CN is a function of the dryness index, $D_I$ for the median condition, CN$_{50}$, as well as for the dry and wet conditions represented by the 25th and 75th quantiles, CN$_{25}$ and CN$_{75}$. These changes are linked to changes in the water balance partitioning given by $(\langle \overline{ET}_0\rangle + \overline{ET}_{1,25})/\langle\overline{R}\rangle$ and $(\langle \overline{ET}_0\rangle + \overline{ET}_{1,75})/\langle\overline{R}\rangle$, where $\overline{ET}_{1,25}$ and $\overline{ET}_{1,75}$ represent 25th and 75th quantiles of lower soil layer ET. Overall dry and wet and dry conditions (encompassing both soil layers) are captured by   $\overline{ET}_{25}/\langle\overline{R}\rangle$ and $\overline{ET}_{75}/\langle\overline{R}\rangle$, where $\overline{ET}_{25}$ and $\overline{ET}_{75}$ denote the 25th and 75th quantiles of total ET from both soil layers. The average $\langle\overline{ET}\rangle/\langle\overline{R}\rangle$ follows the original curve of \citeA{budyko1974climate},  with energy and water limits highlighted by the bold black line. The base model parameters are $\mu = 0.05$, $B_I = 0.7$, $\beta = 0.9$, and $\gamma = 53$, and $\overline{w} = 46.1$ cm to match the original formulation of \citeA{budyko1974climate}.}
\end{figure*}

In the classical formulation of the SCS-CN method, the curve number (CN) is treated as an average or median value, with antecedent wet and dry conditions prescribed through a lookup table indexed by this median \cite{nrcs2004national}. In contrast, the new probabilistic formulation naturally defines antecedent wet and dry conditions as quantiles of the CN PDF, rather than imposing them deterministically. The CN PDF is given by
\begin{align}
p_{\text{CN}}(\text{CN}) = \frac{25400}{\text{CN}^2}p_{\overline{S}_1}\left(\frac{25400}{\text{CN}}-254\right),
\label{eq:pCN}
\end{align}
where for the maximum potential PDF $p_{\overline{S}_1}(\cdot)$ of Eq. (\ref{eq:pS}), parameter units are in millimeters. For instance, the dry condition may correspond to the 25th quantile of this distribution, while the wet condition aligns with the 75th quantile (Fig. \ref{fig:6}). These wet and dry CN values are directly governed by hydroclimatic variability, as characterized by the dryness index, $D_I$. Furthermore, they are inherently linked to variations in the Budyko-type curve, providing a physically consistent representation of how runoff generation responds to changes in water balance partitioning (Fig. \ref{fig:6}). 

The wet and dry CN values are associated with  small change in the water balance partitioning of the Budyko curve, with the wet and dry CN conditions characterized by the ensemble average evapotranspiration of the upper soil layer,  $\langle \overline{ET}_0\rangle$, plus the 25th and 75th quantiles of lower-layer evapotranspiration, $\overline{ET}_{1,25}$ and $\overline{ET}_{1,75}$, respectively (Fig. \ref{fig:6}). A broader wet and dry characterization---encompassing both the CN and initial abstraction---emerges from the ratios  $\overline{ET}_{25}/\langle\overline{R}\rangle$ and $\overline{ET}_{75}/\langle\overline{R}\rangle$, where $\overline{ET}_{25}$ and $\overline{ET}_{75}$ denote the 25th and 75th quantiles of total evapotranspiration from both the upper and lower soil layers. Crucially, changes in the upper soil layer initial abstraction manifest as shifts in the ratio $\overline{ET}/\langle \overline{R}\rangle$ that may exceed the water limit defined by the Budyko curve under average conditions (Fig. \ref{fig:6}).

\section{Watershed Scale Average Water Balance and Budyko Curve}

At the watershed scale,  the ensemble averages of rainfall, $\langle \overline{R}\rangle$, evapotranspiration, $\langle\overline{ET}\rangle$, baseflow, $\langle \overline{Q}_b\rangle$, and river (or stream) flow, $\langle\overline{Q}_f\rangle$, describe the long-term partitioning of water balance components, i.e., 

\begin{align}
\label{eq:ET/R}
\frac{\langle \overline{ET} \rangle}{\langle \overline{R}\rangle} =& D_I \langle \overline{\chi}_0 \rangle + D_I \langle \omega_{\text{PET}}\rangle\langle \overline{\chi}_1\rangle\\
\frac{\langle \overline{Q}_b\rangle}{\langle \overline{Q}_f\rangle} =& \frac{B_I \langle \overline{\chi}_1\rangle}{1-D_I \langle \overline{\chi}_0 \rangle - D_I \langle \omega_{\text{PET}}\rangle \langle \overline{\chi}_1\rangle}, \label{eq:Qb/Qf} 
\end{align}
where the ensemble averages represent an integration over the respective PDFs, e.g., $\langle\overline{ET}\rangle = \int_0^{\infty} \overline{ET} \: p_{\overline{ET}}(\overline{ET}) d\overline{ET}$, and in the case of rainfall,   $\langle \overline{R}\rangle = \lambda\overline{\alpha}$, where $\lambda$ is the rainfall frequency and $\overline{\alpha}$ is the average storm depth. Equation (\ref{eq:ET/R}) describes the partitioning of rainfall between evapotranspiration, $\frac{\langle \overline{ET} \rangle}{\langle \overline{R}\rangle}$, and stream (or river) flow, $1-\frac{\langle \overline{ET} \rangle}{\langle \overline{R}\rangle}$. Equation (\ref{eq:Qb/Qf}) further partitions streamflow between the fraction of baseflow, $\frac{\langle \overline{Q}_b\rangle}{\langle \overline{Q}_f\rangle}$, and the fraction of runoff, $1- \frac{\langle \overline{Q}_b\rangle}{\langle \overline{Q}_f\rangle}$. Together, Eqs. (\ref{eq:ET/R}) and (\ref{eq:Qb/Qf}) provide a framework for understanding the average partitioning of water balance fluxes.

\begin{figure*}
\noindent\includegraphics[width=1\textwidth]{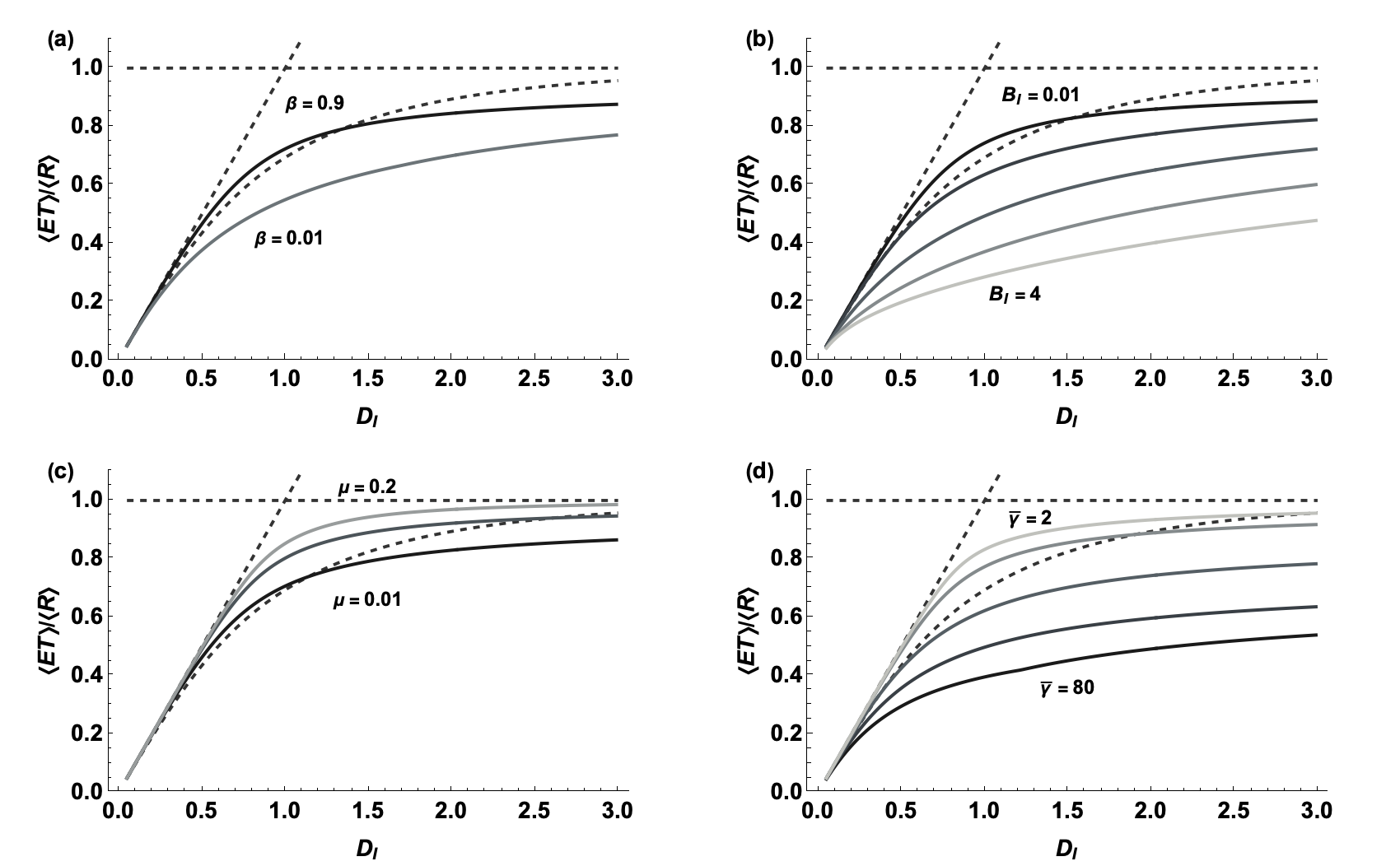}
\centering
\caption{Budyko type curves showing the fraction of precipitation that is evapotranspired as a function of the dryness index, $D_I$ for different values of a) the fraction of hydrologically connected, near-stream area, $\beta$, b) the baseflow index, $B_I = \overline{Q}_{b,\max}/\lambda\overline{\alpha}$, c) soil layer partition fraction, $\mu$, and d) the storage index, $\overline{\gamma}$. Unless stated otherwise the base model parameters are $\overline{\gamma}_1 = 20$, $\mu = 0.01$, $B_I = 0.1$, and $\beta = 0.1$.
\label{fig:Budkyo}
}
\end{figure*}


The fraction $\frac{\langle \overline{ET} \rangle}{\langle \overline{R}\rangle}$ of Eq. (\ref{eq:ET/R}) provides a novel interpretation of Budyko's curve, based on the connection of the different hydrological concepts achieved by the previous stochastic solutions. In fact, the average soil moistures of the upper and lower soil layers, $\langle \overline{\chi}_0 \rangle$ and $\langle\overline{\chi}_1 \rangle$, here depend on the dryness (aridity) index, $D_I$, the storage index, $\overline{\gamma}$, parameters $\beta$ and $\mu$ associated with the extended SCS-CN method, and the baseflow index, $B_I$, which summarizes the baseflow dynamics. When $\mu = 1$, the Budkyo curve of Eq. (\ref{eq:ET/R}) reduces to the stochastic ecohydrological formulation of the Budyko curve, $\frac{\langle \overline{ET} \rangle}{\langle \overline{R}\rangle} = D_I \langle \overline{\chi}_0 \rangle$, as described in \citeA{porporato2004soil}. By incorporating the average soil moisture of the lower soil layer, $\langle \overline{\chi}_1 \rangle$, Eq. (\ref{eq:ET/R}) extends this traditional definition to reflect the combination of stochastic ecohydrology, semi-distributed hydrologic modeling, and event-based SCS-CNx rainfall-runoff partitioning (see Fig. \ref{fig:Budkyo}). 

This framework also provides flexibility in analyzing variations of Budyko-type curves that characterize stream and river flow partitioning. For instance, stream flow partitioning can be examined in terms of the ratio of baseflow to total flow, $\frac{\langle \overline{Q}_b\rangle}{\langle \overline{Q}_F\rangle}$, as defined by Eq. (\ref{eq:Qb/Qf}), or as the ratio of baseflow to rainfall, $\frac{\langle \overline{Q}_b\rangle}{\langle \overline{R}\rangle}$, expressed as $B_I\langle \overline{\chi}_1 \rangle$. These formulations quantify how baseflow variations are influenced not only by the dryness index, $D_I$, but also by the storage index, $\overline{\gamma}$, the fraction of near-stream area, $\beta$, and the partitioning of storage between the upper and lower soil layers as governed by $\mu$. Typically, $\frac{\langle \overline{Q}_b\rangle}{\langle \overline{Q}_f\rangle}$ is referred to as the baseflow index, while $\frac{\langle \overline{Q}_b\rangle}{\langle \overline{R}\rangle}$ is known as the baseflow fraction \cite{gnann2019there}. To maintain consistency with the dryness index, which quantifies potential evapotranspiration relative to rainfall, we refer to $B_I$ as a baseflow index because it characterizes the potential of baseflow relative to rainfall. Consequently, we refer to $\frac{\langle \overline{Q}_b\rangle}{\langle \overline{Q}_f\rangle}$ as a baseflow fraction, as it represents the fraction of total river (or stream) flow that originates from baseflow.

The variance of hydrologic fluxes, particularly the runoff variance (on a storm event basis), $\sigma^2_{\overline{Q}}$, is also critical. Together, the long-term averages and variances of the hydrologic fluxes provide a more holistic representation of the watershed dynamics. Here, the variance of runoff is given by
\begin{align}
\sigma^2_{\overline{Q}} =& (1-\langle \omega_{\lambda}\rangle)\langle \overline{Q}\rangle^2+ \langle \omega_{\lambda}\rangle \int_0^{\infty}(\overline{Q}-\langle\overline{Q}\rangle)^2p_{\overline{Q}_1}(\overline{Q})d\overline{Q},
\label{eq:Qsigma} 
\end{align}
where $\langle \omega_{\lambda}\rangle$, $p_{\overline{Q}_1}(\overline{Q})$, and $\langle\overline{Q}\rangle$ are given by Eqs. (\ref{eq:w_lambda}), (\ref{eq:pQ}), and (\ref{eq:Qavg}), respectively.
Calibrating the model to both the observed average runoff, $\langle \overline{Q}\rangle$, and the observed runoff variance, $\sigma^2_{\overline{Q}}$, aligns the model with the runoff dynamics essential for flood mapping and engineering design efforts.

\section{Representation of Long-term Statistics}

While the statistical model in Eqs. (\ref{eq:pR}) - (\ref{eq:Q1avg}) provides a steady-state probabilistic description of the hydrologic system, it can also represent time-dependent (non-stationary) PDFs integrated over a longer observation period, $T$, which may span multiple seasons to several decades. As such, these PDFs then account for the long-term temporal variability within $T$. During this period, the model is based on the time-averaged values of potential evapotranspiration $\langle \overline{\text{PET}}\rangle$, storm frequency $\langle \lambda \rangle$, and rainfall per storm event $\langle \overline{\alpha}\rangle$ (see \ref{sec:long-term-balance}). Substituting these time-averaged values into the steady-state PDFs yields distributions that reflect the time-averaged probabilistic dynamics over the observation period. Specifically, the substitutions (i.e., $\overline{\gamma}\longrightarrow\langle\overline{\gamma}\rangle = \frac{\langle\overline{\alpha}\rangle}{\overline{w}}$ and $D_I\longrightarrow \langle D_I\rangle = \frac{\langle \overline{\text{PET}} \rangle}{\langle \lambda \rangle \langle\overline{\alpha}\rangle}$) are applied to the steady-state soil moisture PDFs, $p_{\overline{\chi}_0}(\overline{\chi}_0)$ and $p{\overline{\chi}_1}(\overline{\chi}_1)$, as well as to the overall probabilistic description in Eqs. (\ref{eq:pQb}) - (\ref{eq:Q1avg}).

As detailed in \ref{sec:long-term-balance}, the soil moisture PDFs of Eqs. (\ref{eq:px0}) and (\ref{eq:PDFscsCNx}), $p_{\overline{\chi}_0}(\overline{\chi}_0)$ and $p{\overline{\chi}_1}(\overline{\chi}_1)$,  are interpreted as lying between a drier condition, where the non-stationary PDFs, $p_{\overline{\chi}_0}(\overline{\chi}_0;t)$ and $p_{\overline{\chi}_1}(\overline{\chi}_1;t)$, are integrated over the temporal variability of continuous fluxes, and a wetter condition, where they are integrated over the temporal variability of storm frequency.  Similarly, the PDFs of the hydrologic fluxes approximate the non-stationary counterparts integrated over time. For instance, Eqs. (\ref{eq:pQb}) - (\ref{eq:pQ}) now represent $p_{\overline{ET}}(\overline{ET}) =\frac{1}{T}\int_{0}^{\infty}p_{\overline{ET}t}(\overline{ET};t)dt$ and $p_{\overline{Q}}(\overline{Q}) =\int_{0}^{\infty}p_{\overline{Q}t}(\overline{Q};t)p_{t_{\lambda}}(t) dt$, where $p_{t_{\lambda}}(t)=\frac{\lambda(t)}{\langle \lambda \rangle T}$. Because storm variability is already accounted for in $p_{t_{\lambda}}(t)$, the frequency of runoff in turn becomes the average $\langle \lambda \rangle$, rather than the time-varying $\lambda(t)$ \cite{bartlett2018state,bartlett2019jump,bartlett2014excess}. In this framework, the partitioning of the water balance and runoff variance in Eqs. (\ref{eq:ET/R})-(\ref{eq:Qsigma}) then reflect time-averaged values, incorporating seasonal, annual, and multi-year variability.

\section{Model Application and Calibration}
The core model consists of the soil moisture PDFs described in Eqs. (\ref{eq:px0}) and (\ref{eq:PDFscsCNx}); the PDFs of hydrologic fluxes, including baseflow, ET, and runoff, as defined in Eqs. (\ref{eq:pQb})–(\ref{eq:pQ_all}); and the expressions for the long-term averages of water balance partitioning and runoff variance, provided in Eqs. (\ref{eq:ET/R})–(\ref{eq:Qsigma}). This model has nine parameters---five climate parameters derived directly from observational data and four unknown hydrologic parameters calibrated by fitting the model to the data (as discussed in \ref{sec:CalibrationDetails}). When representing long-term statistics, the climate parameters include the long-term average dryness index, $\langle D_I\rangle$, and the long-term average rainfall parameters ($\langle\overline{\alpha}\rangle$, $\langle\overline{\alpha}_{\circ}\rangle$, $\langle\overline{\alpha}_{*}\rangle$) and the weight parameter, $\omega$,  associated with the exponential and mixed exponential rainfall distributions of Eqs. (\ref{eq:pR}) and (\ref{eq:pzm}). The unknown hydrologic parameters are the unit area storage capacity, $\overline{w}$, the fraction, $\mu$, partitioning storage between an upper and lower soil layer, the near-stream area, $\beta$, and the baseflow index, $B_I$.

\begin{figure*}
\noindent\includegraphics[width=\textwidth]{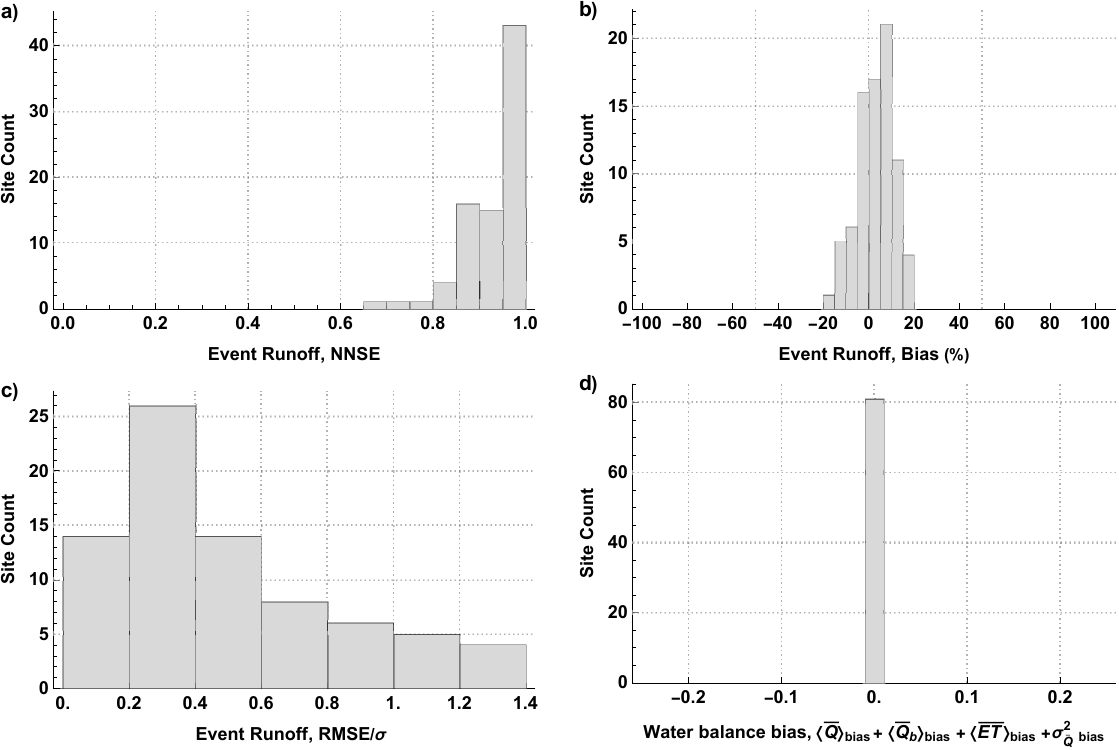}
\centering
\caption{For the 81 USGS gages sites, the count of sites by a) normalized Nash-Sutcliffe efficiencies, b) the bias percent, c) the root mean squared error (RMSE) normalized by one standard deviation and d) the absolute bias of the average daily fluxes of runoff, baseflow, evapotranspiration, and the storm event variance of runoff, which is exactly zero for all sites.}
\label{FigX1}
\end{figure*}

\subsection{Data Processing Steps}

The data processing provides an estimation of the climate parameters of the model and a characterization of the observed water balance required for calibrating the unknown hydrologic model parameters. The data processing is performed on observation data consisting of complete years of record (i.e., there are very few days of missing data over each year) and consists of two steps:
\begin{enumerate}
    \item Calculating the model climatic parameters of the long-term dryness index $\langle D_I \rangle$ from observed $\langle\overline{\text{PET}}\rangle$ and rainfall $\langle\lambda \overline{\alpha}\rangle$, and calculating the long-term average rainfall parameters $\langle\overline{\alpha}\rangle$, $\langle\overline{\alpha}_{\circ}\rangle$, $\langle\overline{\alpha}_{*}\rangle$, and $\omega$ by fitting the PDFs of Eqs. (\ref{eq:pR}) and (\ref{eq:pzm}) to storm event rainfall totals from the entire record.
    \item Calculating the observed water balance partitioning, i.e., $\frac{\langle \overline{Q}_b \rangle}{\langle \overline{Q}_f \rangle}^{\text{obs}}$ and $\frac{\langle \overline{ET} \rangle}{\langle \overline{R} \rangle}^{\text{obs}}$, the observed runoff variance, ${\sigma^2_{\overline{Q}}} ^{\text{obs}}$, and the observed runoff quantiles, $\overline{Q}_p^{\text{obs}}$. Note that the runoff variance and quantiles are based on aggregating the runoff to storm event totals.
\end{enumerate}
These data processing steps were performed for the case study dicussed in this paper (for more details see \ref{sec:CalibrationDetails}).

\subsection{Calibration Steps}
Calibration aligns the observed and modeled water balance dynamics of Eqs. (\ref{eq:ET/R}) - (\ref{eq:Qsigma}), as well as the observed and modeled quantiles of runoff. For a watershed, this is provided by finding the optimal values of the unknown parameters $\overline{w}$, $\beta$, $\mu$, and $B_I$, that are calibrated in two steps: 
\begin{enumerate}
    \item Iterating through a set of $\beta$ value, e.g., $\beta\in\{0.01,.1,.2,.3,.4,.5,.6,.7,.8,.9, 1 \}$, and solving for the three unknowns of $\overline{w}$, $\mu$, and $B_I$ from the system of three equations (\ref{eq:ET/R}) - (\ref{eq:Qsigma}) where the left-hand-side of the respective equations are set equal to the observed values of $\frac{\langle \overline{Q}_b \rangle}{\langle \overline{Q}_f \rangle}^{\text{obs.}}$, $\frac{\langle \overline{ET} \rangle}{\langle \overline{R} \rangle}^{\text{obs.}}$, and ${\sigma^2_{\overline{Q}}} ^{\text{obs.}}$.    
    \item Selecting the optimal value of $\beta$ and associated solution set $\{\overline{w}, \mu, B_I\}$ that minimize the root mean squared error (RMSE) between the observed and modeled quantiles of runoff, i.e., $\overline{Q}_p^{\text{obs.}}\approx \overline{Q}_p^{\text{model}}$, where the modeled runoff quantiles are based on the inverse CDF derived from the runoff PDF of Eq. (\ref{eq:pQ_all}).
\end{enumerate}
The system of equations (\ref{eq:ET/R})- (\ref{eq:Qsigma}) is considered with the following substitutions: $\overline{\gamma}\longrightarrow \frac{\overline{w}}{\langle\overline{\alpha}\rangle}$ and $\overline{\gamma}_1\longrightarrow \frac{\overline{w}(1-\mu)}{\langle\overline{\alpha}\rangle}$; noting the terms $\langle \overline{\chi}_0 \rangle$, $\langle \omega_{\text{PET}}\rangle$, and $\langle \omega_{\lambda}\rangle$ of Eqs. (\ref{eq:x0_avg}) - (\ref{eq:w_PET}) depend on $\overline{w}$ and $\mu$; and noting the terms
$\langle \overline{\chi}_1\rangle$ and $p_{\overline{Q}}(\overline{Q})$ of Eqs. (\ref{eq:SCSCNxAverage}), (\ref{eq:SCSCNaverage}) and (\ref{eq:pQ_all})  depend on $\overline{w}$, $\beta$, $\mu$, and $B_I$. Significantly, with calibration, the model exactly matches both the variance of runoff, ${\sigma^2_{\overline{Q}}} ^{\text{obs.}}$,and the long-term water balance partitioning, $\frac{\langle \overline{Q}_b \rangle}{\langle \overline{Q}_f \rangle}^{\text{obs.}}$ and $\frac{\langle \overline{ET} \rangle}{\langle \overline{R} \rangle}^{\text{obs.}}$.

\begin{figure*}
\noindent\includegraphics[width=\textwidth]{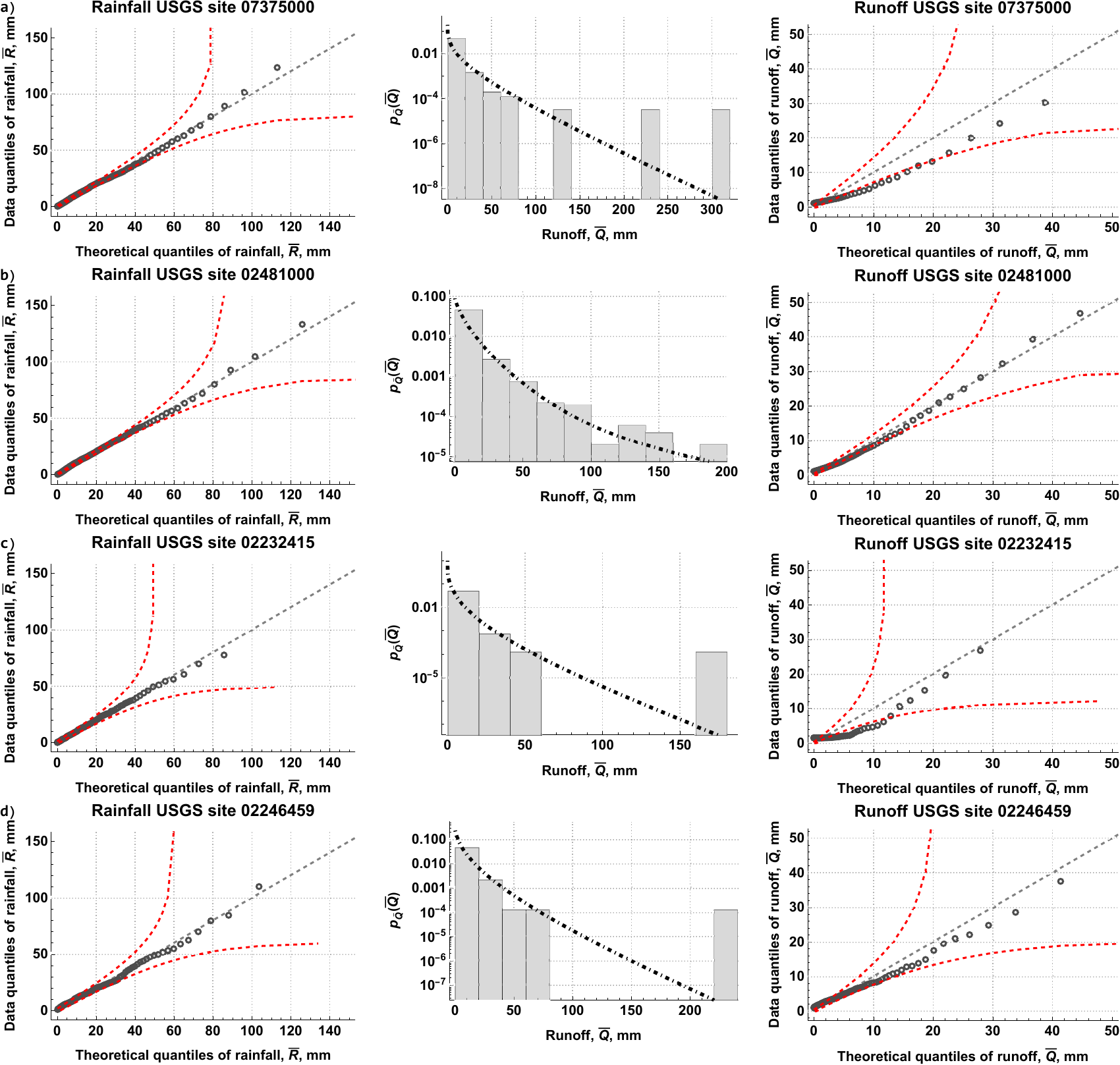}
\centering
\caption{Examples of the rainfall runoff model performance for a) the Tchefuncte River near Folsom, LA (95.50 sq. mi. drainage and 23-yrs. of data), b) the Biloxi River at Wortham, MS (96.2 sq. mi drainage and 34-yrs. of data), c) Taylor Creek Near Cocoa, FL (54.3 sq. mi. drainage and 5-yrs. of data) and d) Cedar River at San Juan Avenue at Jacksonville, FL (22.8 sq. mi. drainage and 6-yrs. of data).  Quantile-quantile (Q-Q) plots compare the theoretical (model) distribution to the data distribution for rainfall and runoff with the 1:1 line (dashed) and 95\% confidence bands (red, curved lines), while a log scale plot compares the theoretical runoff PDF of Eq. (\ref{eq:pQ_all}) (dotted-dashed line) with the empirical runoff PDF (histogram bars).}
\label{FigX2}
\end{figure*}

\subsection{Case Study}

The model was calibrated to 81 USGS gages with drainage areas between 2 and 298 square miles across watersheds in the vicinity of southern Louisiana and the St. Johns River basin in Florida. These watersheds were selected because they support existing compound flooding studies, and future work should expand the model validations to other regions. In addition to the daily flow data at each USGS gage, daily total rainfall and yearly average $\langle\overline{\text{PET}}\rangle$ data were sourced from Daymet and MODIS, respectively; see \ref{sec:CalibrationDetails}. Storm-event rainfall and runoff values were the basis of calibration, with storm event values aggregated from the data as described in \ref{sec:CalibrationDetails}. 

For all watersheds, the calibrated model accurately matched (with zero error) both the long-term water balance and runoff variance, while well capturing the runoff quantiles (Fig. \ref{FigX1}). For the modeled runoff quantiles, performance was assessed using normalized Nash-Sutcliffe Efficiency (NNSE), bias percentage (often called percent bias (PBIAS)), and RMSE normalized by runoff standard deviation (Fig. \ref{FigX1}). These metrics consistently indicated excellent performance, with NNSE values meeting criteria for a well-suited model, and RMSE skewed toward ``very good" to ``good" ranges. Across all 81 sites, the average NNSE was 0.93 while the median NNSE was 0.95. Gage records spanned from 5 to 34 years of data, and generally, gages with longer records had improved calibration accuracy.

Four representative watersheds were selected for detailed comparisons of rainfall and runoff (Fig. \ref{FigX2}). QQ plots show that the model in some cases may underestimate the runoff quantiles. This likely is because the model compensates for the zero runoff of the upper soil layer. A large part of this underestimation could be corrected by a future model reformulation where the upper soil layer produces runoff. Despite limited calibration parameters ($\mu$, $\overline{w}$, $B_I$, and $\beta$), the model performed well and should apply to most gaged watersheds. In future work, the calibrated parameters ($\mu$, $\overline{w}$, $B_I$, and $\beta$) could be linked to static and dynamic watershed characteristics, extending the model’s utility to ungaged watersheds. As the soil moisture PDFs, $p_{\overline{\chi}_0}(\overline{\chi}_0;t)$ and $p_{\overline{\chi}_1}(\overline{\chi}_1)$, represent the time-dependent PDFs, $p_{\overline{\chi}_0}(\overline{\chi}_0;t)$ and $p_{\overline{\chi}_1}(\overline{\chi}_1;t)$, integrated over long-term seasonal variability, the model should be interpreted as an average solution over the seasonal periods of many years.

\begin{figure*}
\noindent\includegraphics[width=\textwidth]{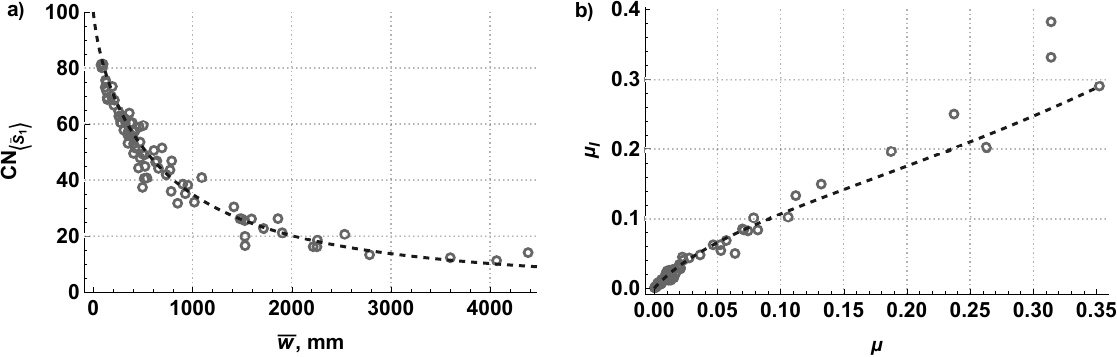}
\centering
\caption{For the 81 case study USGS watersheds, a) the relationship between the CN$_{\langle\overline{S}_1\rangle}$ (i.e., Eq. (\ref{eq:CN}) based on $\langle\overline{S}_1\rangle$) and the storage depth $\overline{w}$ and b) the relationship between the soil layer partitioning parameter, $\mu$, and the initial abstraction ratio, $\mu_I$, based on the ensemble average initial abstraction and maximum potential retention of Eqs. (\ref{eq:Iavg}) and (\ref{eq:Savg}). The dashed lines are the CN and initial abstraction ratio Eqs. (\ref{eq:CN}) and (\ref{eq:InitialAbstraction}) based on $\langle\overline{S}_1\rangle$, $\langle\overline{I}\rangle$, with the median parameter value across all 81 watersheds.}
\label{fig:10}
\end{figure*}

For the 81 watersheds analyzed, we further examined the relationships between model parameters and the initial abstraction ratio, $\mu_I$, as well as the CN$_{\langle\overline{S}_1\rangle}$ based on the average wetness condition of $\overline{S}_1$. Notably, as shown in Figure \ref{fig:10}, CN$_{\langle\overline{S}_1\rangle}$ correlates with the storage depth, $\overline{w}$. Similarly, the initial abstraction ratio, $\mu_I$, exhibits a strong correlation with the soil layer partitioning parameter, $\mu$. This correspondence suggests that the unknown model parameters ($\mu$, $\overline{w}$, $B_I$, and $\beta$) effectively capture the fundamental hydrologic behavior traditionally described by CN and the initial abstraction ratio, $\mu_I$. In particular, the storage depth, $\overline{w}$, and soil layer partitioning, $\mu$, offer physically grounded replacements for these empirical abstractions, providing a more mechanistic understanding of event-based infiltration and runoff dynamics within the broader context of water balance processes.

\begin{figure*}
\noindent\includegraphics[width=\textwidth]{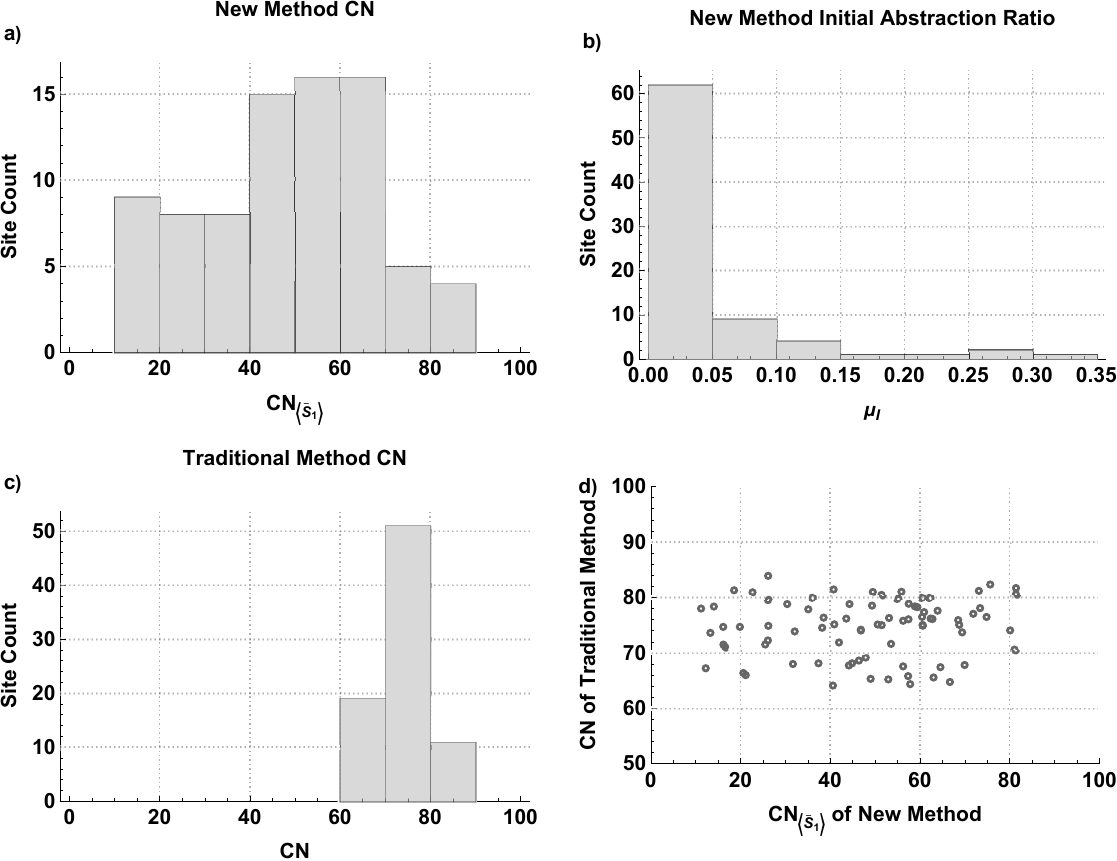}
\centering
\caption{For the 81 case study USGS watersheds, a) the resulting distribution of the calibrated model CN$_{\langle\overline{S}_1\rangle}$, b) the distribution of the calibrated model initial abstraction ratio, $\mu_I$, based on the ensemble average initial abstraction and maximum potential retention of Eqs. (\ref{eq:Iavg}) and (\ref{eq:Savg}), c) the distribution of the classic CN, which (for each watershed) is a spatial average of the 250-meter resolution CN data product of \citeA{jaafar2019gcn250}, and d) the relationship between the CN$_{\langle\overline{S}_1\rangle}$ of the new model and the CN of the classic SCS-CN approach.}
\label{fig:11}
\end{figure*}

The frequency of the observed CN$_{\langle\overline{S}_1\rangle}$ is considered alongside the frequency of observed initial abstraction ratios, $\mu_I$ (Fig. \ref{fig:11}). Across all sites, the average $\mu_I$ is approximately 0.045, aligning well with previous studies that suggest a representative value of 0.05 \cite{yuan2001modified}. A notable departure from conventional practice is observed in the lower bound of CN$_{\langle\overline{S}_1\rangle}$, which extends as low as 10, whereas the classic method constrained CN values to be 40 or higher \cite{ponce1996runoff}. Across the 81 watersheds, CN$_{\langle\overline{S}_1\rangle}$ spans a range of 10 to 85, while the traditional CN framework generally constrains values between 65 and 85 (Fig. \ref{fig:11}). For each watershed, these traditional CN values are a spatial average of the 250-meter resolution CN map based on the classic lookup table approach \cite{jaafar2019gcn250}. While the new model extends the SCS-CN method by incorporating the additional parameter, $\beta$, this alone does not explain the occurrence of CN$_{\langle\overline{S}_1\rangle}$ values well below 40 \cite{bartlett2016beyond}. The case study sites suggest that these lower CN values reflect substantially greater watershed storage, $\overline{w}$, than what is implicitly assumed by a CN of 40 (see Figs. \ref{fig:10} and \ref{fig:11}). Strikingly, no significant correlation is observed between the model CN$_{\langle\overline{S}_1\rangle}$ values and the classic CN values (Fig. \ref{fig:11}). This lack of correlation underscores the fundamental differences and limitations of the classic event-based SCS-CN method in contrast with this new framework, which explicitly accounts for the entire water balance dynamics over both storm event runoff and interstorm processes. 

\section{Discussion}
\label{sec:discussion}

This new model integrates event-based SCS-CNx runoff, semi-distributed hydrology, and stochastic ecohydrology into a unified, analytically tractable statistical model for quantifying the long-term statistics of watershed hydrology.  The model consists of PDFs that may be derived based on the spatial heterogeneity of any semi-distributed model. Here the SCS-CNx method serves as the basis for the model development, with two soil layers: the upper layer accounts for initial abstraction, and the lower layer govens SCS-CNx runoff incorporating spatial variability in soil moisture and storage. The dynamics of both soi layers are upscaled to a unit-area behavior using a mean-field approximation where the collective behavior of point-scale soil moisture converges to an average state. This upscaling approach, when applied to the point process of runoff, recovers the SCS-CNx rainfall-runoff curve. The same upscaling is applied to interstorm fluxes such as evapotranspiration and baseflow. These combined storm-event and inter-storm dynamics form the foundation for analytical PDFs that describe the water balance processes, including soil moisture distributions. While the model is for unmanaged watersheds without irrigation or control structures, future work could expand the statistical model to managed systems. For example, irrigation could be included in the PDF solution  \cite{bartlett2019jump,vico2010traditional,vico2011rainfed1,vico2011rainfed2, porporato2022ecohydrology}.

The probabilistic model PDFs effectively capture the long-term average partitioning of the water balance and runoff variability. This is demonstrated in the application to 81 watersheds, where the model precisely matches the observed runoff variance, accurately reproduces the observed average water balance, and faithfully replicates the overall runoff distributions. Climate parameters such as the dryness index, $\langle D_I \rangle$ and rainfall characteristics can be inferred from watershed-scale observations of $\langle \overline{\text{PET}} \rangle$ and $\langle \overline{\alpha} \lambda \rangle$; see \ref{sec:CalibrationDetails}. However, the hydrology-specific parameters ($\mu$, $\overline{w}$, $B_I$, and $\beta$) are typically unknown and require calibration.

By directly solving for the statistics of hydrological processes, the model reveals the relationships between hydrologic variability, climate statistics, watershed properties, and the magnitude of the continuous interstorm fluxes of baseflow and evapotranspiration. These relationships are crucial for engineering hydrology, particularly in design and planning applications. For example, the model relates the SCS-CN runoff variability to key climatic drivers such as rainfall frequency, $\langle \lambda \rangle$, average rainfall per storm, $\langle \overline{\alpha} \rangle$, and the Budyko dryness index $D_I$, and this allows the SCS-CN method to better address the growing challenges of accounting for runoff and the associated flash flood risks, as influenced by global warming and land-use changes \cite{labat2004evidence,yin2018large,velpuri2013analysis,alipour2020assessing}.

The model PDFs capture the variability of semi-distributed modeling physics, surpassing traditional ecohydrological models that are confined to point or plot scales or that emphasize streamflow without fully accounting for the entire water balance. Building on the principles of stochastic ecohydrology, which link climate, ecology, and hydrology at the point scale, this framework extends these relationships to provide novel insights at the watershed scale. For example, it derives a Budyko curve that incorporates not only the dryness index but also hydrological parameters such as the baseflow coefficient and SCS-CNx model parameters—--relationships that would otherwise remain obscured in the noisy Monte Carlo simulations of traditional stochastic hydrology. This analytical model offers an accessible yet comprehensive approach to quantifying watershed process variability amid the complex interplay of climate and land use changes.  For example, the probabilistic model enables more precise graduated risk assessments for flood mapping that account for climate change and could be the basis of new practitioner tools for evaluating climate adaptation and planning.

\subsection{Advancing the SCS-CN Method}

Despite its origin as a curve-fitting exercise developed on a limited number of small agricultural watersheds, the SCS-CN method has demonstrated remarkable staying power in both research and professional practice \cite{hawkins2014curve, garen2005curve}. This persistence can be attributed not only to its simplicity and ease of use but also to the fact that its underlying representation of rainfall-runoff partitioning captures key features of rainfall-runoff transformation \cite{bartlett2016beyond}. The original SCS-CN method has well-documented deficiencies, notably its inability to accurately represent runoff in forested watersheds, as well as its mis-characterization of ``violent" and ``complacent" runoff responses \cite{hawkins2015complacent,tedela2011runoff,tedela2008evaluation}. These deficiencies have been addressed in the extended version of the SCS-CN method that provides a more accurate representation of the rainfall-runoff response in forested watersheds while also capturing the ``violent" and ``complacent" runoff behaviors \cite{bartlett2016beyond,bartlett2017reply}. Although the performance of the SCS-CNx method has been validated with rank-order rainfall-runoff data, which approximates a rainfall-runoff curve for a time-averaged CN, the SCS-CNx method previously had not been evaluated under a dynamic CN evolving under continuous conditions \cite{michel2005soil,mishra2013soil,mishra2004long,cho2018spatially,wang2019soil,bartlett2016beyond}.

Here we expanded the purview of the SCS-CNx method, including a reinterpretation within a continuous time context that integrates stochastic ecohydrology and semi-distributed modeling. When $\beta = 0$, the model defaults to the rainfall-runoff partitioning of the original SCS-CN method, resulting in three unknown parameters ($\mu$, $\overline{w}$, and $B_I$), as opposed to the two unknowns (the CN and initial abstraction) in the event-based method. This allows accounting for both storm-event and inter-storm dynamics. As a result, the reinterpreted model analytically solves for the CN and initial abstraction values based on the coupled storm and inter-storm properties, as expressed by climate other calibrated hydrologic parameters. While quantifying the CN and initial abstraction was the main challenge in the event-based SCS-CNx method, in this reinterpreted SCS-CNx method, the challenge lies in determining the hydrologic parameter values ($\mu$, $\overline{w}$, $\beta$, and $B_I$) for ungaged locations. 

The new model not only couples the SCS-CNx rainfall-runoff partitioning to continuous fluxes of evapotranspiration and baseflow, but also directly provides the PDFs describing the process statistics. Unlike traditional deterministic hydrological models, which are calibrated to fit a specific time series, this statistical model is calibrated directly to the statistics of the long-term observational data, based on the convergence to a set of parameters (i.e., $\mu$, $\overline{w}$, $B_I$, and $\beta$) over an infinite number of plausible time series trajectories. This approach mitigates over-calibration and provides a robust model response across a wide range of conditions. Furthermore, by calibrating PDFs directly to observed statistics, the model avoids introducing substantial biases into key statistical metrics, which commonly occurs when employing  traditional stochastic hydrology approaches \cite{farmer2016deterministic,kirby1975model,labat2004evidence,lichty1978rainfall}. Accordingly, the new analytical PDFs allow for a clear and more accurate evaluation of the SCS-CNx method rainfall runoff partitioning, specifically ensuring consistency between the water balance partitioning and the runoff statistics.

While the 81 USGS watersheds examined in this study are insufficient to definitively prove a wide ranging applicability of the SCS-CNx method rainfall runoff partitioning, they demonstrate that the method is relevant outside the small agricultural watersheds and forested watersheds that guided the development of the SCS-CN and SCS-CNx methods. This highlights the broad applicability of SCS-CNx method, because these watersheds are predominantly located in low-lying wetland areas and, in some cases, karst-dominated aquifers (e.g., in Florida).

The new probabilistic model is simple with few parameters and consequently setting the model up for a new watershed takes minutes in terms of data processing and calibration. It would be of interest to calibrate the model for all HUC10-to-HUC12-scale USGS watersheds and examine its performance across this broader range. Such work could confirm the wide-ranging applicability of the SCS-CNx method while also providing insights on the unknown (calibrated) hydrologic parameters and how they relate to static watershed attributes. In this sense, the validation data could be foundational to deriving a lookup table for the hydrologic parameters  $\mu$, $\overline{w}$, $\beta$, and $B_I$, in line with the original CN tables.

The model’s probability density functions (PDFs) offer a pathway for advancing the SCS-CN and SCS-CNx methods beyond their traditional reliance on antecedent condition lookup tables, which often fail to account for climate variability. By incorporating coupled storm and interstorm dynamics, the model introduces an analytical PDF for the curve number (CN) that explicitly captures its variability as governed by climate and hydrologic process parameters. This reinterpretation of the SCS-CNx method allows for the selection of a median or average CN based on a complete statistical characterization that reflects both watershed attributes and climate conditions. The explicit linkage of the CN and initial abstraction to climate enhances the accuracy of runoff estimation for engineering design and planning purposes. It also improves runoff predictions for ungaged watersheds, particularly for extreme events critical to flood mapping and management. 

\begin{figure*}
\centering
\includegraphics[width=1.0\linewidth]{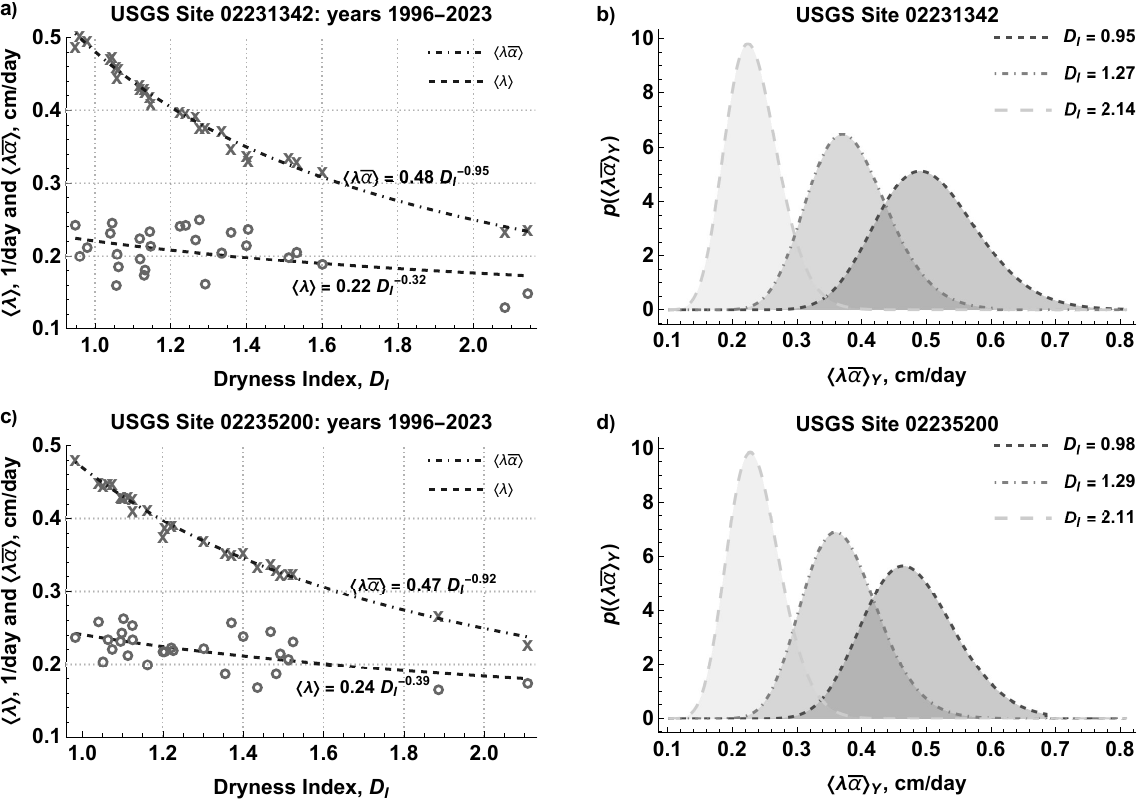}%
\caption{For 27 years of data for two USGS gages in the St. Johns watershed, panels a) and c) show the relationship between the average rainfall intensity, $\langle \alpha \lambda \rangle$, and the rainfall frequency, $\langle \lambda \rangle$ with the dryness index, $D_I$, while panels b) and d) for different $D_I$ show the PDF of the  average rainfall intensity for a single year, $p(\langle \alpha \lambda \rangle_Y)$, derived from the product of annual rainfall frequency and intensity, both modeled as Gamma-distributed random variables based on the averaging of an $n$ number of storm events linked to the  frequency of storms as a function of $D_I$.
\label{fig:lambdaalphaVar}}
\end{figure*}

\subsection{Model Performance under Varying Climate Regimes}

For evaluating the model performance under varying climate, observations may be compared to model results qualified by confidence intervals. For the water balance partitioning, $\frac{\langle \overline{ET} \rangle}{\langle \overline{R} \rangle}$ and $\frac{\langle \overline{Q}_b \rangle}{\langle \overline{Q}_f \rangle}$, these confidence intervals represent annual uncertainty in the frequency, $\langle \lambda\rangle_Y$ and average rainfall per storm, $\langle\overline{\alpha}\rangle_Y$. This uncertainty is inherent in the limited number of storms that occur year, and following the law of large numbers, $\langle \lambda\rangle_Y$ and $\langle\overline{\alpha}\rangle_Y$ only converge to the expected values as the number of repetitions of the year increases. For a given year with a specific dryness index, $D_I$, for that year, the likelihood of these averages is modeled as  $\langle \lambda \rangle_Y \sim \text{Gamma}\left(n, \frac{\langle \lambda \rangle}{n}\right)$ for the rainfall frequency and  $\langle \alpha \rangle_Y \sim \text{Gamma}\left(n, \frac{\langle \overline{\alpha} \lambda \rangle}{\langle \lambda \rangle n}\right)$ for the average rainfall per storm. Here, $\langle \overline{\alpha} \lambda \rangle$ and $\langle \lambda \rangle$ denote the median values for a given $D_I$ (see panels a and c of Fig. \ref{fig:lambdaalphaVar}). These Gamma distributions arise from averaging $n = \langle \lambda \rangle 365$ exponentially distributed variables for inter-storm durations and rainfall amounts, respectively. The resulting PDF $p(\langle \lambda \overline{\alpha} \rangle_Y)$ exhibits lower variance for wetter $D_I$ values and greater variance for drier $D_I$ values (see Fig. \ref{fig:lambdaalphaVar}). From the 95\% confidence interval of $p(\langle \lambda \overline{\alpha} \rangle_Y)$, we derive the 95\% confidence intervals for $\frac{\langle \overline{ET} \rangle}{\langle \overline{R} \rangle}$ and $\frac{\langle \overline{Q}_b \rangle}{\langle \overline{Q}_f \rangle}$.

Once calibrated to determine the parameters $\overline{w}$, $\beta$, $B_I$, and $\mu$, the model offers insights into how year-to-year changes in the dryness index $D_I$ and the storage index $\overline{\gamma}$ affect the watershed water balance. Along these lines, for each year from 1996 to 2023, we compared the modeled ratios $\frac{\langle \overline{ET} \rangle}{\langle \overline{R} \rangle}$ and $\frac{\langle \overline{Q}_b \rangle}{\langle \overline{Q}_f \rangle}$ to observed values from two watershed gages. Over the combine 56 years of record of the two gages, observations of $\frac{\langle \overline{Q}_b \rangle}{\langle \overline{Q}_f \rangle}$ fall within the model 95\% confidence interval for 54 out of 56 years, and for $\frac{\langle \overline{ET} \rangle}{\langle \overline{R} \rangle}$, this holds true for 45 out of 54 years. Overall, the model reasonably captures how year-over-year climate variations (as captured by changes in the dryness and storage indices, $D_I$ and $\langle\overline{\gamma}\rangle$) translate into changes in the dynamics of the water balance, as reflected in the ratios $\frac{\langle \overline{ET} \rangle}{\langle \overline{R} \rangle}$ and $\frac{\langle \overline{Q}_b \rangle}{\langle \overline{Q}_f \rangle}$.

\begin{figure*}
\centering
\includegraphics[width=1.0\linewidth]{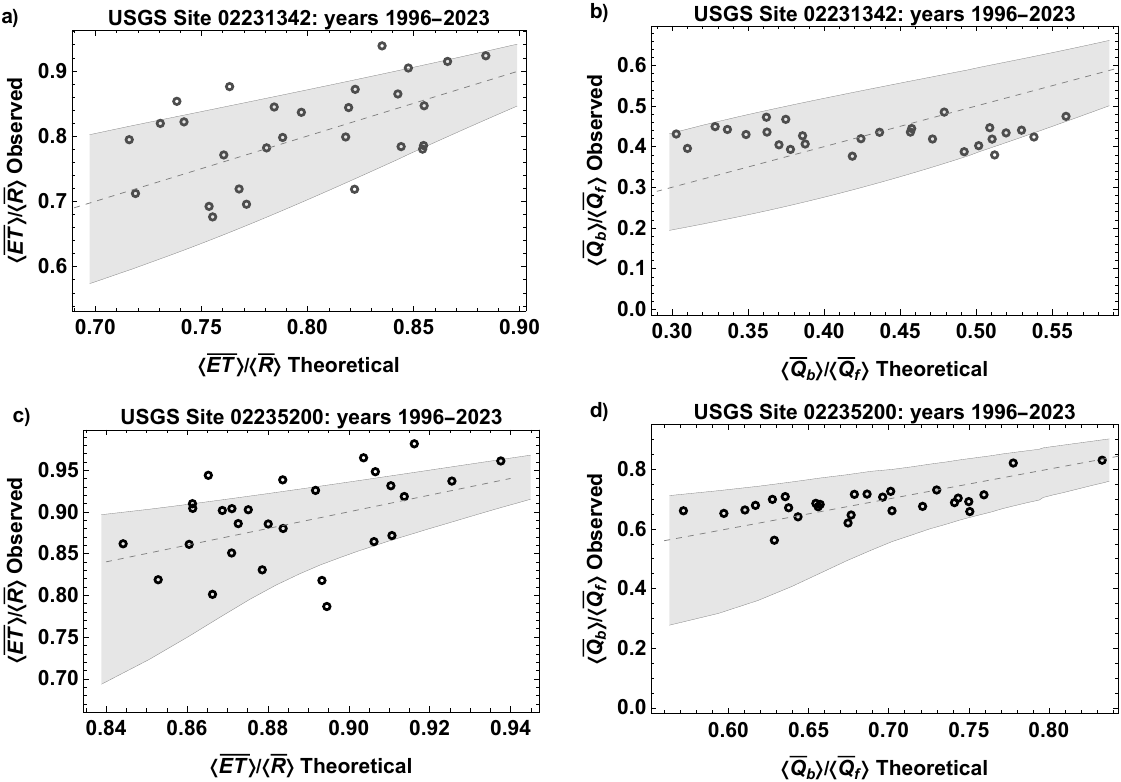}%
\caption{For each of the 28 years of data at each gage, a comparison of the observed water balance to the theoretical values predicted by the calibrated model for a) and c) $\langle \overline{ET}\rangle/\langle\overline{R} \rangle$  and b) and d) $\langle \overline{Q}_b\rangle/\langle \overline{Q}_f \rangle$ where in each case the 95\% confidence band (gray shading) is based on the likelihood of the average rainfall intensity based on the PDF $p(\langle\lambda\overline{\alpha}\rangle_Y)$ for a given dryness index, $D_I$ (see Fig. \ref{fig:lambdaalphaVar}).}
\end{figure*}

\subsection{Novel Semi-distributed Ecohydrology}

Spatially explicit studies of stochastic ecohydrology are made very difficult by the need to combine continuous descriptions of water fluxes in space and time with random variability of the external forcing. For this reason, existing ecohydrological theories mostly focus on the temporal domain \cite{porporato2022ecohydrology}. Here, for the first time, we have added implicitly spatial variability by adopting the approach of semi-distributed hydrology, which allowed us to account for spatial heterogeneity of rainfall and soil water-storage properties. This allowed us to come up with a self-consistent representation of watershed water storage, based on a mean-field approximation, as a single stochastic differential equation, which could then be solved via its Chapman-Kolmogorov formulation (see \ref{sec:dynamics_WC}).

The fact that distributions of water storage and fluxes be obtained at the watershed scale open the door to novel theoretical ecohydrological consideration regarding soil-climate and vegetation interactions. These are crucial for upscaled representation of water, energy and carbon fluxes as well as biogeochemical end ecological processes at such scales. We expect this to be useful also for earth system modeling and global climate models as well \cite{fatichi2015abiotic}.

\section{Concluding Remarks}

This study advances stochastic ecohydrology by extending its framework from the point and plot scale to the watershed scale, introducing analytical probability density functions (PDFs) that describe the long-term, time-averaged variability of water balance dynamics. These PDFs depend on key hydrological and climatic variables, including the long-term averages of rainfall frequency, $\langle \lambda \rangle$, storm total rainfall per event, $\langle \overline{\alpha}\rangle$, and the dryness index, $\langle D_I\rangle$. In a case study of 81 watersheds, the calibrated model accurately captured long-term water balance dynamics and runoff variability, achieving a median normalized Nash–Sutcliffe efficiency of 0.95 for the modeled SCS-CNx runoff quantiles. This excellent performance demonstrates that the  model successfully links the event-based SCS-CNx runoff with continuous inter-storm fluxes in a semi-distributed modeling context, offering precise, watershed-scale insights unattainable through Monte-Carlo approaches.

A key contribution of this framework is its ability to analytically link watershed scale soil moisture variability to climatic, hydrological, and ecological processes. For example, the Budyko-type curve derived from the model accounts not only for the dryness index $D_I$ but also for hydrological parameters such as watershed storage $\overline{w}$, baseflow fraction, $B_I$, and extended SCS-CN parameters, $\mu$ and $\beta$. This offers a detailed statistical accounting of water balance partitioning and variability, enabling applications in both event-based modeling and long-term ecohydrological analysis. The framework is particularly well-suited for future research into the sensitivity of hydrologic extremes to changing climatic and land-use conditions.

The analytical PDFs directly associate the likelihoods of the curve number (CN) and initial abstraction, $\overline{I}$, to climatic factors such as the Budyko dryness index $D_I$, baseflow index, $B_I$, watershed storage, $\overline{w}$, and parameters of the extended SCS-CN method.  While the Budyko dryness index, $D_I$, can be easily derived from observational data, further research is needed to map SCS-CN and hydrological parameters, such as baseflow index, $B_I$,  and watershed storage, $\overline{w}$, to specific watershed attributes like land use and soil type. Leveraging machine learning and remote sensing could streamline this mapping process, thereby enhancing the framework’s applicability to ungaged watersheds. As a modern counterpart to the traditional use of CN lookup tables, the model analytical PDF of the curve number (CN) can guide its selection based on climate factors. Furthermore, the model captures year-over-year changes in water balance partitioning driven by variations in the dryness index. Future research could leverage this model by integrating it with climate change scenarios, land-use planning, and operational assessments, offering a new tool to practitioners for evaluating the risks associated with emerging hydroclimatic changes.
Future research will expand the framework to incorporate time-varying PDFs, enabling operational forecasting of watershed variability under changing climatic conditions.


%
%
%
%
%
%
%

\begin{acknowledgments}
This work was funded through a cooperative endeavor agreement between The Water Institute and the Louisiana Office of Community Development as part of the Louisiana Watershed Initiative (LWI) funded via Catalog of Federal Domestic Assistance (CFDA) 14.228 Grant B-18-DP-22-0001. 
\end{acknowledgments}

%
%
%
%
%
%
%
%
%

\appendix

\section{Stochastic Dynamics of the Lower Soil Layer}
\label{sec:dynamics_WC}
This appendix formulates a stochastic representation of the lower soil layer water balance dynamics by integrating deterministic soil moisture processes with a probabilistic rainfall characteristics. The derivations are based on water balance components normalized by the total storage capacity, which include infiltration inputs, runoff, and losses from baseflow and evapotranspiration. From this water balance, the appendix establishes a probabilistic description of soil moisture variability, incorporating both the SCS-CNx runoff equation and a marked Poisson process representation of deep percolation infiltration (moisture inputs). The resulting evolution of the soil moisture PDF, as described by a Master equation, forms the basis for deriving the steady-state probability density function (PDF) of soil moisture, which is crucial for understanding long-term hydrologic variability and water balance dynamics. This probablistic formulation serve as a key component in linking soil moisture variability to the probabilistic distributions of runoff and evapotranspiration, as discussed in the main study.

The stochastic description of the lower soil layer is provided by the normalized water balance, where the terms of Eq. (\ref{eq:dx1/dt}) are divided by $\overline{w}(1-\mu)$ yielding
\begin{align}
\frac{d\overline{\chi}_1}{dt}= \overline{y}(t)-\overline{q}[\overline{y}(t),\overline{\chi}_1(t)] +\overline{m}_1\left[\overline{\chi}_1(t)\right],
\label{eq:dx1n/dt}
\end{align}
where $\overline{\chi}_1$ fluctuates between 0 and 1 because of normalized deep percolation infiltration, $\overline{y}(t)$, runoff, $\overline{q}[\overline{y}(t),\overline{\chi}_1(t)]$, and normalized losses from baseflow and evapotranspiration represented by
\begin{align}
\overline{m}_1\left(\overline{\chi}_1\right) = -\overline{k}\:\overline{\chi}_1,
\label{eq:mxSCSCN}
\end{align}
where $\overline{k} =\overline{k}_{\text{PET}}+\overline{k}_b$, $\overline{k}_{\text{PET}}=\frac{\overline{\text{PET}} \langle\omega_{\text{PET}}\rangle}{\overline{w}(1-\mu)}$, and $\overline{k}_{b}=\frac{\overline{Q}_{b,\max}}{\overline{w}(1-\mu)}$. The normalized runoff, $\overline{q}$ is retrieved from Eq. (\ref{eq:SCSCNx}) when infiltration and runoff ($\overline{Y}$ and $\overline{Q}$) are substituted with $\overline{y} \ \overline{w} (1-\mu)$ and $\overline{q}\:\overline{w} (1-\mu)$, respectively, i.e.,
\begin{align}
   \overline{q} =  \overline{y}(F_t +(1-F_t)\beta \:\overline{u}_1),
\label{eq:SCS-CNxNorm}
\end{align}
where the fraction of area producing runoff,
\begin{align}
F_t(\overline{u}_1,\overline{\chi}_1)=\frac{\overline{\chi}_1-\overline{u}_1}{1-\overline{u}_1},
\label{eq:Ft|ux}
\end{align}
depends on the evolving soil moisture content, $\overline{\chi}_1$, during the storm event, and the antecedent moisture content, $\overline{u}_1$, at the storm onset. Eq. (\ref{eq:Ft|ux}) is equivalent to the SCS-CN description of $F_t$ that is dependent on rainfall and the maximum potential retention $\overline{S}_1$, which is a value antecedent to the storm event, i.e.,  $\overline{S}_1 = \overline{w}(1-\mu)(1-\overline{u}_1)$; see \ref{sec:Ft_SCS}.

In the framework of the extended SCS-CNx method of Eqs. (\ref{eq:SCS-CNxNorm}) and (\ref{eq:Ft|ux}), the soil moisture, $\overline{\chi}_1$, increases from the intrastorm infiltration as the soil saturates, while $\overline{u}_1$ encapsulates the system memory from prior rainfall events and drydown periods. These representations align, respectively, with the well-established interpretations of `jump processes' in Marcus calculus and Itô calculus \cite{bartlett2018state, marcus1978modeling, marcus1981modeling, ito1944109}. Given that the soil moiture state variable cannot be subject to two distinct interpretations within the same infiltration transition (during a storm), the mathematical formulation must adopt a consistent calculus framework. To provide such consistency, we approximate the antecedent soil moisture $\overline{u}_1$ as a fraction $\vartheta$ of the continuously evolving state, yielding $\overline{u}_1\approx \vartheta\overline{\chi}_1$. This formulation aligns the overall process with the Marcus calculus description of infiltration as a `jump' transition. Consequently, from Eqs. (\ref{eq:SCS-CNxNorm}) and (\ref{eq:Ft|ux}), infiltration ($\overline{y}-\overline{q}$) may be stated as $\overline{y}\:\overline{b}(\overline{\chi}_1)$ with $\overline{b}(\overline{\chi}_1)$ defined as

\begin{align}
\overline{b}(\overline{\chi}_1)=1-\frac{\overline{\chi}_1(1-\vartheta)}{1-\vartheta\overline{\chi}_1}-\left(1-\frac{\overline{\chi}_1(1-\vartheta)}{1-\vartheta\overline{\chi}_1}\right)\beta\vartheta\overline{\chi}_1,
\label{eq:buzStratSCSCN}
\end{align}
where the fraction $\vartheta$ is found via a consistency condition that equates the exact transition probability  of the It\^o `jump' transition to this approximate version of the transition probability of the Marcus `jump' transition (see \ref{sec:consistency}).

For the soil moisture dynamics of Eq. (\ref{eq:dx1n/dt}), a probabilistic description emerges by treating deep percolation infiltration, $\overline{y}$, as a stochastic ``marked" Poisson process, with storm events arriving at rate $\lambda \langle\omega_{\lambda}\rangle$ and each marked by a random deep percolation amount \cite{bartlett2014excess, rodrigueziturbe1999probabilistic}. For this marked Poisson process representation of deep percolation, the dynamics of Eq. (\ref{eq:dx1n/dt}) have a probabilistic representation  governed by a Master equation \cite{cox1977theory}:
\begin{align} 
\partial_t \: p_{\overline{\chi}_1}(\overline{\chi}_1;t) =\partial_{\overline{\chi}_1}\left[\overline{m}_1(\overline{\chi}_1)p_{\overline{\chi}_1}(\overline{\chi}_1;t)\right]-\lambda \langle\omega_{\lambda}\rangle p_{\overline{\chi}_1}(\overline{\chi}_1;t) + \lambda \langle\omega_{\lambda}\rangle \int_{-\infty}^{\infty}\frac{p_{\overline{y}}(\overline{\eta}(\overline{\chi}_1)-\overline{\eta}(\overline{u}_1))}{|\overline{b}(\overline{\chi}_1)|}p_{\overline{\chi}_1}(\overline{u}_1;t)d\overline{u}_1, 
\label{eq:dp/dt} 
\end{align}
where the PDF, $p_{\overline{\chi}_1}(\overline{\chi}_1;t)$, changes over time due to three terms on the right-hand side: (1) a loss in probability density from interstorm losses, $\overline{m}(\overline{\chi}_1)$, in Eq. (\ref{eq:mxSCSCN}), (2) a loss based on the Poisson storm arrival rate, $\lambda \langle\omega_{\lambda}\rangle$, and (3) a gain from ``jumps" to new soil moisture values based on the deep percolation PDF, $p_{\overline{y}}(\overline{y})$, and runoff process, $\overline{b}(\overline{\chi}_1)$, in Eq. (\ref{eq:buzStratSCSCN}), where $\overline{\eta}(\overline{\chi}_1) = \int\frac{1}{\overline{b}(\overline{\chi}_1)}d\overline{\chi}$. For an exponential distribution, $p_{\overline{y}}(\overline{y}) = \overline{\gamma}_1e^{-\overline{\gamma}_1 \overline{y}}$ and under steady-state conditions ($\partial_t \: p_{\overline{\chi}_1}(\overline{\chi}_1;t) = 0$), Eq. (\ref{eq:dp/dt}) has a general solution provided in Eq. (46) of \citeA{bartlett2018state}, which is applied in Section \ref{sec:lowerlayer} to retrieve Eqs. (\ref{eq:PDFscsCNx}) and (\ref{eq:PDFscsCN}).

Both the marked Poisson process and the exponential distribution of rainfall effectively capture observed rainfall patterns across multiple events \cite{bartlett2014excess, rodrigueziturbe2004ecohydrology, rodrigueziturbe1999probabilistic}. In the exponential distribution of Eq. (\ref{eq:pR}), $\overline{\alpha}$ represents the average rainfall per storm event. Interestingly, despite rainfall being censored by the upper soil layer, the resulting deep percolation infiltration also is exponentially distributed with the same average $\overline{\alpha}$ due to the rescaling property of the exponential distribution. This property ensures that truncating the rainfall distribution still produces an exponential distribution for deep percolation, with the same average \cite{bartlett2014excess}. Consequently, the exponential distribution parameter for normalized deep percolation infiltration totals is $\overline{\gamma}_1 = \overline{\gamma}(1-\mu)$.

\section{Fraction of Runoff Producing Area}
\label{sec:Ft_SCS}
This appendix aims to provide a detailed derivation of the fraction of runoff-producing area, $F_t$, in the context of the SCS-CNx runoff model. By re-expressing $F_t$
in terms of antecedent soil moisture and evolving soil moisture during a storm event,  this appendix establishes an alternative form of $F_t$ that has not been used in the traditional SCS-CNx and SCS-CN methods. This alternative formulation offers a new perspective on runoff production and its relationship to soil moisture dynamics, contributing to the broader understanding of hydrological processes discussed in the main paper.

For the SCS-CNx runoff, the fraction of runoff producing area, $F_t = \frac{\overline{y}(1-\beta\overline{u}_1)}{(1-\overline{u}_1)+\overline{y}(1- \beta\overline{u}_1)}$, may be stated as $F_t = \frac{\overline{\chi}_1-\overline{u}_1}{1-\overline{u}_1}$ of Eq. (\ref{eq:Ft|ux}), which is in terms of the antecedent (to the storm event) soil moisture, $\overline{u}_1$, and a soil moisture, $\overline{\chi}_1$, that continuously evolves (i.e., increases) during the storm event. This equivalent, alternative form of $F_t$ of Eq. (\ref{eq:Ft|ux}) is found from $F_t = \frac{\overline{y}(1-\beta\overline{u}_1)}{(1-\overline{u}_1)+\overline{y}(1- \beta\overline{u}_1)}$  with a substitution of normalized infiltration, $\overline{y}$, with $f(\overline{\chi}_1,\overline{u}_1)$. This function $f(\overline{\chi}_1,\overline{u}_1)$ is retrieved by solving for $\overline{y}$ in the equality $\overline{\chi}_1-\overline{u}_1= \overline{y}-\overline{q}$, which relates the change in the soil moisture, $\overline{\chi}_1 -\overline{u}_1$ to the water input to the soil $\overline{y}-\overline{q}$ where $\overline{q}$ of Eq. (\ref{eq:SCS-CNxNorm}) is based on   $F_t = \frac{\overline{y}(1-\beta\overline{u}_1)}{(1-\overline{u}_1)+\overline{y}(1- \beta\overline{u}_1)}$  \cite{bartlett2016beyond}. 

\section{Equivalence of It\^o and Marcus Jump Interpretations}
\label{sec:consistency}

In the context of the SCS-CNx runoff model, this appendix establishes the equivalence between the Itô and Marcus jump interpretations of the maximum potential retention in the SCS-CNx method. The traditional SCS-CN and SCS-CNx methods follow Itô calculus for jump transitions during rainfall events, where runoff is determined by the antecedent soil moisture immediately before a rainfall pulse. However, the fraction of saturated area, $F_t$, of \ref{sec:Ft_SCS} suggests an alternative interpretation based on Marcus calculus. To ensure consistency between these approaches, a fraction $\vartheta$ is introduced, linking the antecedent soil moisture to the continuously evolving soil moisture state. This appendix derives a consistency condition for $\vartheta$ and provides both a numerical solution and an analytical approximation for $\vartheta$, reinforcing the validity of the Marcus calculus interpretation of runoff generation in the SCS-CNx framework.

The original SCS-CN method (Eq. \ref{eq:SCSCN}) and the extended SCS-CNx method (Eq. \ref{eq:SCSCNx}) both model rainfall infiltration based on the It\^o calculus jump transition. In both methods, runoff and infiltration are determined by the soil moisture, $\overline{u}_1$, immediately before the rainfall pulse. However, the steady-state solution of Eq. (\ref{eq:PDFscsCNx}) assumes that the antecedent soil moisture, $\overline{u}_1$, can be expressed as a fraction, $\vartheta$, of the continuously evolving state, i.e., $\overline{u}_1 = \vartheta \overline{\chi}_1$, as shown in Eqs. (\ref{eq:SCS-CNxNorm}) and (\ref{eq:Ft|ux}). Thus, the solution follows the Marcus calculus interpretation of the jump transition. In this approximation, the fraction $\vartheta$ is determined by ensuring consistency between the infiltration models described by the It\^o transition PDF, $W_I(\overline{\chi}_1 | \overline{u}_1)$, and the Marcus transition PDF, $W_S(\overline{\chi}_1 | \overline{u}_1, \vartheta)$, both of which represent the likelihood of soil moisture transitioning from one state to another during the storm event.

To establish this consistency, the fraction $\vartheta$ must satisfy the following condition: 
\begin{align}
\int_0^{\overline{\chi}_{1,\max}}\overline{\chi}_1\int_0^{\overline{\chi}_1}W_I(\overline{\chi}_1|\overline{u}_1,t)p_{\chi_1}(\overline{u}_1,\vartheta)d\overline{u}_1d\overline{\chi}_1=\int_0^{\overline{\chi}_{1,\max}}\overline{\chi}_1\int_0^{\overline{\chi}_1} W_S(\overline{\chi}_1|\overline{u}_1,\vartheta)p_{\overline{\chi}_1}(\overline{u}_1,\vartheta)d\overline{u}_1 d\overline{\chi}_1,
\label{eq:ConsistencyCondition1}
\end{align}
where $p{\overline{\chi}_1}(\overline{\chi}_1, \vartheta)$ is the steady-state PDF of Eq. (\ref{eq:PDFscsCNx}). The left-hand side of this equation represents the average It\^o jump transition, while the right-hand side represents the average Marcus jump transition. In other words, on average, the It\^o and Marcus interpretations of the SCS-CNx infiltration are equivalent. This consistency condition enforces an unbiased representation of the jump infiltration transition within the master equation (\ref{eq:dp/dt}), which governs the evolution of the soil moisture PDF in Eq. (\ref{eq:PDFscsCNx}). 

The respective It\^o and Marcus transition PDFs, $W_I(\overline{\chi}_1|\overline{u}_1)$ and $W_S(\overline{\chi}_1|\overline{u}_1,\vartheta)$, are based on Eqs. (17) and (26) of \citeA{bartlett2018state},
\begin{align}
\label{eq:PDFwxuItoSCSCN}
W_I(\overline{\chi}_1|\overline{u}_1)&=\lambda\frac{(1-\overline{u}_1)^2 }{(1-\overline{\chi}_1)^2 (1-\beta  \overline{u}_1)}\overline{\gamma} e^{-\frac{\overline{\gamma} (1-\overline{u}) (\overline{\chi}_1-\overline{u}_1)}{(1-\overline{\chi}_1) (1-\beta \overline{u}_1)}}\\
W_S(\overline{\chi}_1|\overline{u}_1,\vartheta)&=\lambda\frac{1}{b(\overline{\chi}_1; \vartheta)}\overline{\gamma} e^{-\overline{\gamma} \ln \left[\left(\frac{1-\overline{\chi}_1}{1-\overline{u}_1}\right)^{-\frac{\beta  (1-\vartheta )}{\beta  (1-\beta \vartheta )}} \left(\frac{1-\beta  \overline{\chi}_1 \vartheta }{1-\beta  \overline{u}_1 \vartheta }\right)^{-\frac{1-\beta }{\beta  (1-\beta  \vartheta )}}\right]},
\label{eq:PDFwxuStratSCSCN}
\end{align}
where Eq. (\ref{eq:PDFwxuItoSCSCN}) is based on the  normalized infiltration $\overline{b}(\overline{u}_1,\overline{z}) = \overline{z} - \overline{q}$, as derived from $\overline{q}$ in Eq. (\ref{eq:SCS-CNxNorm}), with $F_t = \frac{\overline{z}(1-\beta \overline{u}_1)}{(1-\overline{u}_1) + \overline{z}(1-\beta \overline{u}_1)}$. In contrast, Eq. (\ref{eq:PDFwxuStratSCSCN}) is based on the normalized infiltration $\overline{z} \cdot \overline{b}(\overline{\chi}_1; \vartheta)$, where $\overline{b}(\overline{\chi}_1; \vartheta)$ is given by Eq. (\ref{eq:buzStratSCSCN}). Both transition PDFs can be expressed in terms of jump transitions by substituting $\overline{\chi}_1 = \Delta \overline{\chi}_1 + \overline{u}_1$. 

\begin{figure*}
\centering
\includegraphics[width=1\linewidth]{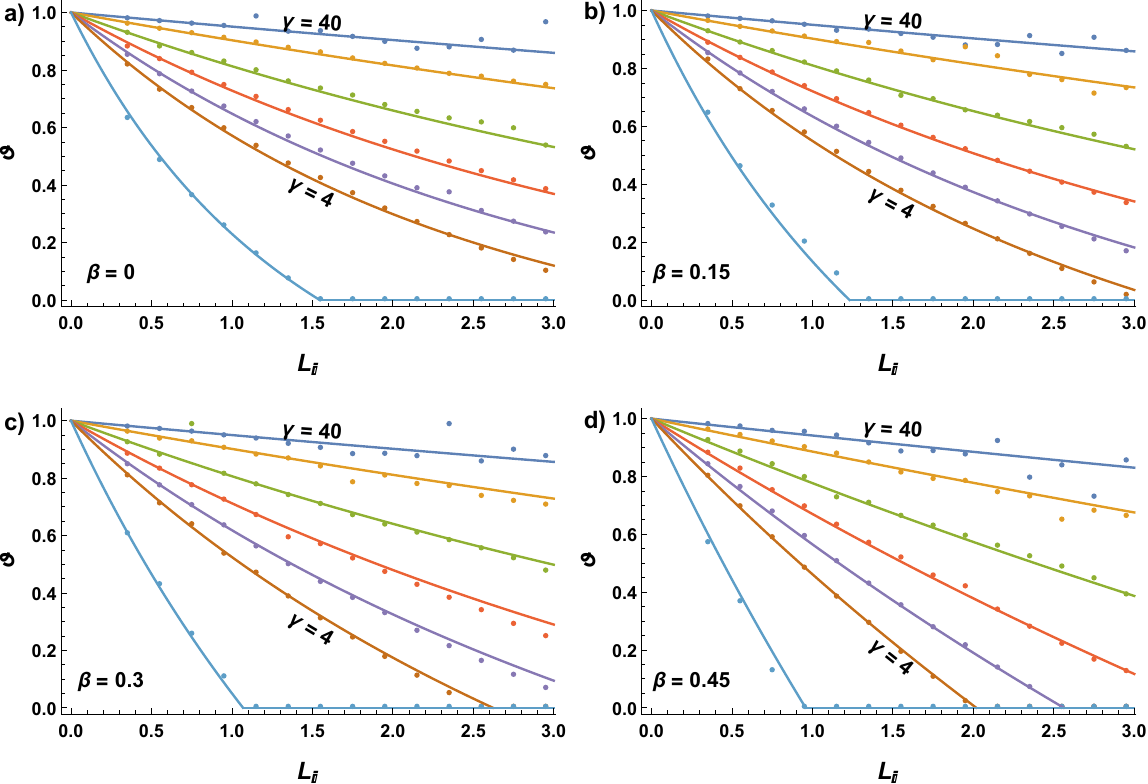}%
\caption{\label{fig:Fig51} Comparison verifying the goodness of fit between the $\vartheta$ calculated numerically from the consistency condition of Eq. (\ref{eq:ConsistencyCondition1}) (dots) and the values of Eq. (\ref{eq:vartheta}) (lines) for a range of parameters. The lines (in ascending order) are for $\overline{\gamma}=2, 4, 5, 6.66, 10, 20, 40$. (Color version available online).}
\end{figure*}
 
Although the fraction $\vartheta$ can be determined numerically from Eq. (\ref{eq:ConsistencyCondition1}), it may also be approximated with an ansatz---an educated estimate of the analytical functional form of $\vartheta$. Based on the initial numerical results, $\vartheta$ decreases with $\frac{L_I}{\overline{\gamma}_1}$, where the decrease is controlled by a power term that adjusts its concavity. An exponential term $e^{-\frac{L_I}{\overline{\gamma}_1}}$ reasonably governs the convexity of this decrease. Numerical solutions suggest that $\vartheta$ can be reasonably expressed as a function of $\overline{\gamma}$, $\beta$, and $L_I$, such that: 
\begin{align}
\vartheta &=
\begin{cases}
\max\left[e^{-2\left(1-\xi_1\beta^2\right)\frac{L_I}{\overline{\gamma}_1 ^{1-\beta^2/2}}}-(\xi_2+\xi_3\beta+\xi_4\beta^{4.5})\frac{L_I}{\overline{\gamma}_1^{2\left(1-\xi_1\beta^2\right)}},0\right], &0\leq \beta \leq 0.5 \\
\max\left[e^{-\frac{L_I}{\overline{\gamma}_1^{7/8}}}-(\xi_7+\xi_8\beta+\xi_9\beta^{13.5})\left(\frac{L_I}{\overline{\gamma}_1}\right)^{1+\xi_5+\xi_6 \beta^{1.5}},0\right], &0.5 <\beta < 1 \\
\max\left[e^{-\frac{L_I}{\overline{\gamma}_1}}-\xi_{10}\:\frac{L_I}{\overline{\gamma}_1},0\right],& \beta =1,
\end{cases}
,
\label{eq:vartheta}
\end{align}
where to minimize the error with the consistency condition of Eq. (\ref{eq:ConsistencyCondition1}), the constants are $\xi_1 = 2$, $\xi_2 = 0.5$, $\xi_3 = 2.75$, $\xi_4 = -16.29$, $\xi_5=-0.2$, $\xi_6=0.566$, $\xi_7=-1.352 $, $\xi_8 =5 $, $\xi_9 =57.1$, and $\xi_{10} =2.66$.  This analytical function (\ref{eq:vartheta}) closely matches the numerical solution for $\vartheta$ across a range of plausible parameter values. As shown in Figure \ref{fig:Fig51}, it provides an excellent fit for different values of $\beta$ across watersheds with limited storage ($\overline{\gamma} = 2$), significant storage ($\overline{\gamma} = 40$), and for both very wet ($L_I = 0$) and very dry ($L_I = 3$) environments.

\section{PDF Normalization Constants, Averages, and Variances}
\label{sec:ConAvgVar}

Here, we provide the soil moisture PDF normalization constants, averages, and variances. For the lower soil layer PDF of Eq. (\ref{eq:PDFscsCNx}), $p_{\overline{\chi}_1}(\overline{\chi}_1)$, the normalization constant is 
\begin{align}
N=&\frac{\Gamma\left(\frac{\overline{\gamma}_1(1-\vartheta)}{1-\beta\vartheta}+1+\frac{\overline{\gamma}_1 }{L_I}\right)}{\Gamma\left(\frac{\overline{\gamma}_1(1-\vartheta)}{1-\beta\vartheta}+1\right)\Gamma\left(\frac{\overline{\gamma}_1 }{L_I}\right)\,_2F_1\left(-\frac{\overline{\gamma}_1(1-\beta)}{\beta(1-\beta\vartheta)},\frac{\overline{\gamma}_1}{L_I};\frac{\overline{\gamma}_1(1-\vartheta)}{1-\beta\vartheta}+1+\frac{\overline{\gamma}_1 }{L_I};\beta\vartheta\right)}, 
\end{align}
the average soil moisture is given by
\begin{align}
\langle\overline{\chi}_1\rangle=\frac{\overline{\gamma}_1 \,_2F_1\left(-\frac{\overline{\gamma}_1(1-\beta)}{\beta(1-\beta\vartheta)},1+\frac{\overline{\gamma}_1}{L_I};\frac{\overline{\gamma}_1(1-\vartheta)}{1-\beta\vartheta}+2+\frac{\overline{\gamma}_1 }{L_I};\beta\vartheta\right)}{\left(\overline{\gamma}_1+L_I(1+\overline{\gamma}_1\frac{(1-\vartheta)}{1-\beta\vartheta})\right)\,_2F_1\left(-\frac{\overline{\gamma}_1(1-\beta)}{\beta(1-\beta\vartheta)},\frac{\overline{\gamma}_1}{L_I};\frac{\overline{\gamma}_1(1-\vartheta)}{1-\beta\vartheta}+1+\frac{\overline{\gamma}_1 }{L_I};\beta\vartheta\right)},
\label{eq:SCSCNxAverage}
\end{align}
and the variance is given by
\begin{align}
\sigma^2_{\overline{\chi}_1}=\frac{\overline{\gamma}(L_I+\overline{\gamma}_1) \,_2F_1\left(-\frac{\overline{\gamma}_1(1-\beta)}{\beta(1-\beta\vartheta)},2+\frac{\overline{\gamma}_1}{L_I};\frac{\overline{\gamma}_1(1-\vartheta)}{1-\beta\vartheta}+3+\frac{\overline{\gamma}_1 }{L_I};\beta\vartheta\right)}{\left(\overline{\gamma}_1+L_I(1+\overline{\gamma}_1\frac{(1-\vartheta)}{1-\beta\vartheta})\right)\left(\overline{\gamma}_1+L_I(2+\overline{\gamma}_1\frac{(1-\vartheta)}{1-\beta\vartheta})\right)\,_2F_1\left(-\frac{\overline{\gamma}_1(1-\beta)}{\beta(1-\beta\vartheta)},\frac{\overline{\gamma}_1}{L_I};\frac{\overline{\gamma}_1(1-\vartheta)}{1-\beta\vartheta}+1+\frac{\overline{\gamma}_1 }{L_I};\beta\vartheta\right)}-\langle\overline{\chi}_1\rangle^2,
\label{eq:SCSCNxVariance}
\end{align}
where $\,_2F_1\left(\cdot,\cdot;\cdot;\cdot\right)$ is the hypergeometric function \cite{abramowitz2012handbook}. Eqs. (\ref{eq:PDFscsCNx}), (\ref{eq:SCSCNxAverage}) and (\ref{eq:SCSCNxVariance}) are for an extended SCS-CNx method  that includes both saturation excess (i.e., threshold excess) runoff as well as pre-threshold runoff (prior to saturation at a point) for a fraction of area $\beta>0$ that is near the stream \cite{bartlett2016beyond}.

In the case where $\beta=0$ and the rainfall-runoff partitioning defaults to the SCS-CN method, the lower soil layer PDF of Eq. (\ref{eq:PDFscsCN}), $p_{\overline{\chi}_1}(\overline{\chi}_1)$, normalization constant is
\begin{align}
N=&\frac{\Gamma\left(\overline{\gamma}_1(1-\vartheta)+1+\frac{\overline{\gamma}_1 }{L_I}\right)}{\Gamma\left(\overline{\gamma}_1(1-\vartheta)+1\right)\Gamma\left(\frac{\overline{\gamma}_1 }{L_I}\right)\,_1F_1\left(\frac{\overline{\gamma}}{L_I};\overline{\gamma}_1(1-\vartheta)+1+\frac{\overline{\gamma}_1 }{L_I};-\overline{\gamma}_1\vartheta\right)},
\end{align}
where $\Gamma(\cdot)$ is the gamma function and  $\,_1F_1\left(\cdot;\cdot;\cdot\right)$ is the confluent hypergeometric function of the 1st kind \cite{abramowitz2012handbook}. For the original SCS-CN method ($\beta=0$), the average is given by
\begin{align}
\langle\overline{\chi}_1\rangle=\frac{\overline{\gamma} \,_1F_1\left(1+\frac{\overline{\gamma}}{L_I};\overline{\gamma}(1-\vartheta)+2+\frac{\overline{\gamma} }{L_I};-\overline{\gamma}\vartheta\right)}{\left(\overline{\gamma}+L_I(1+\overline{\gamma}(1-\vartheta))\right)\,_1F_1\left(\frac{\overline{\gamma}}{L_I};\overline{\gamma}(1-\vartheta)+1+\frac{\overline{\gamma} }{L_I};-\overline{\gamma}\vartheta\right)},
\label{eq:SCSCNaverage}
\end{align}
and the corresponding variance is given by
\begin{align}
\sigma^2_{\overline{\chi}_1}=\frac{\overline{\gamma}(L_I+\overline{\gamma}) \,_1F_1\left(1+\frac{\overline{\gamma}}{L_I};\overline{\gamma}(1-\vartheta)+2+\frac{\overline{\gamma} }{L_I};-\overline{\gamma}\vartheta\right)}{\left(\overline{\gamma}+L_I(1+\overline{\gamma}(1-\vartheta))\right)\left(\overline{\gamma}+L_I(2+\overline{\gamma}(1-\vartheta))\right)\,_1F_1\left(\frac{\overline{\gamma}}{L_I};\overline{\gamma}(1-\vartheta)+1+\frac{\overline{\gamma} }{L_I};-\overline{\gamma}\vartheta\right)}-\langle\overline{\chi}\rangle^2,
\label{eq:SCSCNvariance}
\end{align}
where $\,_1F_1\left(\cdot;\cdot;\cdot\right)$ is the confluent hypergeometric function of the 1st kind \cite{abramowitz2012handbook}.

\section{Derivation of the Runoff and ET PDFs}
\label{sec:PDF_derivations}

This appendix derives the PDFs for runoff and evapotranspiration based on the physical processes governing the soil moisture dynamics. Explicit runoff and ET distributions are formulated by linking deterministic runoff and ET dynamics to the stochastic nature of rainfall and soil moisture variability. The derivations incorporate the antecedent soil moisture PDF, rainfall statistics, and governing equations such as the SCS-CNx runoff equation and the PET-based evapotranspiration formulation. These dynamics establish a multivariate PDF for analyzing runoff and evapotranspiration variability and serve as key components in the long-term water balance analysis presented in the main study.

The runoff PDF is based on the antecedent soil moisture PDF of either Eq. (\ref{eq:PDFscsCNx}) or (\ref{eq:PDFscsCN}), the SCS-CNx equation (\ref{eq:SCSCNx}), and on a mixed exponential PDF of rainfall,
These components may be represented as a joint distribution
\begin{align}
p_{\overline{Q}\:\hat{\overline{R}}\:\overline{\chi}_1}(\overline{Q},\overline{R},\overline{u}_1) = \delta\left(\overline{R} F_t+\overline{R}(1-F_t)\beta \overline{u}_1)-\overline{Q}\right)\hat{p}_{\overline{R}}(\overline{R})p_{\overline{\chi}_1}(\overline{u}_1),
\label{eq:pQRx}
\end{align}
where $p_{\hat{\overline{R}}}(\overline{R})$ is the PDF of rainfall, $p_{\overline{\chi}_1}(\overline{u}_1)$ is the PDF of the soil moisture, and the conditional PDF $p_{\overline{Q}|\hat{\overline{R}}\: \overline{u}_1}(\overline{Q}|\overline{R},\overline{u}_1)$ is represented by a Dirac delta function $\delta(\cdot)$ and the SCS-CNx runoff equation of Eq. (\ref{eq:SCSCNx}) with $F_t = \frac{\overline{R}(1-P_I)}{\overline{S}+\overline{R}(1-P_I)}$. Note that $\overline{u}_1$ represents the soil moisture immediately antecedent to the storm event and the PDF $p_{\overline{\chi}_1}(\overline{u}_1)$ is given by either Eqs. (\ref{eq:PDFscsCNx}) or (\ref{eq:PDFscsCN}) because infiltration events are modeled by a homogeneous Poisson process at a constant rate \cite<see>{bartlett2014excess,daly2007intertime}. Integrating Eq. (\ref{eq:pQRx}) over rainfall, $\overline{R}$, and the antecedent soil moisture, $\overline{u}_1$, retrieves the univariate runoff PDF of Eq. (\ref{eq:pQ}).

The evapotranspiration PDF is defined with similar logic and is based on the upper and lower soil layer PDFs and the expression for evapotranspiration, i.e., $\overline{ET} = \overline{\text{PET}}\overline{\chi}_0 +\overline{\text{PET}} \omega_{\text{PET}}(\overline{\chi}_0)\overline{\chi}_1$, where $\omega_{\text{PET}}(\overline{\chi}_0)=1-\overline{\chi}_0$. A joint PDF may be based on these components, i.e., 

\begin{align}
p_{\overline{ET}\:\overline{\chi}_0\:\overline{\chi}_1} (\overline{ET},\overline{\chi}_0,\overline{\chi}_1)= \delta\left(\overline{\text{PET}}\overline{\chi}_0+\overline{\text{PET}}\omega_{\text{PET}}(\overline{\chi}_0) \overline{\chi}_1-\overline{ET}\right)p_{\overline{\chi}_0}(\overline{\chi}_0)p_{\overline{\chi}_1}(\overline{\chi}_1),
\end{align}
where the soil moisture PDFs are given by Eqs. (\ref{eq:px0}), (\ref{eq:PDFscsCNx}), and (\ref{eq:PDFscsCN}) and the conditional PDF $p_{\overline{ET}|\overline{\chi}_0\: \overline{\chi}_1}(\overline{ET}|\overline{\chi}_0,\overline{\chi}_1)$ is represented the Dirac delta function $\delta(\cdot)$. Integrating first over $\overline{\chi}_1$ and then over $\overline{\chi}_0$ yields the evapotranspiration PDF of Eq. (\ref{eq:pET}).

\section{Master Equation: Long-Term Water Balance Statistics}
\label{sec:long-term-balance}

This appendix presents a detailed derivation of long-term water balance statistics based on a modified version of the master equation governing the evolution of the soil moisture PDF. By integrating non-stationary probability density functions (PDFs) over extended observation periods, this appendix demonstrates how time-invariant PDFs (such as the steady-state PDFs of this model) capture hydrologic variability across seasonal, annual, and multi-year timescales. The formulations extend the steady-state solutions from the main text, showing how they may emerge as time-averaged approximations of transient dynamics. This approach provides a framework for approximating long-term water balance statistics by reconciling transient hydrologic variability with time-averaged statistical representations.

The steady-state solutions given in Eqs. (\ref{eq:px0}), (\ref{eq:PDFscsCNx}), and (\ref{eq:PDFscsCN}) can also be interpreted as time-dependent (non-stationary) probability density functions (PDFs) integrated over an extended observation period, $T$, which may span multiple seasons to several decades. These PDFs encapsulate long-term temporal variability within $T$, capturing fluctuations across seasonal, annual, and multi-year scales. Over this period, the PDFs depend on the long-term average statistics of rainfall frequency, potential evapotranspiration, and the distributions of normalized infiltration and rainfall amounts. These quantities are defined as:

\begin{align}
\langle \lambda \rangle &= \int_0^{T} \frac{1}{T}\lambda(t)dt \\
\langle \overline{\text{PET}} \rangle &= \int_0^{T} \frac{1}{T}\text{PET}(t)dt \\
p_{\overline{y}}(\overline{y}) &= \int_0^T  p_{\overline{y}t}(\overline{y};t)p_{t_{\lambda}}(t)dt \\
p_{\overline{z}}(\overline{z}) &= \int_0^T  p_{\overline{z}t}(\overline{z};t)p_{t_{\lambda}}(t)dt 
\label{eq:pzt}
\end{align}
where $\langle \lambda\rangle$ and $\langle \overline{\text{PET}}\rangle$ respectively represent the time averaged storm frequency and potential evapotranspiration, $\langle \alpha\rangle$ is the average of the assumed exponential distributions for infiltration and rainfall, and $p_{t_{\lambda}}(t)=\frac{\lambda(t)}{\langle \lambda \rangle T}$ represents the likelihood (relative probability density) of the storm events at time $t$ within the total observation period, $T$. 

By making the following substitutions: $\lambda \longrightarrow \langle \lambda \rangle$, $\overline{\gamma} \longrightarrow \langle \overline{\gamma} \rangle = \frac{\langle \overline{\alpha} \rangle}{\overline{w}}$, and $D_I \longrightarrow \langle D_I \rangle = \frac{\langle \overline{\text{PET}} \rangle}{\langle \lambda \rangle \langle \alpha \rangle}$,   the model PDFs of Eqs. (\ref{eq:px0})-(\ref{eq:pQ}) approximate the non-stationary PDFs integrated over the observation period $T$. Similarly, the rainfall PDFs of Eqs. (\ref{eq:pR}) and (\ref{eq:pzm}) now follow Eq. (\ref{eq:pzt}), i.e., $p_{\overline{R}}(\overline{R}) = \int_0^T p_{\overline{R}t}(\overline{R};t)p_{t_{\lambda}}(t)dt$. With these substitutions, the soil moisture PDFs of Eqs. (\ref{eq:px0}) and (\ref{eq:PDFscsCNx}) represent a distribution between two limiting cases of 1) drier interstorm conditions respresented by $p_{\overline{\chi}_{(\cdot),C}}(\overline{\chi})$ and 2) wetter storm event conditions represented by $p_{\overline{\chi}_{(\cdot),\lambda}}(\overline{\chi})$, i.e.,  $p_{\overline{\chi}_{(\cdot),C}}(\overline{\chi}) \leq p_{\overline{\chi}_{(\cdot)}}(\overline{\chi}) \leq  p_{\overline{\chi}_{(\cdot),\lambda}}(\overline{\chi})$, where $_{(\cdot)}$ is a placeholder that is 0 for the upper soil layer and 1 for the lower soil layer. These limiting cases are expressed as
\begin{align}
\label{eq:pxC}
p_{\overline{\chi}_{(\cdot),C}}(\overline{\chi}) &= \int_0^T p_{\overline{\chi}_{(\cdot)}}(\overline{\chi};t) p_{t_{C}}(t) dt \\
p_{\overline{\chi}_{(\cdot),\lambda}}(\overline{\chi}) &= \int_0^T p_{\overline{\chi}_{(\cdot)}}(\overline{\chi};t)p_{t_{\lambda}}(t) dt
\label{eq:pxl}
\end{align}
where $p_{t_{C}}(t)$ quantifies the relative activity of continuous processes at time, $t$, relative to the overall observation period, $T$.  For the upper soil layer $p_{t_{C}}(t)=\frac{\overline{\text{PET}}(t)}{\langle \overline{\text{PET}} \rangle T}$ and for the lower soil layer $p_{t_{C}}(t)=\frac{\overline{\text{PET}}(t)\langle\omega_{\text{PET}}\rangle+\overline{Q}_{b,\max}}{(\langle \overline{\text{PET}} \rangle \langle\omega_{\text{PET}}\rangle +\overline{Q}_{b,\max})T}$, where $\overline{\text{PET}}(t)$ is the time varying potential evapotranspiration The two-layer soil representation provides flexibility, allowing the water balance to be preserved on average even if variability in individual components is not fully captured within the approximated solutions between $p_{\overline{\chi}_{(\cdot),C}}(\overline{\chi})$ and $p_{\overline{\chi}_{(\cdot),\lambda}}(\overline{\chi})$.

To illustrate how the steady-state solution, incorporating the specified substitutions, captures long-term hydrologic dynamics, we derive this PDF using a time-integrated form of the master equation over the observation period $T$.  For the lower soil layer, this version of the master equation is  based on $p_{\overline{\chi}_{(\cdot),C}}(\overline{\chi})$ and $p_{\overline{\chi}_{(\cdot),\lambda}}(\overline{\chi})$ and is given by:
\begin{align}
0 =& \frac{d}{d\overline{\chi}_1}\bigg[\overbrace{\frac{\langle\overline{\text{PET}}\rangle \langle\omega_{\text{PET}}\rangle+\overline{Q}_{b,\max}}{\overline{w}(1-\mu)} \overline{\chi}_1\:}^{\overline{m}_1(\overline{\chi}_1)}p_{\overline{\chi}_{1,C}}(\overline{\chi}_1)\bigg]-\langle\lambda\rangle  \langle\omega_{\lambda}\rangle p_{\overline{\chi}_{1,\lambda}}(\overline{\chi}_1)\nonumber \\
&+ 
\langle\lambda\rangle \langle\omega_{\lambda}\rangle \int_{-\infty}^{\infty} \frac{ p_{\overline{y}}(\overline{\eta}(\overline{\chi}_1)-\overline{\eta}(\overline{u}_1))}{|\overline{b}(\overline{\chi}_1)|}p_{\overline{\chi}_{1,\lambda}}(\overline{u}_1)d\overline{u}_1,
\label{eq:px1lt0}
\end{align}
where the continuous loss consists of the time-averaged component, $\frac{\langle\overline{\text{PET}}\rangle \langle\omega_{\text{PET}}\rangle+\overline{Q}_{b,\max}}{\overline{w}(1-\mu)}$. Here, for the assumed linear losses to baseflow and evapotranspiration, note that $p_{\overline{\chi}_{1,C}}(\overline{\chi}_1) = \frac{\langle\overline{\text{PET}}\rangle}{\langle\overline{\text{PET}}\rangle+\overline{Q}_{b,\max}}p_{\overline{\chi}_{\text{PET}}}(\overline{\chi}_1)+\frac{\overline{Q}_{b,\max}}{\langle\overline{\text{PET}}\rangle+\overline{Q}_{b,\max}}p_{\overline{\chi}_1}(\overline{\chi}_1)$, where $p_{\overline{\chi}_{\text{PET}}}(\overline{\chi}_1)=\int_{-\infty}^{\infty}\frac{\text{PET}(t)}{T\langle\overline{\text{PET}}\rangle}p_{\overline{\chi}_1}(\overline{\chi}_1;t)dt$ and $p_{\overline{\chi}_1}(\overline{\chi}_1)=\int_{-\infty}^{\infty}\frac{1}{T}p_{\overline{\chi}_1}(\overline{\chi}_1;t)dt$. To solve Eq. (\ref{eq:px1lt0}), we substitute  $p_{\overline{\chi}_{1,C}}(\overline{\chi}_1)\longrightarrow p_{\overline{\chi}_1}(\overline{\chi}_1)$ and $p_{\overline{\chi}_{1,\lambda}}(\overline{\chi}_1)\longrightarrow p_{\overline{\chi}_1}(\overline{\chi}_1)$. This yields a solution for $p_{\overline{\chi}_1}(\overline{\chi}_1)$  that retains the functional form of the steady-state solutions given in Eqs. (\ref{eq:px0}), (\ref{eq:PDFscsCNx}), and (\ref{eq:PDFscsCN}). However, rather than representing a strictly steady-state process, this solution approximates the integration of the non-stationary dynamics over the observation period. Specifically, $p_{\overline{\chi}_1}(\overline{\chi}_1)$ lies between the two limiting cases defined in Eqs. (\ref{eq:pxC}) and (\ref{eq:pxl}), such that $p_{\overline{\chi}_{1,C}}(\overline{\chi}_1) \leq p_{\overline{\chi}_1}(\overline{\chi}_1) \leq  p_{\overline{\chi}_{1,\lambda}}(\overline{\chi}_1)$

Equation (\ref{eq:px1lt0}) is derived by taking the time average of both sides of the following master equation. This is achieved by multiplying by $1/T$ and integrating from 0 to $T$:

\begin{align}
\partial_t \: p_{\overline{\chi}_1}(\overline{\chi}_1;t) =& \partial_{\overline{\chi}_1}\left[\frac{\langle\overline{\text{PET}}\rangle \langle\omega_{\text{PET}}\rangle+\overline{Q}_{b,\max}}{\overline{w}(1-\mu)} \overline{\chi}_1 \: T p_{\overline{\chi}_1}(\overline{\chi}_1;t)p_{t_{C}}(t)\right]-\langle\lambda\rangle T \langle\omega_{\lambda}\rangle p_{\overline{\chi}_1}(\overline{\chi}_1;t)p_{t_{\lambda}}(t)  \nonumber \\
&+ 
\langle\lambda\rangle T \langle\omega_{\lambda}\rangle \int_{-\infty}^{\infty}\int_{-\infty}^{\infty} \frac{ \delta(\overline{\eta}(\overline{\chi}_1))-\overline{\eta}(\overline{u}_1)-y)}{|\overline{b}(\overline{\chi}_1)|}p_{\overline{y}}(y)p_{\overline{\chi}_{\lambda}}(\overline{u}_1)p_{t_{\lambda}|\overline{y}\:\overline{u}_1}(t|\overline{y},\overline{u}_1) d\overline{y} d\overline{u}_1, 
\label{eq:px1lt1}
\end{align}
where for this average on the left-hand-side, we  reasonably assume that $p_{\overline{\chi}_1}(\overline{\chi}_1;T)- p_{\overline{\chi}_1}(\overline{\chi}_1;0) = 0$ since the process is approximately cyclical. The familiar form of the master equation is recovered by simplifying Eq. (\ref{eq:px1lt1}) using these substitutions:  $p_{\overline{y}}(\overline{y})p_{\overline{\chi}_{\lambda}}(\overline{u}_1)p_{t_{\lambda}|\overline{y}\:\overline{u}_1}(t|\overline{y},\overline{u}) \longrightarrow p_{\overline{y}t}(\overline{y};t)p_{\overline{u}_1t}(\overline{u}_1;t)p_{t_{\lambda}}(t)$, $p_{t_{\lambda}}(t)\longrightarrow\frac{\lambda(t)}{\langle \lambda \rangle T}$, $p_{t_{C}}(t)\longrightarrow\frac{\overline{\text{PET}}(t)\langle\omega_{\text{PET}}\rangle+\overline{Q}_{b,\max}}{(\langle \overline{\text{PET}} \rangle\langle\omega_{\text{PET}}\rangle +\overline{Q}_{b,\max})T}$, and $\frac{\text{PET}(t)\langle\omega_{\text{PET}}\rangle+\overline{Q}_{b,\max}}{\overline{w}(1-\mu)}\overline{\chi}_1 \longrightarrow \overline{m}_1(\overline{\chi}_1,t) $, and integrating over $\overline{y}$ \cite{bartlett2018state}, i.e., 

\begin{align}
\partial_t \: p_{\overline{\chi}_1}(\overline{\chi}_1;t)  =& \frac{d}{d\overline{\chi}_1}\left[\overline{m}_1(\overline{\chi}_1;t) p_{\overline{\chi}_1}(\overline{\chi}_1;t)\right]-\lambda(t)\langle\omega_{\lambda}\rangle p_{\overline{\chi}_1}(\overline{\chi}_1;t) \nonumber\\
&+  \lambda(t)\langle\omega_{\lambda}\rangle\int_{-\infty}^{\infty} \frac{ p_{\overline{y}t}(\overline{\eta}(\overline{\chi}_1)-\overline{\eta}(\overline{u}_1)-\overline{y};t)}{|\overline{b}(\overline{\chi}_1)|}p_{\overline{\chi}_{1}}(\overline{u}_1,t)d\overline{u}_1.
\label{eq:dpt/dt}
\end{align}
This result of Eq. (\ref{eq:dpt/dt}) is the previous master equation (\ref{eq:dp/dt}), but now with an explicit characterization of $\lambda$ and $p_{\overline{y}}(\overline{y})$ as time varying.

\section{Data Processing Details}
\label{sec:CalibrationDetails}

This appendix provides details on the data processing methods used to organize and transform the watershed data for calibrating the probabilistic model to USGS streamflow data across different regions. It outlines the classification of USGS gages based on Hydrologic Unit Code (HUC-8) watersheds, the retrieval and processing of rainfall and evapotranspiration datasets, and the procedures for baseflow separation and runoff aggregation. These steps form the foundation for calculating key hydrological metrics, including the observed runoff quantiles and variance, which are central to the analyses presented in the main paper.

\begin{figure*}
\centering
\includegraphics[width=1\linewidth]{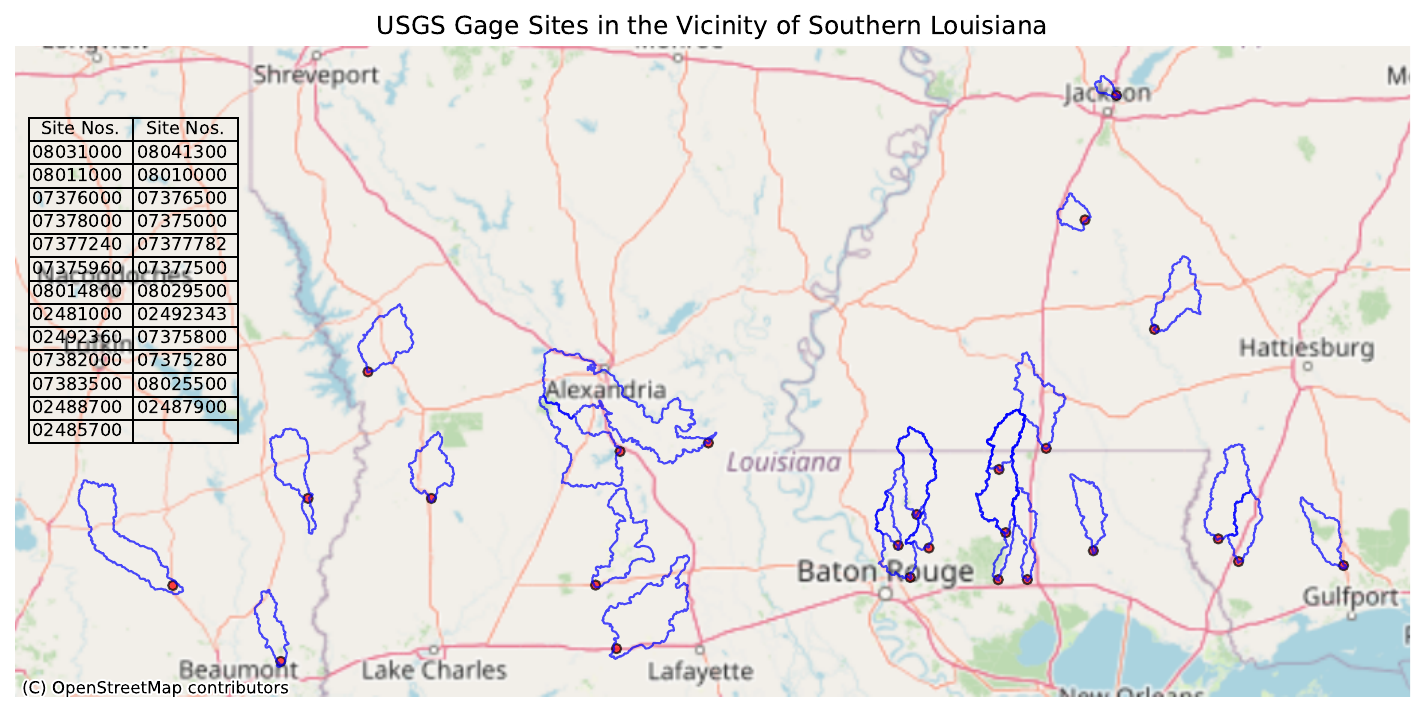}%
\caption{\label{fig:LAsites} The 25 USGS gage site locations with the corresponding watersheds boundaries.}
\end{figure*}

\begin{figure*}
\centering
\includegraphics[width=3in]{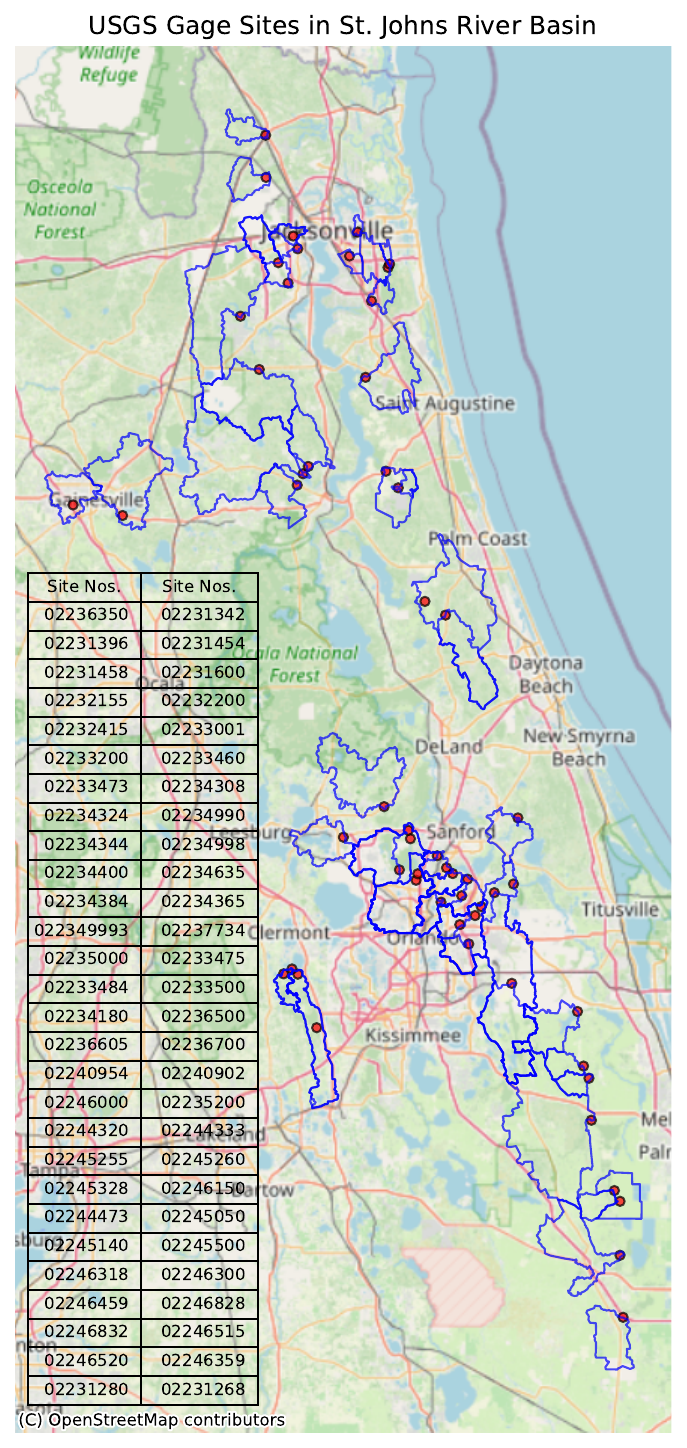}%
\caption{\label{fig:FLsites} The 56 USGS gage site locations with the corresponding watersheds boundaries}
\end{figure*}

To systematically organize and analyze streamflow data across different regions, we categorized USGS gages based on the Hydrologic Unit Code (HUC) system, developed by the USGS. This system provides a hierarchical framework for classifying watersheds into nested levels of hydrologic units. HUC 8 refers to subbasin-scale watersheds identified by an eight-digit code. Of the 81 USGS gages , the 25 gages in or near southern Louisiana spanned the HUC 8 watersheds of 12010005, 12020006, 08080201, 08070203, 08070202, 08090201, 08080204, 03170009, 03180004, 08080102, 08070205, 03180003, and 03180002. The remaining 56 gages from the St. Johns River watershed in Florida were from the HUC 8 watersheds of 03070205, 03080103, 03080101, and 03080102. Within these HUC 8 watersheds, we considered all USGS gages with drainage areas between 2 and 298 square miles, excluding gages on watersheds with control structures or with tidal influence (see Figs. \ref{fig:LAsites} and \ref{fig:FLsites}).

For the watershed of each gage, the daily rainfall time series was retrieved from Daymet at 1000 (km) pixel resolution, while the potential evapotranspiration $\langle\overline{\text{PET}}\rangle$ was retrieved from the Terra Moderate Resolution Imaging Spectroradiometer (MODIS) MOD16A3GF Version 6.1 Evapotranspiration/Latent Heat Flux (ET/LE) product---a year-end gap-filled yearly composite dataset produced at 500 meter (m) pixel resolution.  USGS gage daily flow data was gap filled for up to 5 days using logarithmic interpolation to account for flow recession. Baseflow separation was performed with 7 different methods: Lyne and Hollick (LH), Chapman, Chapman and
Maxwell (CM), Boughton, Furey, Eckhardt, EWMA, and Willems methods. The method with the highest Kling–Gupta efficiency was used to determine baseflow used in the calibration \cite{xie2020evaluation}.

The observed evapotranspiration, $\langle \overline{ET} \rangle^{\text{obs.}}$, is estimated as the difference between the annual rainfall and the annual flow volume. The ratio $\langle \overline{Q}_b \rangle/\langle \overline{Q}_f\rangle^{\text{obs}}$ is obtained from a baseflow separation using one of several available methods (e.g., the best performing method may be selected, as discussed in  \cite{xie2020evaluation}). The baseflow separation also yields the total runoff for each storm event, which is the basis for the observed quantiles of runoff, ${\overline{Q}_p}^{\text{obs.}}$, and the observed variance of runoff, ${\sigma^2_{\overline{Q}}}^{\text{obs.}}$.

Daily rainfall values were aggregated into storm events by combining two consecutive days of rainfall as one event. A third day's rainfall was included in the total if it was less than 25\% of the combined rainfall from the previous two days. Runoff aggregation began based on whether there was runoff the day before the rainfall event. If no runoff occurred the day before, the aggregation started on the same day as the rainfall event. However, if runoff was present before the rainfall event, the aggregation started the day after. Runoff was accumulated using the baseflow separation time series until zero runoff was detected after at least three consecutive days, or until the next rainfall event occurred. These storm-event-based rainfall and runoff values were then used to calculate long-term average rainfall, $\langle\overline{R}\rangle$, the observed variance of the runoff, $\sigma^2_{\overline{Q}}$, and the runoff quantiles $\overline{Q}_p$.
 
\bibliography{References}

\begin{thebibliography}{}

\bibitem [\protect \citeauthoryear {%
Abramowitz%
\ \BBA {} Stegun%
}{%
Abramowitz%
\ \BBA {} Stegun%
}{%
{\protect \APACyear {2012}}%
}]{%
abramowitz2012handbook}
\APACinsertmetastar {%
abramowitz2012handbook}%
\begin{APACrefauthors}%
Abramowitz, M.%
\BCBT {}\ \BBA {} Stegun, I\BPBI A.%
\end{APACrefauthors}%
\unskip\
\newblock
\APACrefYear{2012}.
\newblock
\APACrefbtitle {Handbook of mathematical functions: with formulas, graphs, and mathematical tables} {Handbook of mathematical functions: with formulas, graphs, and mathematical tables}.
\newblock
\APACaddressPublisher{}{Courier Dover Publications}.
\PrintBackRefs{\CurrentBib}

\bibitem [\protect \citeauthoryear {%
Alipour%
, Ahmadalipour%
\BCBL {}\ \BBA {} Moradkhani%
}{%
Alipour%
\ \protect \BOthers {.}}{%
{\protect \APACyear {2020}}%
}]{%
alipour2020assessing}
\APACinsertmetastar {%
alipour2020assessing}%
\begin{APACrefauthors}%
Alipour, A.%
, Ahmadalipour, A.%
\BCBL {}\ \BBA {} Moradkhani, H.%
\end{APACrefauthors}%
\unskip\
\newblock
\APACrefYearMonthDay{2020}{}{}.
\newblock
{\BBOQ}\APACrefatitle {Assessing flash flood hazard and damages in the southeast United States} {Assessing flash flood hazard and damages in the southeast united states}.{\BBCQ}
\newblock
\APACjournalVolNumPages{Journal of Flood Risk Management}{13}{2}{e12605}.
\PrintBackRefs{\CurrentBib}

\bibitem [\protect \citeauthoryear {%
Aoki%
\ \protect \BOthers {.}}{%
Aoki%
\ \protect \BOthers {.}}{%
{\protect \APACyear {2014}}%
}]{%
aoki2014nonequilibrium}
\APACinsertmetastar {%
aoki2014nonequilibrium}%
\begin{APACrefauthors}%
Aoki, H.%
, Tsuji, N.%
, Eckstein, M.%
, Kollar, M.%
, Oka, T.%
\BCBL {}\ \BBA {} Werner, P.%
\end{APACrefauthors}%
\unskip\
\newblock
\APACrefYearMonthDay{2014}{}{}.
\newblock
{\BBOQ}\APACrefatitle {Nonequilibrium dynamical mean-field theory and its applications} {Nonequilibrium dynamical mean-field theory and its applications}.{\BBCQ}
\newblock
\APACjournalVolNumPages{Reviews of Modern Physics}{86}{2}{779--837}.
\PrintBackRefs{\CurrentBib}

\bibitem [\protect \citeauthoryear {%
Bartlett%
, Daly%
, McDonnell%
, Parolari%
\BCBL {}\ \BBA {} Porporato%
}{%
Bartlett%
\ \protect \BOthers {.}}{%
{\protect \APACyear {2015}}%
}]{%
bartlett2014excess}
\APACinsertmetastar {%
bartlett2014excess}%
\begin{APACrefauthors}%
Bartlett, M\BPBI S.%
, Daly, E.%
, McDonnell, J\BPBI J.%
, Parolari, A\BPBI J.%
\BCBL {}\ \BBA {} Porporato, A.%
\end{APACrefauthors}%
\unskip\
\newblock
\APACrefYearMonthDay{2015}{}{}.
\newblock
{\BBOQ}\APACrefatitle {Stochastic rainfall-runoff model with explicit soil moisture dynamics} {Stochastic rainfall-runoff model with explicit soil moisture dynamics}.{\BBCQ}
\newblock
\APACjournalVolNumPages{Proceedings of the Royal Society A: Mathematical, Physical and Engineering Science}{471}{2183}{20150389}.
\PrintBackRefs{\CurrentBib}

\bibitem [\protect \citeauthoryear {%
Bartlett%
, Parolari%
, McDonnell%
\BCBL {}\ \BBA {} Porporato%
}{%
Bartlett%
\ \protect \BOthers {.}}{%
{\protect \APACyear {2017}}%
}]{%
bartlett2017reply}
\APACinsertmetastar {%
bartlett2017reply}%
\begin{APACrefauthors}%
Bartlett, M\BPBI S.%
, Parolari, A\BPBI J.%
, McDonnell, J.%
\BCBL {}\ \BBA {} Porporato, A.%
\end{APACrefauthors}%
\unskip\
\newblock
\APACrefYearMonthDay{2017}{}{}.
\newblock
{\BBOQ}\APACrefatitle {Reply to comment by {F}red {L}. {O}gden et al. on ‘‘{B}eyond the {SCS-CN} method: {A} theoretical framework for spatially lumped rainfall-runoff response,’’} {Reply to comment by {F}red {L}. {O}gden et al. on ‘‘{B}eyond the {SCS-CN} method: {A} theoretical framework for spatially lumped rainfall-runoff response,’’}.{\BBCQ}
\newblock
\APACjournalVolNumPages{Water Resources Research}{53}{7}{6351-6354}.
\PrintBackRefs{\CurrentBib}

\bibitem [\protect \citeauthoryear {%
Bartlett%
, Parolari%
, McDonnell%
\BCBL {}\ \BBA {} Porporato%
}{%
Bartlett%
\ \protect \BOthers {.}}{%
{\protect \APACyear {2016}}%
{\protect \APACexlab {{\protect \BCnt {1}}}}}]{%
bartlett2016beyond}
\APACinsertmetastar {%
bartlett2016beyond}%
\begin{APACrefauthors}%
Bartlett, M\BPBI S.%
, Parolari, A\BPBI J.%
, McDonnell, J\BPBI J.%
\BCBL {}\ \BBA {} Porporato, A.%
\end{APACrefauthors}%
\unskip\
\newblock
\APACrefYearMonthDay{2016{\protect \BCnt {1}}}{}{}.
\newblock
{\BBOQ}\APACrefatitle {Beyond the {SCS-CN} method: {A} theoretical framework for spatially lumped rainfall-runoff response} {Beyond the {SCS-CN} method: {A} theoretical framework for spatially lumped rainfall-runoff response}.{\BBCQ}
\newblock
\APACjournalVolNumPages{Water Resources Research}{52}{6}{4608--4627}.
\PrintBackRefs{\CurrentBib}

\bibitem [\protect \citeauthoryear {%
Bartlett%
, Parolari%
, McDonnell%
\BCBL {}\ \BBA {} Porporato%
}{%
Bartlett%
\ \protect \BOthers {.}}{%
{\protect \APACyear {2016}}%
{\protect \APACexlab {{\protect \BCnt {2}}}}}]{%
bartlett2015unified2}
\APACinsertmetastar {%
bartlett2015unified2}%
\begin{APACrefauthors}%
Bartlett, M\BPBI S.%
, Parolari, A\BPBI J.%
, McDonnell, J\BPBI J.%
\BCBL {}\ \BBA {} Porporato, A.%
\end{APACrefauthors}%
\unskip\
\newblock
\APACrefYearMonthDay{2016{\protect \BCnt {2}}}{}{}.
\newblock
{\BBOQ}\APACrefatitle {Framework for event-based semidistributed modeling that unifies the {SCS}-{CN} method, {VIC}, {PDM}, and {TOPMODEL}} {Framework for event-based semidistributed modeling that unifies the {SCS}-{CN} method, {VIC}, {PDM}, and {TOPMODEL}}.{\BBCQ}
\newblock
\APACjournalVolNumPages{Water Resources Research}{52}{9}{7036--7052}.
\PrintBackRefs{\CurrentBib}

\bibitem [\protect \citeauthoryear {%
Bartlett%
\ \BBA {} Porporato%
}{%
Bartlett%
\ \BBA {} Porporato%
}{%
{\protect \APACyear {2018}}%
}]{%
bartlett2018state}
\APACinsertmetastar {%
bartlett2018state}%
\begin{APACrefauthors}%
Bartlett, M\BPBI S.%
\BCBT {}\ \BBA {} Porporato, A.%
\end{APACrefauthors}%
\unskip\
\newblock
\APACrefYearMonthDay{2018}{}{}.
\newblock
{\BBOQ}\APACrefatitle {State dependent jump processes, Ito-Stratonovich interpretations, transient, and potential solutions} {State dependent jump processes, ito-stratonovich interpretations, transient, and potential solutions}.{\BBCQ}
\newblock
\APACjournalVolNumPages{Physical Review E}{}{}{}.
\PrintBackRefs{\CurrentBib}

\bibitem [\protect \citeauthoryear {%
Bartlett%
, Porporato%
\BCBL {}\ \BBA {} Rondoni%
}{%
Bartlett%
\ \protect \BOthers {.}}{%
{\protect \APACyear {2019}}%
}]{%
bartlett2019jump}
\APACinsertmetastar {%
bartlett2019jump}%
\begin{APACrefauthors}%
Bartlett, M\BPBI S.%
, Porporato, A.%
\BCBL {}\ \BBA {} Rondoni, L.%
\end{APACrefauthors}%
\unskip\
\newblock
\APACrefYearMonthDay{2019}{}{}.
\newblock
{\BBOQ}\APACrefatitle {Jump processes with deterministic and stochastic controls} {Jump processes with deterministic and stochastic controls}.{\BBCQ}
\newblock
\APACjournalVolNumPages{Physical Review E}{100}{4}{042133}.
\PrintBackRefs{\CurrentBib}

\bibitem [\protect \citeauthoryear {%
Bender%
, Heenen%
\BCBL {}\ \BBA {} Reinhard%
}{%
Bender%
\ \protect \BOthers {.}}{%
{\protect \APACyear {2003}}%
}]{%
bender2003self}
\APACinsertmetastar {%
bender2003self}%
\begin{APACrefauthors}%
Bender, M.%
, Heenen, P\BHBI H.%
\BCBL {}\ \BBA {} Reinhard, P\BHBI G.%
\end{APACrefauthors}%
\unskip\
\newblock
\APACrefYearMonthDay{2003}{}{}.
\newblock
{\BBOQ}\APACrefatitle {Self-consistent mean-field models for nuclear structure} {Self-consistent mean-field models for nuclear structure}.{\BBCQ}
\newblock
\APACjournalVolNumPages{Reviews of Modern Physics}{75}{1}{121}.
\PrintBackRefs{\CurrentBib}

\bibitem [\protect \citeauthoryear {%
Beven%
}{%
Beven%
}{%
{\protect \APACyear {2012}}%
}]{%
beven2012rainfall}
\APACinsertmetastar {%
beven2012rainfall}%
\begin{APACrefauthors}%
Beven, K.%
\end{APACrefauthors}%
\unskip\
\newblock
\APACrefYear{2012}.
\newblock
\APACrefbtitle {Rainfall-Runoff Modelling: The Primer} {Rainfall-runoff modelling: The primer}.
\newblock
\APACaddressPublisher{}{Wiley}.
\PrintBackRefs{\CurrentBib}

\bibitem [\protect \citeauthoryear {%
Beven%
}{%
Beven%
}{%
{\protect \APACyear {2021}}%
}]{%
beven2021issues}
\APACinsertmetastar {%
beven2021issues}%
\begin{APACrefauthors}%
Beven, K.%
\end{APACrefauthors}%
\unskip\
\newblock
\APACrefYearMonthDay{2021}{}{}.
\newblock
{\BBOQ}\APACrefatitle {Issues in generating stochastic observables for hydrological models} {Issues in generating stochastic observables for hydrological models}.{\BBCQ}
\newblock
\APACjournalVolNumPages{Hydrological Processes}{35}{6}{e14203}.
\PrintBackRefs{\CurrentBib}

\bibitem [\protect \citeauthoryear {%
Beven%
\ \BBA {} Kirkby%
}{%
Beven%
\ \BBA {} Kirkby%
}{%
{\protect \APACyear {1979}}%
}]{%
beven1979physically}
\APACinsertmetastar {%
beven1979physically}%
\begin{APACrefauthors}%
Beven, K.%
\BCBT {}\ \BBA {} Kirkby, M.%
\end{APACrefauthors}%
\unskip\
\newblock
\APACrefYearMonthDay{1979}{}{}.
\newblock
{\BBOQ}\APACrefatitle {A physically based, variable contributing area model of basin hydrology/Un mod{\`e}le {\`a} base physique de zone d'appel variable de l'hydrologie du bassin versant} {A physically based, variable contributing area model of basin hydrology/un mod{\`e}le {\`a} base physique de zone d'appel variable de l'hydrologie du bassin versant}.{\BBCQ}
\newblock
\APACjournalVolNumPages{Hydrological Sciences Journal}{24}{1}{43--69}.
\PrintBackRefs{\CurrentBib}

\bibitem [\protect \citeauthoryear {%
Bl{\"o}schl%
\ \BBA {} Montanari%
}{%
Bl{\"o}schl%
\ \BBA {} Montanari%
}{%
{\protect \APACyear {2010}}%
}]{%
bloschl2010climate}
\APACinsertmetastar {%
bloschl2010climate}%
\begin{APACrefauthors}%
Bl{\"o}schl, G.%
\BCBT {}\ \BBA {} Montanari, A.%
\end{APACrefauthors}%
\unskip\
\newblock
\APACrefYearMonthDay{2010}{}{}.
\newblock
{\BBOQ}\APACrefatitle {Climate change impacts—throwing the dice?} {Climate change impacts—throwing the dice?}{\BBCQ}
\newblock
\APACjournalVolNumPages{Hydrological Processes: An International Journal}{24}{3}{374--381}.
\PrintBackRefs{\CurrentBib}

\bibitem [\protect \citeauthoryear {%
Botter%
, Peratoner%
, Porporato%
, Rodriguez-Iturbe%
\BCBL {}\ \BBA {} Rinaldo%
}{%
Botter%
, Peratoner%
\BCBL {}\ \protect \BOthers {.}}{%
{\protect \APACyear {2007}}%
}]{%
botter2007signatures}
\APACinsertmetastar {%
botter2007signatures}%
\begin{APACrefauthors}%
Botter, G.%
, Peratoner, F.%
, Porporato, A.%
, Rodriguez-Iturbe, I.%
\BCBL {}\ \BBA {} Rinaldo, A.%
\end{APACrefauthors}%
\unskip\
\newblock
\APACrefYearMonthDay{2007}{}{}.
\newblock
{\BBOQ}\APACrefatitle {Signatures of large-scale soil moisture dynamics on streamflow statistics across US climate regimes} {Signatures of large-scale soil moisture dynamics on streamflow statistics across us climate regimes}.{\BBCQ}
\newblock
\APACjournalVolNumPages{Water resources research}{43}{11}{}.
\PrintBackRefs{\CurrentBib}

\bibitem [\protect \citeauthoryear {%
Botter%
, Porporato%
, Rodriguez-Iturbe%
\BCBL {}\ \BBA {} Rinaldo%
}{%
Botter%
, Porporato%
\BCBL {}\ \protect \BOthers {.}}{%
{\protect \APACyear {2007}}%
}]{%
botter2007basin}
\APACinsertmetastar {%
botter2007basin}%
\begin{APACrefauthors}%
Botter, G.%
, Porporato, A.%
, Rodriguez-Iturbe, I.%
\BCBL {}\ \BBA {} Rinaldo, A.%
\end{APACrefauthors}%
\unskip\
\newblock
\APACrefYearMonthDay{2007}{}{}.
\newblock
{\BBOQ}\APACrefatitle {Basin-scale soil moisture dynamics and the probabilistic characterization of carrier hydrologic flows: Slow, leaching-prone components of the hydrologic response} {Basin-scale soil moisture dynamics and the probabilistic characterization of carrier hydrologic flows: Slow, leaching-prone components of the hydrologic response}.{\BBCQ}
\newblock
\APACjournalVolNumPages{Water resources research}{43}{2}{{W02}{417}}.
\PrintBackRefs{\CurrentBib}

\bibitem [\protect \citeauthoryear {%
Brutsaert%
}{%
Brutsaert%
}{%
{\protect \APACyear {2005}}%
}]{%
brutsaert2005hydrology}
\APACinsertmetastar {%
brutsaert2005hydrology}%
\begin{APACrefauthors}%
Brutsaert, W.%
\end{APACrefauthors}%
\unskip\
\newblock
\APACrefYear{2005}.
\newblock
\APACrefbtitle {Hydrology: An Introduction} {Hydrology: An introduction}.
\newblock
\APACaddressPublisher{}{Cambridge University Press}.
\newblock
\begin{APACrefURL} \url{https://books.google.com/books?id=yX\_xS55xxyoC} \end{APACrefURL}
\PrintBackRefs{\CurrentBib}

\bibitem [\protect \citeauthoryear {%
Budyko%
}{%
Budyko%
}{%
{\protect \APACyear {1974}}%
}]{%
budyko1974climate}
\APACinsertmetastar {%
budyko1974climate}%
\begin{APACrefauthors}%
Budyko, M.%
\end{APACrefauthors}%
\unskip\
\newblock
\APACrefYear{1974}.
\newblock
\APACrefbtitle {Climate and Life} {Climate and life}.
\newblock
\APACaddressPublisher{}{Academic Press}.
\PrintBackRefs{\CurrentBib}

\bibitem [\protect \citeauthoryear {%
Cho%
\ \BBA {} Engel%
}{%
Cho%
\ \BBA {} Engel%
}{%
{\protect \APACyear {2018}}%
}]{%
cho2018spatially}
\APACinsertmetastar {%
cho2018spatially}%
\begin{APACrefauthors}%
Cho, Y.%
\BCBT {}\ \BBA {} Engel, B\BPBI A.%
\end{APACrefauthors}%
\unskip\
\newblock
\APACrefYearMonthDay{2018}{}{}.
\newblock
{\BBOQ}\APACrefatitle {Spatially distributed long-term hydrologic simulation using a continuous SCS CN method-based hybrid hydrologic model} {Spatially distributed long-term hydrologic simulation using a continuous scs cn method-based hybrid hydrologic model}.{\BBCQ}
\newblock
\APACjournalVolNumPages{Hydrological Processes}{32}{7}{904--922}.
\PrintBackRefs{\CurrentBib}

\bibitem [\protect \citeauthoryear {%
Cox%
\ \BBA {} Miller%
}{%
Cox%
\ \BBA {} Miller%
}{%
{\protect \APACyear {1977}}%
}]{%
cox1977theory}
\APACinsertmetastar {%
cox1977theory}%
\begin{APACrefauthors}%
Cox, D\BPBI R.%
\BCBT {}\ \BBA {} Miller, H\BPBI D.%
\end{APACrefauthors}%
\unskip\
\newblock
\APACrefYear{1977}.
\newblock
\APACrefbtitle {The theory of stochastic processes} {The theory of stochastic processes}\ (\BVOL~134).
\newblock
\APACaddressPublisher{}{CRC Press}.
\PrintBackRefs{\CurrentBib}

\bibitem [\protect \citeauthoryear {%
Daly%
\ \BBA {} Porporato%
}{%
Daly%
\ \BBA {} Porporato%
}{%
{\protect \APACyear {2007}}%
}]{%
daly2007intertime}
\APACinsertmetastar {%
daly2007intertime}%
\begin{APACrefauthors}%
Daly, E.%
\BCBT {}\ \BBA {} Porporato, A.%
\end{APACrefauthors}%
\unskip\
\newblock
\APACrefYearMonthDay{2007}{}{}.
\newblock
{\BBOQ}\APACrefatitle {Intertime jump statistics of state-dependent Poisson processes} {Intertime jump statistics of state-dependent poisson processes}.{\BBCQ}
\newblock
\APACjournalVolNumPages{Physical Review E}{75}{1}{{011}{119}}.
\PrintBackRefs{\CurrentBib}

\bibitem [\protect \citeauthoryear {%
Dooge%
}{%
Dooge%
}{%
{\protect \APACyear {1986}}%
}]{%
dooge1986looking}
\APACinsertmetastar {%
dooge1986looking}%
\begin{APACrefauthors}%
Dooge, J\BPBI C.%
\end{APACrefauthors}%
\unskip\
\newblock
\APACrefYearMonthDay{1986}{}{}.
\newblock
{\BBOQ}\APACrefatitle {Looking for hydrologic laws} {Looking for hydrologic laws}.{\BBCQ}
\newblock
\APACjournalVolNumPages{Water Resources Research}{22}{9S}{46S--58S}.
\PrintBackRefs{\CurrentBib}

\bibitem [\protect \citeauthoryear {%
Duethmann%
, Bl{\"o}schl%
\BCBL {}\ \BBA {} Parajka%
}{%
Duethmann%
\ \protect \BOthers {.}}{%
{\protect \APACyear {2020}}%
}]{%
duethmann2020does}
\APACinsertmetastar {%
duethmann2020does}%
\begin{APACrefauthors}%
Duethmann, D.%
, Bl{\"o}schl, G.%
\BCBL {}\ \BBA {} Parajka, J.%
\end{APACrefauthors}%
\unskip\
\newblock
\APACrefYearMonthDay{2020}{}{}.
\newblock
{\BBOQ}\APACrefatitle {Why does a conceptual hydrological model fail to correctly predict discharge changes in response to climate change?} {Why does a conceptual hydrological model fail to correctly predict discharge changes in response to climate change?}{\BBCQ}
\newblock
\APACjournalVolNumPages{Hydrology and Earth System Sciences}{24}{7}{3493--3511}.
\PrintBackRefs{\CurrentBib}

\bibitem [\protect \citeauthoryear {%
Farmer%
\ \BBA {} Vogel%
}{%
Farmer%
\ \BBA {} Vogel%
}{%
{\protect \APACyear {2016}}%
}]{%
farmer2016deterministic}
\APACinsertmetastar {%
farmer2016deterministic}%
\begin{APACrefauthors}%
Farmer, W\BPBI H.%
\BCBT {}\ \BBA {} Vogel, R\BPBI M.%
\end{APACrefauthors}%
\unskip\
\newblock
\APACrefYearMonthDay{2016}{}{}.
\newblock
{\BBOQ}\APACrefatitle {On the deterministic and stochastic use of hydrologic models} {On the deterministic and stochastic use of hydrologic models}.{\BBCQ}
\newblock
\APACjournalVolNumPages{Water Resources Research}{52}{7}{5619--5633}.
\PrintBackRefs{\CurrentBib}

\bibitem [\protect \citeauthoryear {%
Fatichi%
\ \protect \BOthers {.}}{%
Fatichi%
\ \protect \BOthers {.}}{%
{\protect \APACyear {2015}}%
}]{%
fatichi2015abiotic}
\APACinsertmetastar {%
fatichi2015abiotic}%
\begin{APACrefauthors}%
Fatichi, S.%
, Katul, G\BPBI G.%
, Ivanov, V\BPBI Y.%
, Pappas, C.%
, Paschalis, A.%
, Consolo, A.%
\BDBL {}Burlando, P.%
\end{APACrefauthors}%
\unskip\
\newblock
\APACrefYearMonthDay{2015}{}{}.
\newblock
{\BBOQ}\APACrefatitle {Abiotic and biotic controls of soil moisture spatiotemporal variability and the occurrence of hysteresis} {Abiotic and biotic controls of soil moisture spatiotemporal variability and the occurrence of hysteresis}.{\BBCQ}
\newblock
\APACjournalVolNumPages{Water Resources Research}{51}{5}{3505--3524}.
\PrintBackRefs{\CurrentBib}

\bibitem [\protect \citeauthoryear {%
Feldmann%
\ \BBA {} Whitt%
}{%
Feldmann%
\ \BBA {} Whitt%
}{%
{\protect \APACyear {1997}}%
}]{%
feldmann1997fitting}
\APACinsertmetastar {%
feldmann1997fitting}%
\begin{APACrefauthors}%
Feldmann, A.%
\BCBT {}\ \BBA {} Whitt, W.%
\end{APACrefauthors}%
\unskip\
\newblock
\APACrefYearMonthDay{1997}{}{}.
\newblock
{\BBOQ}\APACrefatitle {Fitting mixtures of exponentials to long-tail distributions to analyze network performance models} {Fitting mixtures of exponentials to long-tail distributions to analyze network performance models}.{\BBCQ}
\newblock
\BIn{} \APACrefbtitle {INFOCOM'97. Sixteenth Annual Joint Conference of the IEEE Computer and Communications Societies. Driving the Information Revolution., Proceedings IEEE} {Infocom'97. sixteenth annual joint conference of the ieee computer and communications societies. driving the information revolution., proceedings ieee}\ (\BVOL~3, \BPGS\ 1096--1104).
\PrintBackRefs{\CurrentBib}

\bibitem [\protect \citeauthoryear {%
Freeze%
\ \BBA {} Harlan%
}{%
Freeze%
\ \BBA {} Harlan%
}{%
{\protect \APACyear {1969}}%
}]{%
freeze1969blueprint}
\APACinsertmetastar {%
freeze1969blueprint}%
\begin{APACrefauthors}%
Freeze, R\BPBI A.%
\BCBT {}\ \BBA {} Harlan, R.%
\end{APACrefauthors}%
\unskip\
\newblock
\APACrefYearMonthDay{1969}{}{}.
\newblock
{\BBOQ}\APACrefatitle {Blueprint for a physically-based, digitally-simulated hydrologic response model} {Blueprint for a physically-based, digitally-simulated hydrologic response model}.{\BBCQ}
\newblock
\APACjournalVolNumPages{Journal of Hydrology}{9}{3}{237--258}.
\PrintBackRefs{\CurrentBib}

\bibitem [\protect \citeauthoryear {%
Garen%
\ \BBA {} Moore%
}{%
Garen%
\ \BBA {} Moore%
}{%
{\protect \APACyear {2005}}%
}]{%
garen2005curve}
\APACinsertmetastar {%
garen2005curve}%
\begin{APACrefauthors}%
Garen, D\BPBI C.%
\BCBT {}\ \BBA {} Moore, D\BPBI S.%
\end{APACrefauthors}%
\unskip\
\newblock
\APACrefYearMonthDay{2005}{}{}.
\newblock
{\BBOQ}\APACrefatitle {Curve number hydrology in water quality modeling: uses, abuses, and future directions} {Curve number hydrology in water quality modeling: uses, abuses, and future directions}.{\BBCQ}
\newblock
\APACjournalVolNumPages{JAWRA Journal of the American Water Resources Association}{41}{2}{377--388}.
\PrintBackRefs{\CurrentBib}

\bibitem [\protect \citeauthoryear {%
Gnann%
, Woods%
\BCBL {}\ \BBA {} Howden%
}{%
Gnann%
\ \protect \BOthers {.}}{%
{\protect \APACyear {2019}}%
}]{%
gnann2019there}
\APACinsertmetastar {%
gnann2019there}%
\begin{APACrefauthors}%
Gnann, S\BPBI J.%
, Woods, R\BPBI A.%
\BCBL {}\ \BBA {} Howden, N\BPBI J.%
\end{APACrefauthors}%
\unskip\
\newblock
\APACrefYearMonthDay{2019}{}{}.
\newblock
{\BBOQ}\APACrefatitle {Is there a baseflow Budyko curve?} {Is there a baseflow budyko curve?}{\BBCQ}
\newblock
\APACjournalVolNumPages{Water Resources Research}{55}{4}{2838--2855}.
\PrintBackRefs{\CurrentBib}

\bibitem [\protect \citeauthoryear {%
Grimaldi%
, Nardi%
, Piscopia%
, Petroselli%
\BCBL {}\ \BBA {} Apollonio%
}{%
Grimaldi%
\ \protect \BOthers {.}}{%
{\protect \APACyear {2021}}%
}]{%
grimaldi2021continuous}
\APACinsertmetastar {%
grimaldi2021continuous}%
\begin{APACrefauthors}%
Grimaldi, S.%
, Nardi, F.%
, Piscopia, R.%
, Petroselli, A.%
\BCBL {}\ \BBA {} Apollonio, C.%
\end{APACrefauthors}%
\unskip\
\newblock
\APACrefYearMonthDay{2021}{}{}.
\newblock
{\BBOQ}\APACrefatitle {Continuous hydrologic modelling for design simulation in small and ungauged basins: A step forward and some tests for its practical use} {Continuous hydrologic modelling for design simulation in small and ungauged basins: A step forward and some tests for its practical use}.{\BBCQ}
\newblock
\APACjournalVolNumPages{Journal of Hydrology}{595}{}{125664}.
\PrintBackRefs{\CurrentBib}

\bibitem [\protect \citeauthoryear {%
Hawkins%
}{%
Hawkins%
}{%
{\protect \APACyear {2014}}%
}]{%
hawkins2014curve}
\APACinsertmetastar {%
hawkins2014curve}%
\begin{APACrefauthors}%
Hawkins, R\BPBI H.%
\end{APACrefauthors}%
\unskip\
\newblock
\APACrefYearMonthDay{2014}{}{}.
\newblock
{\BBOQ}\APACrefatitle {Curve Number Method: Time to Think Anew?} {Curve number method: Time to think anew?}{\BBCQ}
\newblock
\APACjournalVolNumPages{Journal of Hydrologic Engineering}{}{}{}.
\PrintBackRefs{\CurrentBib}

\bibitem [\protect \citeauthoryear {%
Hawkins%
, Ward%
\BCBL {}\ \BBA {} Woodward%
}{%
Hawkins%
\ \protect \BOthers {.}}{%
{\protect \APACyear {2015}}%
}]{%
hawkins2015complacent}
\APACinsertmetastar {%
hawkins2015complacent}%
\begin{APACrefauthors}%
Hawkins, R\BPBI H.%
, Ward, T\BPBI J.%
\BCBL {}\ \BBA {} Woodward, D\BPBI E.%
\end{APACrefauthors}%
\unskip\
\newblock
\APACrefYearMonthDay{2015}{}{}.
\newblock
{\BBOQ}\APACrefatitle {The Complacent-Violent Runoff: A Departure from Traditional Behavior} {The complacent-violent runoff: A departure from traditional behavior}.{\BBCQ}
\newblock
\APACjournalVolNumPages{Watershed Management 2015}{}{}{169}.
\PrintBackRefs{\CurrentBib}

\bibitem [\protect \citeauthoryear {%
Hotta%
}{%
Hotta%
}{%
{\protect \APACyear {2003}}%
}]{%
Hotta2003}
\APACinsertmetastar {%
Hotta2003}%
\begin{APACrefauthors}%
Hotta, T.%
\end{APACrefauthors}%
\unskip\
\newblock
\APACrefYearMonthDay{2003}{}{}.
\newblock
{\BBOQ}\APACrefatitle {Mean-Field Approximation} {Mean-field approximation}.{\BBCQ}
\newblock
\BIn{} \APACrefbtitle {Nanoscale Phase Separation and Colossal Magnetoresistance: The Physics of Manganites and Related Compounds} {Nanoscale phase separation and colossal magnetoresistance: The physics of manganites and related compounds}\ (\BPGS\ 157--167).
\newblock
\APACaddressPublisher{Berlin, Heidelberg}{Springer Berlin Heidelberg}.
\newblock
\begin{APACrefDOI} \doi{10.1007/978-3-662-05244-0_8} \end{APACrefDOI}
\PrintBackRefs{\CurrentBib}

\bibitem [\protect \citeauthoryear {%
It{\^o}%
}{%
It{\^o}%
}{%
{\protect \APACyear {1944}}%
}]{%
ito1944109}
\APACinsertmetastar {%
ito1944109}%
\begin{APACrefauthors}%
It{\^o}, K.%
\end{APACrefauthors}%
\unskip\
\newblock
\APACrefYearMonthDay{1944}{}{}.
\newblock
{\BBOQ}\APACrefatitle {109. stochastic integral} {109. stochastic integral}.{\BBCQ}
\newblock
\APACjournalVolNumPages{Proceedings of the Imperial Academy}{20}{8}{519--524}.
\PrintBackRefs{\CurrentBib}

\bibitem [\protect \citeauthoryear {%
Jaafar%
, Ahmad%
\BCBL {}\ \BBA {} El~Beyrouthy%
}{%
Jaafar%
\ \protect \BOthers {.}}{%
{\protect \APACyear {2019}}%
}]{%
jaafar2019gcn250}
\APACinsertmetastar {%
jaafar2019gcn250}%
\begin{APACrefauthors}%
Jaafar, H\BPBI H.%
, Ahmad, F\BPBI A.%
\BCBL {}\ \BBA {} El~Beyrouthy, M.%
\end{APACrefauthors}%
\unskip\
\newblock
\APACrefYearMonthDay{2019}{}{}.
\newblock
\APACrefbtitle {GCN250, new global gridded curve numbers for hydrologic modeling and design, Scientific Data, 6, 1--9.} {Gcn250, new global gridded curve numbers for hydrologic modeling and design, scientific data, 6, 1--9.}
\PrintBackRefs{\CurrentBib}

\bibitem [\protect \citeauthoryear {%
Kavetski%
, Kuczera%
\BCBL {}\ \BBA {} Franks%
}{%
Kavetski%
\ \protect \BOthers {.}}{%
{\protect \APACyear {2003}}%
}]{%
kavetski2003semidistributed}
\APACinsertmetastar {%
kavetski2003semidistributed}%
\begin{APACrefauthors}%
Kavetski, D.%
, Kuczera, G.%
\BCBL {}\ \BBA {} Franks, S\BPBI W.%
\end{APACrefauthors}%
\unskip\
\newblock
\APACrefYearMonthDay{2003}{}{}.
\newblock
{\BBOQ}\APACrefatitle {Semidistributed hydrological modeling: {A} ``saturation path'' perspective on {TOPMODEL} and {VIC}} {Semidistributed hydrological modeling: {A} ``saturation path'' perspective on {TOPMODEL} and {VIC}}.{\BBCQ}
\newblock
\APACjournalVolNumPages{Water resources research}{39}{9}{}.
\PrintBackRefs{\CurrentBib}

\bibitem [\protect \citeauthoryear {%
H\BPBI J.~Kim%
, Sidle%
\BCBL {}\ \BBA {} Moore%
}{%
H\BPBI J.~Kim%
\ \protect \BOthers {.}}{%
{\protect \APACyear {2005}}%
}]{%
kim2005shallow}
\APACinsertmetastar {%
kim2005shallow}%
\begin{APACrefauthors}%
Kim, H\BPBI J.%
, Sidle, R\BPBI C.%
\BCBL {}\ \BBA {} Moore, R\BPBI D.%
\end{APACrefauthors}%
\unskip\
\newblock
\APACrefYearMonthDay{2005}{}{}.
\newblock
{\BBOQ}\APACrefatitle {Shallow lateral flow from a forested hillslope: Influence of antecedent wetness} {Shallow lateral flow from a forested hillslope: Influence of antecedent wetness}.{\BBCQ}
\newblock
\APACjournalVolNumPages{Catena}{60}{3}{293--306}.
\PrintBackRefs{\CurrentBib}

\bibitem [\protect \citeauthoryear {%
J.~Kim%
, Johnson%
, Cifelli%
, Thorstensen%
\BCBL {}\ \BBA {} Chandrasekar%
}{%
J.~Kim%
\ \protect \BOthers {.}}{%
{\protect \APACyear {2019}}%
}]{%
kim2019assessment}
\APACinsertmetastar {%
kim2019assessment}%
\begin{APACrefauthors}%
Kim, J.%
, Johnson, L.%
, Cifelli, R.%
, Thorstensen, A.%
\BCBL {}\ \BBA {} Chandrasekar, V.%
\end{APACrefauthors}%
\unskip\
\newblock
\APACrefYearMonthDay{2019}{}{}.
\newblock
{\BBOQ}\APACrefatitle {Assessment of antecedent moisture condition on flood frequency: An experimental study in Napa River Basin, CA} {Assessment of antecedent moisture condition on flood frequency: An experimental study in napa river basin, ca}.{\BBCQ}
\newblock
\APACjournalVolNumPages{Journal of Hydrology: Regional Studies}{26}{}{100629}.
\PrintBackRefs{\CurrentBib}

\bibitem [\protect \citeauthoryear {%
Kirby%
}{%
Kirby%
}{%
{\protect \APACyear {1975}}%
}]{%
kirby1975model}
\APACinsertmetastar {%
kirby1975model}%
\begin{APACrefauthors}%
Kirby, W.%
\end{APACrefauthors}%
\unskip\
\newblock
\APACrefYearMonthDay{1975}{}{}.
\newblock
{\BBOQ}\APACrefatitle {Model smoothing effect diminishes simulated flood peak variance} {Model smoothing effect diminishes simulated flood peak variance}.{\BBCQ}
\newblock
\APACjournalVolNumPages{Am. Geophys. Union Trans}{56}{6}{361}.
\PrintBackRefs{\CurrentBib}

\bibitem [\protect \citeauthoryear {%
Koutsoyiannis%
}{%
Koutsoyiannis%
}{%
{\protect \APACyear {2020}}%
}]{%
koutsoyiannis2020revisiting}
\APACinsertmetastar {%
koutsoyiannis2020revisiting}%
\begin{APACrefauthors}%
Koutsoyiannis, D.%
\end{APACrefauthors}%
\unskip\
\newblock
\APACrefYearMonthDay{2020}{}{}.
\newblock
{\BBOQ}\APACrefatitle {Revisiting the global hydrological cycle: is it intensifying?} {Revisiting the global hydrological cycle: is it intensifying?}{\BBCQ}
\newblock
\APACjournalVolNumPages{Hydrology and Earth System Sciences}{24}{8}{3899--3932}.
\PrintBackRefs{\CurrentBib}

\bibitem [\protect \citeauthoryear {%
Labat%
, Godd{\'e}ris%
, Probst%
\BCBL {}\ \BBA {} Guyot%
}{%
Labat%
\ \protect \BOthers {.}}{%
{\protect \APACyear {2004}}%
}]{%
labat2004evidence}
\APACinsertmetastar {%
labat2004evidence}%
\begin{APACrefauthors}%
Labat, D.%
, Godd{\'e}ris, Y.%
, Probst, J\BPBI L.%
\BCBL {}\ \BBA {} Guyot, J\BPBI L.%
\end{APACrefauthors}%
\unskip\
\newblock
\APACrefYearMonthDay{2004}{}{}.
\newblock
{\BBOQ}\APACrefatitle {Evidence for global runoff increase related to climate warming} {Evidence for global runoff increase related to climate warming}.{\BBCQ}
\newblock
\APACjournalVolNumPages{Advances in water resources}{27}{6}{631--642}.
\PrintBackRefs{\CurrentBib}

\bibitem [\protect \citeauthoryear {%
Liang%
, Lettenmaier%
, Wood%
\BCBL {}\ \BBA {} Burges%
}{%
Liang%
\ \protect \BOthers {.}}{%
{\protect \APACyear {1994}}%
}]{%
liang1994simple}
\APACinsertmetastar {%
liang1994simple}%
\begin{APACrefauthors}%
Liang, X.%
, Lettenmaier, D\BPBI P.%
, Wood, E\BPBI F.%
\BCBL {}\ \BBA {} Burges, S\BPBI J.%
\end{APACrefauthors}%
\unskip\
\newblock
\APACrefYearMonthDay{1994}{}{}.
\newblock
{\BBOQ}\APACrefatitle {A simple hydrologically based model of land surface water and energy fluxes for general circulation models} {A simple hydrologically based model of land surface water and energy fluxes for general circulation models}.{\BBCQ}
\newblock
\APACjournalVolNumPages{Journal of Geophysical Research: Atmospheres (1984--2012)}{99}{D7}{14415--14428}.
\PrintBackRefs{\CurrentBib}

\bibitem [\protect \citeauthoryear {%
Liang%
\ \BBA {} Xie%
}{%
Liang%
\ \BBA {} Xie%
}{%
{\protect \APACyear {2001}}%
}]{%
liang2001new}
\APACinsertmetastar {%
liang2001new}%
\begin{APACrefauthors}%
Liang, X.%
\BCBT {}\ \BBA {} Xie, Z.%
\end{APACrefauthors}%
\unskip\
\newblock
\APACrefYearMonthDay{2001}{}{}.
\newblock
{\BBOQ}\APACrefatitle {A new surface runoff parameterization with subgrid-scale soil heterogeneity for land surface models} {A new surface runoff parameterization with subgrid-scale soil heterogeneity for land surface models}.{\BBCQ}
\newblock
\APACjournalVolNumPages{Advances in Water Resources}{24}{9-10}{1173--1193}.
\PrintBackRefs{\CurrentBib}

\bibitem [\protect \citeauthoryear {%
Lichty%
\ \BBA {} Liscum%
}{%
Lichty%
\ \BBA {} Liscum%
}{%
{\protect \APACyear {1978}}%
}]{%
lichty1978rainfall}
\APACinsertmetastar {%
lichty1978rainfall}%
\begin{APACrefauthors}%
Lichty, R\BPBI W.%
\BCBT {}\ \BBA {} Liscum, F.%
\end{APACrefauthors}%
\unskip\
\newblock
\APACrefYear{1978}.
\newblock
\APACrefbtitle {A rainfall-runoff modeling procedure for improving estimates of T-year (annual) floods for small drainage basins} {A rainfall-runoff modeling procedure for improving estimates of t-year (annual) floods for small drainage basins}\ (\BVOL~78)\ (\BNUM~7).
\newblock
\APACaddressPublisher{}{Department of the Interior, Geological Survey}.
\PrintBackRefs{\CurrentBib}

\bibitem [\protect \citeauthoryear {%
Lin%
}{%
Lin%
}{%
{\protect \APACyear {2012}}%
}]{%
lin2012hydropedology}
\APACinsertmetastar {%
lin2012hydropedology}%
\begin{APACrefauthors}%
Lin, H.%
\end{APACrefauthors}%
\unskip\
\newblock
\APACrefYear{2012}.
\newblock
\APACrefbtitle {Hydropedology: Synergistic Integration of Soil Science and Hydrology} {Hydropedology: Synergistic integration of soil science and hydrology}.
\newblock
\APACaddressPublisher{}{Academic Press}.
\newblock
\begin{APACrefURL} \url{https://books.google.com/books?id=pmLnQ2Dr-jAC} \end{APACrefURL}
\PrintBackRefs{\CurrentBib}

\bibitem [\protect \citeauthoryear {%
S.~Marcus%
}{%
S.~Marcus%
}{%
{\protect \APACyear {1978}}%
}]{%
marcus1978modeling}
\APACinsertmetastar {%
marcus1978modeling}%
\begin{APACrefauthors}%
Marcus, S.%
\end{APACrefauthors}%
\unskip\
\newblock
\APACrefYearMonthDay{1978}{}{}.
\newblock
{\BBOQ}\APACrefatitle {Modeling and analysis of stochastic differential equations driven by point processes} {Modeling and analysis of stochastic differential equations driven by point processes}.{\BBCQ}
\newblock
\APACjournalVolNumPages{IEEE Transactions on Information theory}{24}{2}{164--172}.
\PrintBackRefs{\CurrentBib}

\bibitem [\protect \citeauthoryear {%
S\BPBI I.~Marcus%
}{%
S\BPBI I.~Marcus%
}{%
{\protect \APACyear {1981}}%
}]{%
marcus1981modeling}
\APACinsertmetastar {%
marcus1981modeling}%
\begin{APACrefauthors}%
Marcus, S\BPBI I.%
\end{APACrefauthors}%
\unskip\
\newblock
\APACrefYearMonthDay{1981}{}{}.
\newblock
{\BBOQ}\APACrefatitle {Modeling and approximation of stochastic differential equations driven by semimartingales} {Modeling and approximation of stochastic differential equations driven by semimartingales}.{\BBCQ}
\newblock
\APACjournalVolNumPages{Stochastics: An International Journal of Probability and Stochastic Processes}{4}{3}{223--245}.
\PrintBackRefs{\CurrentBib}

\bibitem [\protect \citeauthoryear {%
Michel%
, Andr{\'e}assian%
\BCBL {}\ \BBA {} Perrin%
}{%
Michel%
\ \protect \BOthers {.}}{%
{\protect \APACyear {2005}}%
}]{%
michel2005soil}
\APACinsertmetastar {%
michel2005soil}%
\begin{APACrefauthors}%
Michel, C.%
, Andr{\'e}assian, V.%
\BCBL {}\ \BBA {} Perrin, C.%
\end{APACrefauthors}%
\unskip\
\newblock
\APACrefYearMonthDay{2005}{}{}.
\newblock
{\BBOQ}\APACrefatitle {Soil conservation service curve number method: How to mend a wrong soil moisture accounting procedure?} {Soil conservation service curve number method: How to mend a wrong soil moisture accounting procedure?}{\BBCQ}
\newblock
\APACjournalVolNumPages{Water Resources Research}{41}{2}{}.
\PrintBackRefs{\CurrentBib}

\bibitem [\protect \citeauthoryear {%
Milly%
}{%
Milly%
}{%
{\protect \APACyear {1993}}%
}]{%
milly1993analytic}
\APACinsertmetastar {%
milly1993analytic}%
\begin{APACrefauthors}%
Milly, P.%
\end{APACrefauthors}%
\unskip\
\newblock
\APACrefYearMonthDay{1993}{}{}.
\newblock
{\BBOQ}\APACrefatitle {An analytic solution of the stochastic storage problem applicable to soil water} {An analytic solution of the stochastic storage problem applicable to soil water}.{\BBCQ}
\newblock
\APACjournalVolNumPages{Water Resources Research}{29}{11}{3755--3758}.
\PrintBackRefs{\CurrentBib}

\bibitem [\protect \citeauthoryear {%
Mishra%
\ \BBA {} Singh%
}{%
Mishra%
\ \BBA {} Singh%
}{%
{\protect \APACyear {2004}}%
}]{%
mishra2004long}
\APACinsertmetastar {%
mishra2004long}%
\begin{APACrefauthors}%
Mishra, S\BPBI K.%
\BCBT {}\ \BBA {} Singh, V\BPBI P.%
\end{APACrefauthors}%
\unskip\
\newblock
\APACrefYearMonthDay{2004}{}{}.
\newblock
{\BBOQ}\APACrefatitle {Long-term hydrological simulation based on the Soil Conservation Service curve number} {Long-term hydrological simulation based on the soil conservation service curve number}.{\BBCQ}
\newblock
\APACjournalVolNumPages{Hydrological Processes}{18}{7}{1291--1313}.
\PrintBackRefs{\CurrentBib}

\bibitem [\protect \citeauthoryear {%
Mishra%
\ \BBA {} Singh%
}{%
Mishra%
\ \BBA {} Singh%
}{%
{\protect \APACyear {2013}}%
}]{%
mishra2013soil}
\APACinsertmetastar {%
mishra2013soil}%
\begin{APACrefauthors}%
Mishra, S\BPBI K.%
\BCBT {}\ \BBA {} Singh, V\BPBI P.%
\end{APACrefauthors}%
\unskip\
\newblock
\APACrefYear{2013}.
\newblock
\APACrefbtitle {Soil conservation service curve number (SCS-CN) methodology} {Soil conservation service curve number (scs-cn) methodology}\ (\BVOL~42).
\newblock
\APACaddressPublisher{}{Springer Science \& Business Media}.
\PrintBackRefs{\CurrentBib}

\bibitem [\protect \citeauthoryear {%
Moore%
}{%
Moore%
}{%
{\protect \APACyear {1985}}%
}]{%
moore1985probability}
\APACinsertmetastar {%
moore1985probability}%
\begin{APACrefauthors}%
Moore, R.%
\end{APACrefauthors}%
\unskip\
\newblock
\APACrefYearMonthDay{1985}{}{}.
\newblock
{\BBOQ}\APACrefatitle {The probability-distributed principle and runoff production at point and basin scales} {The probability-distributed principle and runoff production at point and basin scales}.{\BBCQ}
\newblock
\APACjournalVolNumPages{Hydrological Sciences Journal}{30}{2}{273--297}.
\PrintBackRefs{\CurrentBib}

\bibitem [\protect \citeauthoryear {%
Pathiraja%
, Westra%
\BCBL {}\ \BBA {} Sharma%
}{%
Pathiraja%
\ \protect \BOthers {.}}{%
{\protect \APACyear {2012}}%
}]{%
pathiraja2012continuous}
\APACinsertmetastar {%
pathiraja2012continuous}%
\begin{APACrefauthors}%
Pathiraja, S.%
, Westra, S.%
\BCBL {}\ \BBA {} Sharma, A.%
\end{APACrefauthors}%
\unskip\
\newblock
\APACrefYearMonthDay{2012}{}{}.
\newblock
{\BBOQ}\APACrefatitle {Why continuous simulation? The role of antecedent moisture in design flood estimation} {Why continuous simulation? the role of antecedent moisture in design flood estimation}.{\BBCQ}
\newblock
\APACjournalVolNumPages{Water Resources Research}{48}{6}{}.
\PrintBackRefs{\CurrentBib}

\bibitem [\protect \citeauthoryear {%
Ponce%
\ \BBA {} Hawkins%
}{%
Ponce%
\ \BBA {} Hawkins%
}{%
{\protect \APACyear {1996}}%
}]{%
ponce1996runoff}
\APACinsertmetastar {%
ponce1996runoff}%
\begin{APACrefauthors}%
Ponce, V\BPBI M.%
\BCBT {}\ \BBA {} Hawkins, R\BPBI H.%
\end{APACrefauthors}%
\unskip\
\newblock
\APACrefYearMonthDay{1996}{}{}.
\newblock
{\BBOQ}\APACrefatitle {Runoff curve number: Has it reached maturity?} {Runoff curve number: Has it reached maturity?}{\BBCQ}
\newblock
\APACjournalVolNumPages{Journal of hydrologic engineering}{1}{1}{11--19}.
\PrintBackRefs{\CurrentBib}

\bibitem [\protect \citeauthoryear {%
Porporato%
, Daly%
\BCBL {}\ \BBA {} Rodriguez-Iturbe%
}{%
Porporato%
\ \protect \BOthers {.}}{%
{\protect \APACyear {2004}}%
}]{%
porporato2004soil}
\APACinsertmetastar {%
porporato2004soil}%
\begin{APACrefauthors}%
Porporato, A.%
, Daly, E.%
\BCBL {}\ \BBA {} Rodriguez-Iturbe, I.%
\end{APACrefauthors}%
\unskip\
\newblock
\APACrefYearMonthDay{2004}{}{}.
\newblock
{\BBOQ}\APACrefatitle {Soil water balance and ecosystem response to climate change} {Soil water balance and ecosystem response to climate change}.{\BBCQ}
\newblock
\APACjournalVolNumPages{The American Naturalist}{164}{5}{625--632}.
\PrintBackRefs{\CurrentBib}

\bibitem [\protect \citeauthoryear {%
Porporato%
, Laio%
, Ridolfi%
\BCBL {}\ \BBA {} Rodriguez-Iturbe%
}{%
Porporato%
\ \protect \BOthers {.}}{%
{\protect \APACyear {2001}}%
}]{%
porporato2001plants}
\APACinsertmetastar {%
porporato2001plants}%
\begin{APACrefauthors}%
Porporato, A.%
, Laio, F.%
, Ridolfi, L.%
\BCBL {}\ \BBA {} Rodriguez-Iturbe, I.%
\end{APACrefauthors}%
\unskip\
\newblock
\APACrefYearMonthDay{2001}{}{}.
\newblock
{\BBOQ}\APACrefatitle {Plants in water-controlled ecosystems: active role in hydrologic processes and response to water stress: III. Vegetation water stress} {Plants in water-controlled ecosystems: active role in hydrologic processes and response to water stress: Iii. vegetation water stress}.{\BBCQ}
\newblock
\APACjournalVolNumPages{Advances in Water Resources}{24}{7}{725--744}.
\PrintBackRefs{\CurrentBib}

\bibitem [\protect \citeauthoryear {%
Porporato%
\ \BBA {} Yin%
}{%
Porporato%
\ \BBA {} Yin%
}{%
{\protect \APACyear {2022}}%
}]{%
porporato2022ecohydrology}
\APACinsertmetastar {%
porporato2022ecohydrology}%
\begin{APACrefauthors}%
Porporato, A.%
\BCBT {}\ \BBA {} Yin, J.%
\end{APACrefauthors}%
\unskip\
\newblock
\APACrefYear{2022}.
\newblock
\APACrefbtitle {Ecohydrology: Dynamics of life and water in the critical zone} {Ecohydrology: Dynamics of life and water in the critical zone}.
\newblock
\APACaddressPublisher{}{Cambridge University Press}.
\PrintBackRefs{\CurrentBib}

\bibitem [\protect \citeauthoryear {%
Rallison%
\ \BBA {} Miller%
}{%
Rallison%
\ \BBA {} Miller%
}{%
{\protect \APACyear {1982}}%
}]{%
rallison1982past}
\APACinsertmetastar {%
rallison1982past}%
\begin{APACrefauthors}%
Rallison, R\BPBI E.%
\BCBT {}\ \BBA {} Miller, N.%
\end{APACrefauthors}%
\unskip\
\newblock
\APACrefYearMonthDay{1982}{}{}.
\newblock
{\BBOQ}\APACrefatitle {Past, present, and future {SCS} runoff procedure} {Past, present, and future {SCS} runoff procedure}.{\BBCQ}
\newblock
\BIn{} \APACrefbtitle {Rainfall-runoff relationship/proceedings, International Symposium on Rainfall-Runoff Modeling held May 18-21, 1981 at Mississippi State University, Mississippi State, Mississippi, USA/edited by VP Singh.} {Rainfall-runoff relationship/proceedings, international symposium on rainfall-runoff modeling held may 18-21, 1981 at mississippi state university, mississippi state, mississippi, usa/edited by vp singh.}
\PrintBackRefs{\CurrentBib}

\bibitem [\protect \citeauthoryear {%
Rigby%
\ \BBA {} Porporato%
}{%
Rigby%
\ \BBA {} Porporato%
}{%
{\protect \APACyear {2006}}%
}]{%
rigby2006simplified}
\APACinsertmetastar {%
rigby2006simplified}%
\begin{APACrefauthors}%
Rigby, J.%
\BCBT {}\ \BBA {} Porporato, A.%
\end{APACrefauthors}%
\unskip\
\newblock
\APACrefYearMonthDay{2006}{}{}.
\newblock
{\BBOQ}\APACrefatitle {Simplified stochastic soil-moisture models: a look at infiltration} {Simplified stochastic soil-moisture models: a look at infiltration}.{\BBCQ}
\newblock
\APACjournalVolNumPages{Hydrology and Earth System Sciences}{10}{}{861--871}.
\PrintBackRefs{\CurrentBib}

\bibitem [\protect \citeauthoryear {%
Rodr{\'\i}guez-Iturbe%
\ \BBA {} Porporato%
}{%
Rodr{\'\i}guez-Iturbe%
\ \BBA {} Porporato%
}{%
{\protect \APACyear {2004}}%
}]{%
rodrigueziturbe2004ecohydrology}
\APACinsertmetastar {%
rodrigueziturbe2004ecohydrology}%
\begin{APACrefauthors}%
Rodr{\'\i}guez-Iturbe, I.%
\BCBT {}\ \BBA {} Porporato, A.%
\end{APACrefauthors}%
\unskip\
\newblock
\APACrefYear{2004}.
\newblock
\APACrefbtitle {Ecohydrology of water-controlled ecosystems: soil moisture and plant dynamics} {Ecohydrology of water-controlled ecosystems: soil moisture and plant dynamics}.
\newblock
\APACaddressPublisher{}{Cambridge University Press}.
\PrintBackRefs{\CurrentBib}

\bibitem [\protect \citeauthoryear {%
Rodriguez-Iturbe%
, Porporato%
, Ridolfi%
, Isham%
\BCBL {}\ \BBA {} Cox%
}{%
Rodriguez-Iturbe%
\ \protect \BOthers {.}}{%
{\protect \APACyear {1999}}%
}]{%
rodrigueziturbe1999probabilistic}
\APACinsertmetastar {%
rodrigueziturbe1999probabilistic}%
\begin{APACrefauthors}%
Rodriguez-Iturbe, I.%
, Porporato, A.%
, Ridolfi, L.%
, Isham, V.%
\BCBL {}\ \BBA {} Cox, D\BPBI R.%
\end{APACrefauthors}%
\unskip\
\newblock
\APACrefYearMonthDay{1999}{}{}.
\newblock
{\BBOQ}\APACrefatitle {Probabilistic modelling of water balance at a point: The role of climate, soil and vegetation} {Probabilistic modelling of water balance at a point: The role of climate, soil and vegetation}.{\BBCQ}
\newblock
\APACjournalVolNumPages{Proceedings of the Royal Society of London. Series A: Mathematical, Physical and Engineering Sciences}{455}{1990}{3789--3805}.
\PrintBackRefs{\CurrentBib}

\bibitem [\protect \citeauthoryear {%
Semenova%
\ \BBA {} Beven%
}{%
Semenova%
\ \BBA {} Beven%
}{%
{\protect \APACyear {2015}}%
}]{%
semenova2015barriers}
\APACinsertmetastar {%
semenova2015barriers}%
\begin{APACrefauthors}%
Semenova, O.%
\BCBT {}\ \BBA {} Beven, K.%
\end{APACrefauthors}%
\unskip\
\newblock
\APACrefYearMonthDay{2015}{}{}.
\newblock
{\BBOQ}\APACrefatitle {Barriers to progress in distributed hydrological modelling} {Barriers to progress in distributed hydrological modelling}.{\BBCQ}
\newblock
\APACjournalVolNumPages{Hydrological Processes}{29}{8}{2074--2078}.
\PrintBackRefs{\CurrentBib}

\bibitem [\protect \citeauthoryear {%
Sidle%
\ \protect \BOthers {.}}{%
Sidle%
\ \protect \BOthers {.}}{%
{\protect \APACyear {2000}}%
}]{%
sidle2000stormflow}
\APACinsertmetastar {%
sidle2000stormflow}%
\begin{APACrefauthors}%
Sidle, R\BPBI C.%
, Tsuboyama, Y.%
, Noguchi, S.%
, Hosoda, I.%
, Fujieda, M.%
\BCBL {}\ \BBA {} Shimizu, T.%
\end{APACrefauthors}%
\unskip\
\newblock
\APACrefYearMonthDay{2000}{}{}.
\newblock
{\BBOQ}\APACrefatitle {Stormflow generation in steep forested headwaters: a linked hydrogeomorphic paradigm} {Stormflow generation in steep forested headwaters: a linked hydrogeomorphic paradigm}.{\BBCQ}
\newblock
\APACjournalVolNumPages{Hydrological Processes}{14}{3}{369--385}.
\PrintBackRefs{\CurrentBib}

\bibitem [\protect \citeauthoryear {%
N.~Tedela%
, McCutcheon%
, Rasmussen%
\BCBL {}\ \BBA {} Tollner%
}{%
N.~Tedela%
\ \protect \BOthers {.}}{%
{\protect \APACyear {2008}}%
}]{%
tedela2008evaluation}
\APACinsertmetastar {%
tedela2008evaluation}%
\begin{APACrefauthors}%
Tedela, N.%
, McCutcheon, S.%
, Rasmussen, T.%
\BCBL {}\ \BBA {} Tollner, W.%
\end{APACrefauthors}%
\unskip\
\newblock
\APACrefYearMonthDay{2008}{}{}.
\newblock
{\BBOQ}\APACrefatitle {Evaluation and Improvements of the Curve Number Method of Hydrological Analysis on Selected Forested Watersheds of Georgia} {Evaluation and improvements of the curve number method of hydrological analysis on selected forested watersheds of georgia}.{\BBCQ}
\newblock
\APACjournalVolNumPages{Report submitted to Georgia Water Resources Institute}{}{}{}.
\PrintBackRefs{\CurrentBib}

\bibitem [\protect \citeauthoryear {%
N\BPBI H.~Tedela%
\ \protect \BOthers {.}}{%
N\BPBI H.~Tedela%
\ \protect \BOthers {.}}{%
{\protect \APACyear {2011}}%
}]{%
tedela2011runoff}
\APACinsertmetastar {%
tedela2011runoff}%
\begin{APACrefauthors}%
Tedela, N\BPBI H.%
, McCutcheon, S\BPBI C.%
, Rasmussen, T\BPBI C.%
, Hawkins, R\BPBI H.%
, Swank, W\BPBI T.%
, Campbell, J\BPBI L.%
\BDBL {}Tollner, E\BPBI W.%
\end{APACrefauthors}%
\unskip\
\newblock
\APACrefYearMonthDay{2011}{}{}.
\newblock
{\BBOQ}\APACrefatitle {Runoff Curve Numbers for 10 small forested watersheds in the mountains of the Eastern United States} {Runoff curve numbers for 10 small forested watersheds in the mountains of the eastern united states}.{\BBCQ}
\newblock
\APACjournalVolNumPages{Journal of Hydrologic Engineering}{}{}{}.
\PrintBackRefs{\CurrentBib}

\bibitem [\protect \citeauthoryear {%
Thomas%
, Vogel%
, Kroll%
\BCBL {}\ \BBA {} Famiglietti%
}{%
Thomas%
\ \protect \BOthers {.}}{%
{\protect \APACyear {2013}}%
}]{%
thomas2013estimation}
\APACinsertmetastar {%
thomas2013estimation}%
\begin{APACrefauthors}%
Thomas, B\BPBI F.%
, Vogel, R\BPBI M.%
, Kroll, C\BPBI N.%
\BCBL {}\ \BBA {} Famiglietti, J\BPBI S.%
\end{APACrefauthors}%
\unskip\
\newblock
\APACrefYearMonthDay{2013}{}{}.
\newblock
{\BBOQ}\APACrefatitle {Estimation of the base flow recession constant under human interference} {Estimation of the base flow recession constant under human interference}.{\BBCQ}
\newblock
\APACjournalVolNumPages{Water Resources Research}{49}{11}{7366--7379}.
\PrintBackRefs{\CurrentBib}

\bibitem [\protect \citeauthoryear {%
{USDA National Resources Conservation Service}%
}{%
{USDA National Resources Conservation Service}%
}{%
{\protect \APACyear {2004}}%
}]{%
nrcs2004national}
\APACinsertmetastar {%
nrcs2004national}%
\begin{APACrefauthors}%
{USDA National Resources Conservation Service}.%
\end{APACrefauthors}%
\unskip\
\newblock
\APACrefYearMonthDay{2004}{}{}.
\newblock
{\BBOQ}\APACrefatitle {National Engineering Handbook: Part 630--Hydrology} {National engineering handbook: Part 630--hydrology}.{\BBCQ}
\newblock
\APACjournalVolNumPages{USDA Natural Resource Conservation Service: Washington, DC, USA}{}{}{}.
\PrintBackRefs{\CurrentBib}

\bibitem [\protect \citeauthoryear {%
Velpuri%
\ \BBA {} Senay%
}{%
Velpuri%
\ \BBA {} Senay%
}{%
{\protect \APACyear {2013}}%
}]{%
velpuri2013analysis}
\APACinsertmetastar {%
velpuri2013analysis}%
\begin{APACrefauthors}%
Velpuri, N.%
\BCBT {}\ \BBA {} Senay, G.%
\end{APACrefauthors}%
\unskip\
\newblock
\APACrefYearMonthDay{2013}{}{}.
\newblock
{\BBOQ}\APACrefatitle {Analysis of long-term trends (1950--2009) in precipitation, runoff and runoff coefficient in major urban watersheds in the United States} {Analysis of long-term trends (1950--2009) in precipitation, runoff and runoff coefficient in major urban watersheds in the united states}.{\BBCQ}
\newblock
\APACjournalVolNumPages{Environmental Research Letters}{8}{2}{024020}.
\PrintBackRefs{\CurrentBib}

\bibitem [\protect \citeauthoryear {%
Vico%
\ \BBA {} Porporato%
}{%
Vico%
\ \BBA {} Porporato%
}{%
{\protect \APACyear {2010}}%
}]{%
vico2010traditional}
\APACinsertmetastar {%
vico2010traditional}%
\begin{APACrefauthors}%
Vico, G.%
\BCBT {}\ \BBA {} Porporato, A.%
\end{APACrefauthors}%
\unskip\
\newblock
\APACrefYearMonthDay{2010}{}{}.
\newblock
{\BBOQ}\APACrefatitle {Traditional and microirrigation with stochastic soil moisture} {Traditional and microirrigation with stochastic soil moisture}.{\BBCQ}
\newblock
\APACjournalVolNumPages{Water Resources Research}{46}{3}{W03509}.
\PrintBackRefs{\CurrentBib}

\bibitem [\protect \citeauthoryear {%
Vico%
\ \BBA {} Porporato%
}{%
Vico%
\ \BBA {} Porporato%
}{%
{\protect \APACyear {2011}}%
{\protect \APACexlab {{\protect \BCnt {1}}}}}]{%
vico2011rainfed1}
\APACinsertmetastar {%
vico2011rainfed1}%
\begin{APACrefauthors}%
Vico, G.%
\BCBT {}\ \BBA {} Porporato, A.%
\end{APACrefauthors}%
\unskip\
\newblock
\APACrefYearMonthDay{2011{\protect \BCnt {1}}}{}{}.
\newblock
{\BBOQ}\APACrefatitle {From rainfed agriculture to stress-avoidance irrigation: I. A generalized irrigation scheme with stochastic soil moisture} {From rainfed agriculture to stress-avoidance irrigation: I. a generalized irrigation scheme with stochastic soil moisture}.{\BBCQ}
\newblock
\APACjournalVolNumPages{Advances in Water Resources}{34}{2}{263--271}.
\PrintBackRefs{\CurrentBib}

\bibitem [\protect \citeauthoryear {%
Vico%
\ \BBA {} Porporato%
}{%
Vico%
\ \BBA {} Porporato%
}{%
{\protect \APACyear {2011}}%
{\protect \APACexlab {{\protect \BCnt {2}}}}}]{%
vico2011rainfed2}
\APACinsertmetastar {%
vico2011rainfed2}%
\begin{APACrefauthors}%
Vico, G.%
\BCBT {}\ \BBA {} Porporato, A.%
\end{APACrefauthors}%
\unskip\
\newblock
\APACrefYearMonthDay{2011{\protect \BCnt {2}}}{}{}.
\newblock
{\BBOQ}\APACrefatitle {From rainfed agriculture to stress-avoidance irrigation: II. Sustainability, crop yield, and profitability} {From rainfed agriculture to stress-avoidance irrigation: Ii. sustainability, crop yield, and profitability}.{\BBCQ}
\newblock
\APACjournalVolNumPages{Advances in Water Resources}{34}{2}{272--281}.
\PrintBackRefs{\CurrentBib}

\bibitem [\protect \citeauthoryear {%
Wang%
\ \protect \BOthers {.}}{%
Wang%
\ \protect \BOthers {.}}{%
{\protect \APACyear {2019}}%
}]{%
wang2019soil}
\APACinsertmetastar {%
wang2019soil}%
\begin{APACrefauthors}%
Wang, Y.%
, Jiang, R.%
, Xie, J.%
, Zhao, Y.%
, Yan, D.%
\BCBL {}\ \BBA {} Yang, S.%
\end{APACrefauthors}%
\unskip\
\newblock
\APACrefYearMonthDay{2019}{}{}.
\newblock
{\BBOQ}\APACrefatitle {Soil and water assessment tool (SWAT) model: A systemic review} {Soil and water assessment tool (swat) model: A systemic review}.{\BBCQ}
\newblock
\APACjournalVolNumPages{Journal of Coastal Research}{93}{SI}{22--30}.
\PrintBackRefs{\CurrentBib}

\bibitem [\protect \citeauthoryear {%
Wasko%
, Nathan%
\BCBL {}\ \BBA {} Peel%
}{%
Wasko%
\ \protect \BOthers {.}}{%
{\protect \APACyear {2020}}%
}]{%
wasko2020changes}
\APACinsertmetastar {%
wasko2020changes}%
\begin{APACrefauthors}%
Wasko, C.%
, Nathan, R.%
\BCBL {}\ \BBA {} Peel, M\BPBI C.%
\end{APACrefauthors}%
\unskip\
\newblock
\APACrefYearMonthDay{2020}{}{}.
\newblock
{\BBOQ}\APACrefatitle {Changes in antecedent soil moisture modulate flood seasonality in a changing climate} {Changes in antecedent soil moisture modulate flood seasonality in a changing climate}.{\BBCQ}
\newblock
\APACjournalVolNumPages{Water Resources Research}{56}{3}{e2019WR026300}.
\PrintBackRefs{\CurrentBib}

\bibitem [\protect \citeauthoryear {%
Wood%
, Lettenmaier%
\BCBL {}\ \BBA {} Zartarian%
}{%
Wood%
\ \protect \BOthers {.}}{%
{\protect \APACyear {1992}}%
}]{%
wood1992land}
\APACinsertmetastar {%
wood1992land}%
\begin{APACrefauthors}%
Wood, E\BPBI F.%
, Lettenmaier, D\BPBI P.%
\BCBL {}\ \BBA {} Zartarian, V\BPBI G.%
\end{APACrefauthors}%
\unskip\
\newblock
\APACrefYearMonthDay{1992}{}{}.
\newblock
{\BBOQ}\APACrefatitle {A land-surface hydrology parameterization with subgrid variability for general circulation models} {A land-surface hydrology parameterization with subgrid variability for general circulation models}.{\BBCQ}
\newblock
\APACjournalVolNumPages{Journal of Geophysical Research: Atmospheres}{97}{D3}{2717--2728}.
\PrintBackRefs{\CurrentBib}

\bibitem [\protect \citeauthoryear {%
Xie%
\ \protect \BOthers {.}}{%
Xie%
\ \protect \BOthers {.}}{%
{\protect \APACyear {2020}}%
}]{%
xie2020evaluation}
\APACinsertmetastar {%
xie2020evaluation}%
\begin{APACrefauthors}%
Xie, J.%
, Liu, X.%
, Wang, K.%
, Yang, T.%
, Liang, K.%
\BCBL {}\ \BBA {} Liu, C.%
\end{APACrefauthors}%
\unskip\
\newblock
\APACrefYearMonthDay{2020}{}{}.
\newblock
{\BBOQ}\APACrefatitle {Evaluation of typical methods for baseflow separation in the contiguous United States} {Evaluation of typical methods for baseflow separation in the contiguous united states}.{\BBCQ}
\newblock
\APACjournalVolNumPages{Journal of Hydrology}{583}{}{124628}.
\PrintBackRefs{\CurrentBib}

\bibitem [\protect \citeauthoryear {%
Yin%
\ \protect \BOthers {.}}{%
Yin%
\ \protect \BOthers {.}}{%
{\protect \APACyear {2018}}%
}]{%
yin2018large}
\APACinsertmetastar {%
yin2018large}%
\begin{APACrefauthors}%
Yin, J.%
, Gentine, P.%
, Zhou, S.%
, Sullivan, S\BPBI C.%
, Wang, R.%
, Zhang, Y.%
\BCBL {}\ \BBA {} Guo, S.%
\end{APACrefauthors}%
\unskip\
\newblock
\APACrefYearMonthDay{2018}{}{}.
\newblock
{\BBOQ}\APACrefatitle {Large increase in global storm runoff extremes driven by climate and anthropogenic changes} {Large increase in global storm runoff extremes driven by climate and anthropogenic changes}.{\BBCQ}
\newblock
\APACjournalVolNumPages{Nature communications}{9}{1}{4389}.
\PrintBackRefs{\CurrentBib}

\bibitem [\protect \citeauthoryear {%
Yuan%
, Mitchell%
, Hirschi%
\BCBL {}\ \BBA {} Cooke%
}{%
Yuan%
\ \protect \BOthers {.}}{%
{\protect \APACyear {2001}}%
}]{%
yuan2001modified}
\APACinsertmetastar {%
yuan2001modified}%
\begin{APACrefauthors}%
Yuan, Y.%
, Mitchell, J.%
, Hirschi, M.%
\BCBL {}\ \BBA {} Cooke, R.%
\end{APACrefauthors}%
\unskip\
\newblock
\APACrefYearMonthDay{2001}{}{}.
\newblock
{\BBOQ}\APACrefatitle {Modified {SCS} curve number method for predicting subsurface drainage flow} {Modified {SCS} curve number method for predicting subsurface drainage flow}.{\BBCQ}
\newblock
\APACjournalVolNumPages{Transactions of the ASAE}{44}{6}{1673--1682}.
\PrintBackRefs{\CurrentBib}

\bibitem [\protect \citeauthoryear {%
Zehe%
\ \BBA {} Bl{\"o}schl%
}{%
Zehe%
\ \BBA {} Bl{\"o}schl%
}{%
{\protect \APACyear {2004}}%
}]{%
zehe2004predictability}
\APACinsertmetastar {%
zehe2004predictability}%
\begin{APACrefauthors}%
Zehe, E.%
\BCBT {}\ \BBA {} Bl{\"o}schl, G.%
\end{APACrefauthors}%
\unskip\
\newblock
\APACrefYearMonthDay{2004}{}{}.
\newblock
{\BBOQ}\APACrefatitle {Predictability of hydrologic response at the plot and catchment scales: Role of initial conditions} {Predictability of hydrologic response at the plot and catchment scales: Role of initial conditions}.{\BBCQ}
\newblock
\APACjournalVolNumPages{Water Resources Research}{40}{10}{}.
\PrintBackRefs{\CurrentBib}

\end{thebibliography}

\end{document}